\documentclass[10pt]{article}
\usepackage[margin=0.6in,columnsep=0.35in]{geometry}
\usepackage[style=numeric-comp,sorting=none,giveninits=true]{biblatex}
\let\citep\autocite%
\let\citet\textcite%
\AtBeginBibliography{\footnotesize}

\usepackage{microtype}
\usepackage{graphicx}
\usepackage{caption}
\usepackage{subcaption}
\usepackage{booktabs}
\usepackage[dvipsnames]{xcolor}
\usepackage{xr-hyper}
\PassOptionsToPackage{hyphens}{url}
\usepackage[colorlinks,citecolor=Gray,urlcolor=Gray,linkcolor=Blue,pdftex]{hyperref}
\usepackage{makecell}
\usepackage{multirow}
\usepackage{chemformula}
\usepackage{braket}

\usepackage{dcolumn}
\usepackage{amsmath}
\usepackage{amssymb}
\usepackage{mathtools}
\usepackage{amsthm}
\usepackage{bm}
\usepackage{siunitx}
\usepackage[tiny]{titlesec}
\usepackage[
    justification=justified,
    singlelinecheck=true,
]{caption}
\DeclareCaptionLabelSeparator{pipe}{ $|$ }
\newcommand{\rv}{\mathbf{r}}
\newcommand{\Rv}{\mathbf{R}}
\newcommand{\Av}{\mathbf{A}}
\newcommand{\xv}{\mathbf{x}}

\newcommand{\thetav}{{\boldsymbol{\theta}}}

\newcommand{\E}{\mathbb{E}}
\newcommand{\nelec}{E}
\newcommand{\norb}{O}
\newcommand{\nnuc}{N}
\newcommand{\elecrep}{T}
\newcommand{\orbrep}{Q}
\newcommand{\energy}{\mathcal{E}}
\newcommand{\molconf}{\mathbf{M}}%

\hypersetup{pdftitle={An ab initio foundation model of wavefunctions that accurately describes chemical bond breaking}}

\makeatletter
\def\blfootnote{\xdef\@thefnmark{}\@footnotetext}
\makeatother

\begin{document}

\begin{center}
~\\[0.5em]
{\LARGE\bfseries
    An ab initio foundation model of wavefunctions that \\[0.3em]
    accurately describes chemical bond breaking
}
\vspace{1.5em}

\newcommand{\maybecomma}{,}
\renewcommand{\author}[2][]{#2${}^{\text{#1}}$\maybecomma}

\author[\(\dagger\),1,*]{Adam Foster}
\author[\(\dagger\),1,2]{Zeno Sch\"{a}tzle}
\author[\(\dagger\),1,2]{P. Bern\'{a}t Szab\'{o}}
\author[\(\dagger\),1,a]{Lixue Cheng}
\author[1,b]{Jonas K\"{o}hler} \\
\author[1,c]{Gino Cassella}
\author[1,b]{Nicholas Gao}
\author[1,d]{Jiawei Li}
\author[1,2,*]{Frank No\'{e}}\renewcommand{\maybecomma}{}
\author[1,*]{Jan Hermann}
\vspace{1em}

$^{\dagger}$These authors contributed equally to this work.
\vspace{0.5em}

{
\itshape
$^1$Microsoft Research AI for Science\hspace{1.5em}
$^2$Freie Universit\"{a}t Berlin
}
\vspace{0.5em}

$^*$Email: \texttt{\small\{\href{mailto:adam.e.foster@microsoft.com}{adam.e.foster},\href{mailto:franknoe@microsoft.com}{franknoe},\href{mailto:jan.hermann@microsoft.com}{jan.hermann}\}@microsoft.com}
\vspace{1.5em}

\begin{minipage}{0.85\linewidth}
\small
\paragraph{Abstract}
Reliable description of bond breaking remains a major challenge for quantum chemistry due to the multireference character of the electronic structure in dissociating species.
Multireference methods in particular suffer from large computational cost, which under the normal paradigm has to be paid anew for each system at a full price, ignoring commonalities in electronic structure across molecules.
Quantum Monte Carlo with deep neural networks uniquely offers to exploit such commonalities by pretraining transferable wavefunction models, but all such attempts were so far limited in scope.
Here, we bring this paradigm to fruition with Orbformer, a transferable wavefunction model pretrained on 22,000 equilibrium and dissociating structures that can be fine-tuned on unseen molecules reaching an accuracy--cost ratio rivalling classical multireference methods.
On established benchmarks as well as more challenging bond dissociations and Diels--Alder reactions, Orbformer is the only method that consistently converges to chemical accuracy (1\,kcal/mol).
This work turns the idea of amortizing the cost of solving the Schrödinger equation over many molecules into a practical approach in quantum chemistry.
\end{minipage}
\end{center}
\vspace{1em}

\blfootnote{
\hspace{-1em}
$^\mathrm{a}$Current address: \textit{The Hong Kong University of Science and Technology}\hspace{1em}
$^\mathrm{b}$Current address: \textit{CuspAI}\hspace{1em}
$^\mathrm{c}$Current address: \textit{Imperial College London}\hspace{1em}
$^\mathrm{d}$Current address: \textit{Tsinghua University}

}

\section{Introduction}
\label{sec:introduction}

A major impediment to predicting and optimizing the properties of molecules and materials in silico is the computational cost of accurate electronic structure calculations.
Ab initio methods that explicitly represent the many-body electronic wavefunction \citep{schrodinger1926undulatory} are the most accurate,  %
but also the most computationally demanding.
Further complicating the picture is the dividing line between weakly and strongly correlated systems. 
For some systems, e.g.~most molecules close to their equilibrium geometry, the electronic wavefunction is weakly correlated, allowing the use of `single-reference' methods such as coupled cluster \citep{KucharskiJCP98} to compute the wavefunction to very high accuracy.
Yet the weakly correlated systems represent only an island in the space of all chemistry and one whose exact boundary is not precisely known. %
For strongly correlated systems, there is no de facto standard wavefunction method, and calculations often require expert study and individual tuning for each system.
Worse still, many multireference methods face a more unfavourable cost scaling than their single-reference counterparts.
A scalable method for multireference calculations would therefore be of immense value, not only for making direct predictions, but also for calibrating faster and more approximate methods to handle strongly correlated phenomena such as many chemical reactions and much of transition-metal chemistry.

Ab initio methods are called such because they do not use any information other than the first-principles electronic Schrödinger equation and a choice of numerical approximations.
But they can also be characterized by each single calculation being done from scratch, with no computation shared between calculations on similar molecules, despite chemistry owning much of its success to the recognition of recurring patterns in the electronic structure of molecules.
The key insight is that these two qualities can be decoupled, and this can be used to drastically lower the cost of ab initio methods.
A method derived exclusively from first principles can still share computation between similar molecules, not only within simultaneous calculations but also over time, effectively storing and reusing recurring computation patterns.
One might worry that the results could be then affected by the choice of such molecules, but this can be rendered moot if the sharing affects only the rate at which the method arrives at a solution, not the solution itself.
It is this shift in paradigm for ab initio methods that is the topic of this work.
This new paradigm is also not entirely without precedent---many traditional methods are iterative in nature and use various initial guesses, but these never go beyond the most rudimentary consideration of electronic structure and are typically built with heuristics rather than from first principles.
In contrast, the kind of foundation model considered here is fully ab initio, has the same complexity as any final solution it yields, and even before any computation is done on a new molecule, it already encodes its fully correlated electronic structure at a semi-quantitative level.
Throughout this work, we use the term foundation model in the specific sense of a single wavefunction model that is pretrained across a broad range of molecules and then serves as a reusable starting point that is adapted to each new system by fine-tuning, rather than in the looser sense of an all-purpose model that is applied to any task without further training.

How can such foundation models be constructed?
This challenge has been picked up by
a new entrant on the quantum chemistry scene, deep quantum Monte Carlo \citep{hermann2023ab} (deep QMC), which represents the wavefunction with a deep neural network and trains it towards the ground state by the principle of energy minimization. %
The earliest work in this field for molecular systems  \citep{hermann2020deep,pfau2020ab} achieved essentially exact solutions on small molecules.
Various new network architectures \citep{spencer_better_2020,glehn2022self,gerard_2022,SchatzleJCP23,lin_explicitly_2023} and  extensions \citep{gao2021abinitio,li_fermionic_2022,wilson2023neural,cassella2023discovering,kim2024neural,li2024computational,szabo2024improved,pfau2024accurate,cheng2025highly} have since been proposed.
Successful applications to strongly correlated molecules have been particularly encouraging \citep{scherbela_accurate_2025}.
Whilst the theoretical cost scaling of this approach with system size is favourable, its large prefactor has prevented wider adoption for practical calculations.
But thanks to its reliance on deep neural networks, deep QMC is ideally suited to profit from the new paradigm of computation sharing and cost amortization between molecules.
Recent work has explored such new ideas by employing parameter sharing to capture common features in the wavefunctions corresponding to different molecules and geometries \citep{scherbela2022solving, gao2021abinitio, gao2023generalizing, gao2024neural, schatzle2025ab, linteau2025universal}, as well as pretraining a wavefunction model on up to 700 molecular structures before fine-tuning \citep{scherbela2023variational,scherbela2024towards,rende2025foundation}.
However, we argue that a critical threshold of scale in this endeavour has not yet been crossed.
Achieving practical generalization needs scale in the training throughput, size and diversity of the molecular dataset covered, and an architecture that can properly make use of this training signal.
With a view to establishing deep QMC as a practical method, one must also take great care when comparing with existing quantum chemistry methods.
We believe that the cost--accuracy trade-off must be assessed first and foremost, and that deep QMC has the greatest potential to bring value to quantum chemistry in the area of multireference problems that are challenging for established methods.

In this work, we introduce Orbformer, a neural network wavefunction ansatz.
Orbformer operates across molecules of different sizes, compositions, and geometries, but can attain or surpass the same energy accuracy as state-of-the-art single-point ans\"{a}tze \citep{glehn2022self}.
Orbformer is designed to learn composable features that generalize over different molecules. %
Modifications of the sampling and variational training procedures allow pretraining a model across 22,000 molecular structures with high geometric and compositional diversity.
Orbformer achieves excellent agreement with experimental results on transition states of a Diels--Alder reaction.
Compared to classical quantum chemistry methods including DFT, NEVPT2, MRCI, and MRCC on five bond dissociation curves, Orbformer demonstrates favourable trade-off between cost and accuracy, has a built-in indicator of uncertainty in the form of the variance of the local energies, and can be systematically improved with more fine-tuning, whereas converging classical methods is not always straightforward.
Compared to earlier deep QMC approaches,
our innovations reduce the cost to reach chemical accuracy by about two orders of magnitude with respect to best single-point calculations, and result in Orbformer being more accurate and benefiting more from pretraining with respect to earlier chemically transferable ans\"{a}tze.
The model learns recognisable features of electronic structure directly from the variational principle, such as the existence of exactly two localized `core' orbitals per carbon atom, without these being hardcoded into the model architecture.
Orbformer, though by no means the end of the story, marks significant progress towards a model that generalizes the electronic structure of molecules.
Given its favourable cost--accuracy trade-off for multireference problems, Orbformer is a useful tool for quantum chemistry applications today.

\section{Results}
\label{sec:results}
\subsection{Orbformer: a foundation model for quantum chemistry}
\label{sec:orbformer}

\begin{figure*}[ht]
    \centering
    \includegraphics[width=\linewidth,clip]{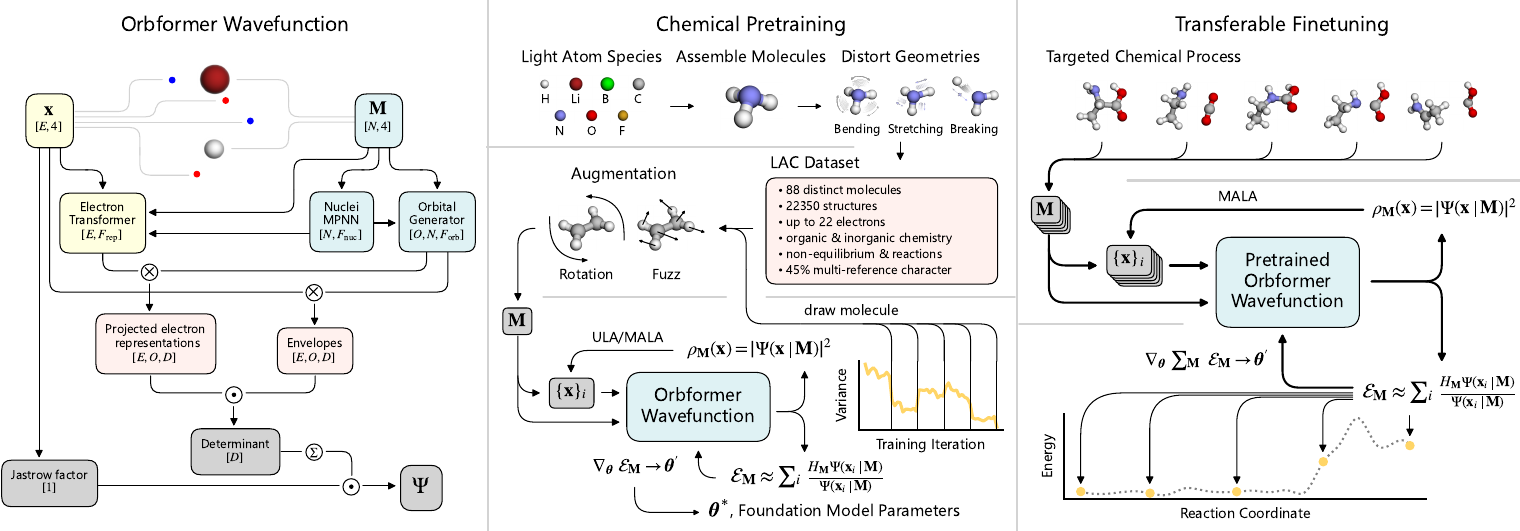}
    \caption{\textbf{Orbformer pretraining and fine-tuning.} The Orbformer wavefunction takes nuclear positions and charges, $\molconf$, and electron positions and spins, $\xv$, as inputs to generate generalized Slater matrices.
    Linear combinations of these generalized Slater determinants, combined with an electron--electron Jastrow factor, give a chemically transferable wave function, $\Psi$, that can generalize across systems of varying size. Orbformer is pretrained on a diverse dataset of small molecules and fragments, yielding a foundation model for non-equilibrium quantum chemistry. To achieve accuracies of 1 kcal/mol, the pretrained Orbformer model is then jointly fine-tuned on a process-specific dataset. Finally, the energy and other observables are evaluated for each of the involved molecular geometries.}
    \label{fig:summary}
\end{figure*}

We introduce Orbformer, a chemically transferable wavefunction ansatz that simultaneously approximates solutions to the electronic Schrödinger equation for a wide range of molecules.
The model accepts as inputs a molecular configuration, $\molconf$, and the spatial-spin co-ordinates of electrons, $\xv$.
Treating $\molconf$ as an input to the wavefunction network, rather than a parameter describing an optimization problem, is at the heart of our approach to sharing computation between similar molecules, and is to the best of our knowledge an option unique to deep QMC \citep{gao2023generalizing,scherbela2024towards}.
Orbformer is constructed  from a Jastrow factor and a sum of generalized Slater determinants
\begin{equation}
    \label{eq:orbformer-simplified}
    \Psi(\xv \mid \molconf) = e^{J(\xv)}\sum_{d} \det\left[ A^d (\xv \mid \molconf) \right]
\end{equation}
ensuring that, for every $\molconf$, the model as a function of $\xv$ is a valid anti-symmetric wavefunction.

The major components of the Orbformer ansatz are: an Electron Transformer, which processes electron features through a sequence of attention layers; and the Orbital Generator, which dynamically creates generalized orbitals for $\molconf$, each consisting of an envelope and a projection of the electron representations (Fig.~\ref{fig:summary} left).
A guiding principle in the design of the ansatz is locality---the interactions between any two particles are carefully crafted to decay with increasing distance and the Orbital Generator is constrained to produce localized orbitals.
We show that this facilitates the re-use of orbitals when similar chemical environments arise in different molecules.
Full details of our architecture are given in Sec.~\ref{sec:methods:architecture}.

We train the Orbformer wavefunction 
to minimize the expectation value of the Hamiltonian operator
\begin{equation}
    \E_{\molconf \sim p_\text{train}(\molconf), \xv\sim |\Psi(\xv\mid \molconf)|^2} \left[ \frac{\hat{H}_\molconf \Psi(\xv \mid \molconf)}{\Psi(\xv \mid \molconf)} \right],
\end{equation}
where the expectation is taken over molecular configurations sampled from a training distribution and electron co-ordinates sampled from the unnormalized distribution defined by the current model. %
This self-generative scheme does not require any external input data apart from the choice of molecular training distribution.
Our set-up generalizes standard variational Monte Carlo (VMC) by addition of an outer expectation over molecular configurations.

\begin{table*}[t]
    \centering
    \caption{Phases of Orbformer chemical pretraining using the Light Atom Curriculum (LAC) dataset.}
    \label{tab:curriculum-main}
    \resizebox{\textwidth}{!}{
    \begin{tabular}{rlrrrrr}
      Phase & Data & Steps &  $\molconf$ batch size & Electrons per $\molconf$ & A100 hours (est.) \\
      \hline
      1a  & LAC $\le 10$ electrons, bend \& stretch geometries & 200k & 8 & 1024 & 320 \\
      1b  & LAC $\le 10$ electrons, all geometries & 200k & 8 & 1024 & 320 \\
      2   & All LAC   & 400k & 16 & 1024 & 4000 \\
    \end{tabular}
    }
\end{table*}

Orbformer's amortization of computational cost arises from the combination of chemical pretraining and transferable fine-tuning.
During pretraining, we present the model with an extremely diverse set of molecules and train variationally to encode the electronic structure of a swathe of chemical space to a semi-quantitative level.
We specifically created a pretraining dataset of 22,350 molecular configurations
with up to 24 electrons composed of \ch{H}, \ch{Li}, \ch{B}, \ch{C}, \ch{N}, \ch{O}, \ch{F} in various geometric configurations, as illustrated in Fig.~\ref{fig:summary} (centre).
This Light Atom Curriculum (LAC) dataset is specifically designed to involve multireference out-of-equilibrium geometries.
To maximize speed and improve stability, we pretrain in three phases of increasing size/complexity, as detailed in Table~\ref{tab:curriculum-main}. 
We introduce innovations in the deep QMC training protocol to aid efficient, large-scale training, which we detail in Sec.~\ref{sec:method}.
The same set of model parameters can be transferred from smaller to larger systems, and there is no baked-in maximum number of particles. In this paper, our largest pretraining system had 24 electrons, whereas the largest fine-tuning system had 106 electrons.

To attain high-accuracy wavefunctions, we start from the pretrained model, and fine-tune to convergence. 
We do not resort to separate fine-tuning for every structure; instead, we group target structures together and fine-tune them using a single set of model parameters (Fig.~\ref{fig:summary} right). 
For example, we routinely fine-tune all geometries with the same set of nuclei together. 
The amortization from this approach can lead to a cost reduction proportional to the number of geometries, over and above the benefits of pretraining.

\subsection{Diels--Alder reaction pathways}
\label{sec:diels-alder}

\begin{figure*}[ht]
    \centering
    \includegraphics[width=0.88\linewidth]{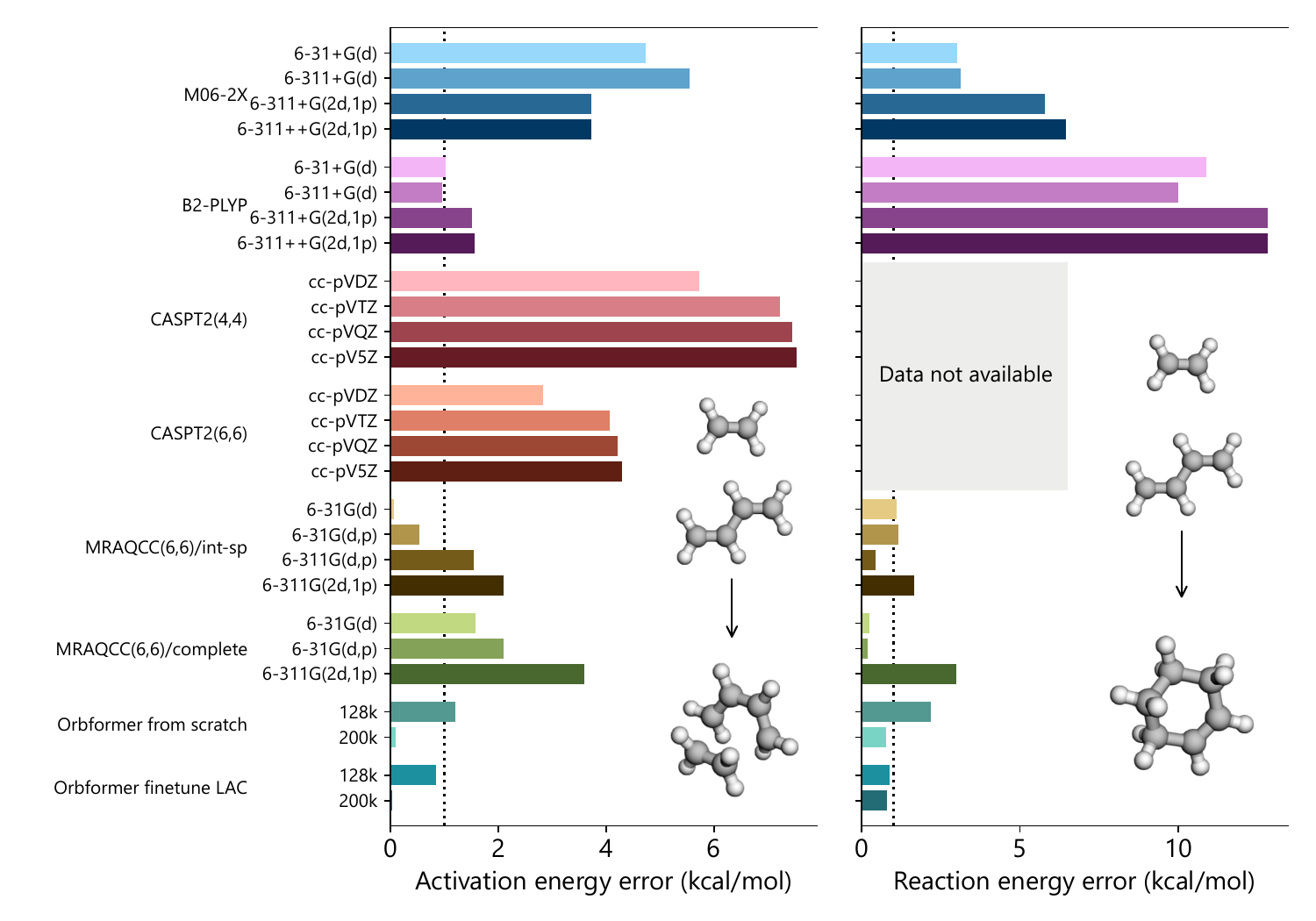}
    \caption{\textbf{Accuracy for activation and reaction energies of the iconic Diels--Alder reaction.}
    Activation energies use the transition state of the concerted pathway.
    The experimental values are computed from the forward and backward activation enthalpies measured by Rowley \cite{rowley1951} and Uchiyama \cite{uchiyama1964thermal}, respectively, applying zero-point and thermal corrections computed at the B3LYP/6-31G** level \cite{lischka2004}.
    The M06-2X and B2-PLYP density functional results are taken from the work of Cui et al.~\citep{cui2014thorough}, complete active space second order perturbation theory values are obtained from Scarborough \citep{scarborough2015active} but are not published for the reaction energy, while multireference coupled cluster results were computed by Lischka and coworkers \citep{lischka2004}, from where the molecular geometries utilized for the Orbformer calculations are also taken.
    The colour gradients represent results obtained with increasingly large single-particle basis sets or QMC steps, as applicable.
    The multireference coupled cluster results employed an active space composed of six orbitals and six electrons, with two sets of configuration state functions: a complete one, and a smaller one built by applying the interacting space restriction from reference configurations with the same symmetry as the electronic state \citep{lischka2004}. 
    }
    \label{fig:diels-alder-activation}
\end{figure*}

We compare Orbformer predictions to experimental data on a practically relevant chemical problem. 
The task of characterizing the mechanism of the iconic Diels--Alder reaction \cite{diels1928synthesen} (cycloaddition of ethene and butadiene) is chosen for this purpose, as its exact mechanism was subject to debate until recently \cite{cui2014thorough,scarborough2015active}, and its accurate description requires careful treatment of both static and dynamic correlation \cite{scarborough2015active,lischka2004}.

Two major pathways have been identified along which the reaction might proceed: a concerted one where the two bonds between the reactants form synchronously, and a two-step mechanism involving biradical intermediates, where the bonds form sequentially \cite{cui2014thorough}.
Using the concerted transition state, Orbformer converges to almost perfect agreement with the experimental activation energy (Fig.~\ref{fig:diels-alder-activation}).
Chemical pretraining improves the convergence rate, despite the transition state being well outside of the pretraining distribution at 46 electrons and involving the simultaneous breaking of two bonds.
Of the traditional wavefunction based methods, only multireference coupled cluster achieves results agreeing with experiment, but only if a small basis set or an artificially restricted set of configuration state functions are employed.
For the two-step pathway, Orbformer predicts an activation energy that is about 9.9 kcal/mol higher than the experimentally measured activation energy (see Supplementary Note~\ref{sec:app:additional-results:da-results} for detailed results) clearly indicating that this is not the favoured mechanism.
Our predicted energy difference between the two pathways is in good agreement with the current theoretical and experimental consensus \cite{cui2014thorough,uchiyama1964thermal}.
For the overall reaction energy, we also found that Orbformer reliably converges to within 1 kcal/mol of the experimental value. For this calculation, where strong correlation is expected to be less important, classical methods fare better, but still exhibit a troubling sensitivity to the choice of basis set. 

These results show that Orbformer is a reliable and highly accurate method for dealing with transition state calculations in realistic settings. For additional benchmarking against experiments, we compared Orbformer predictions against an experimental reference for \ch{N2} dissociation (Supplementary Note~\ref{sec:app:additional-results:n2}) and against experimental atomization energies of small molecules (Supplementary Note~\ref{sec:app:additional-results:atomization}). In both cases, Orbformer matches extremely well with experimental references.

\begin{figure*}[t]
    \centering
    \includegraphics[width=\textwidth]{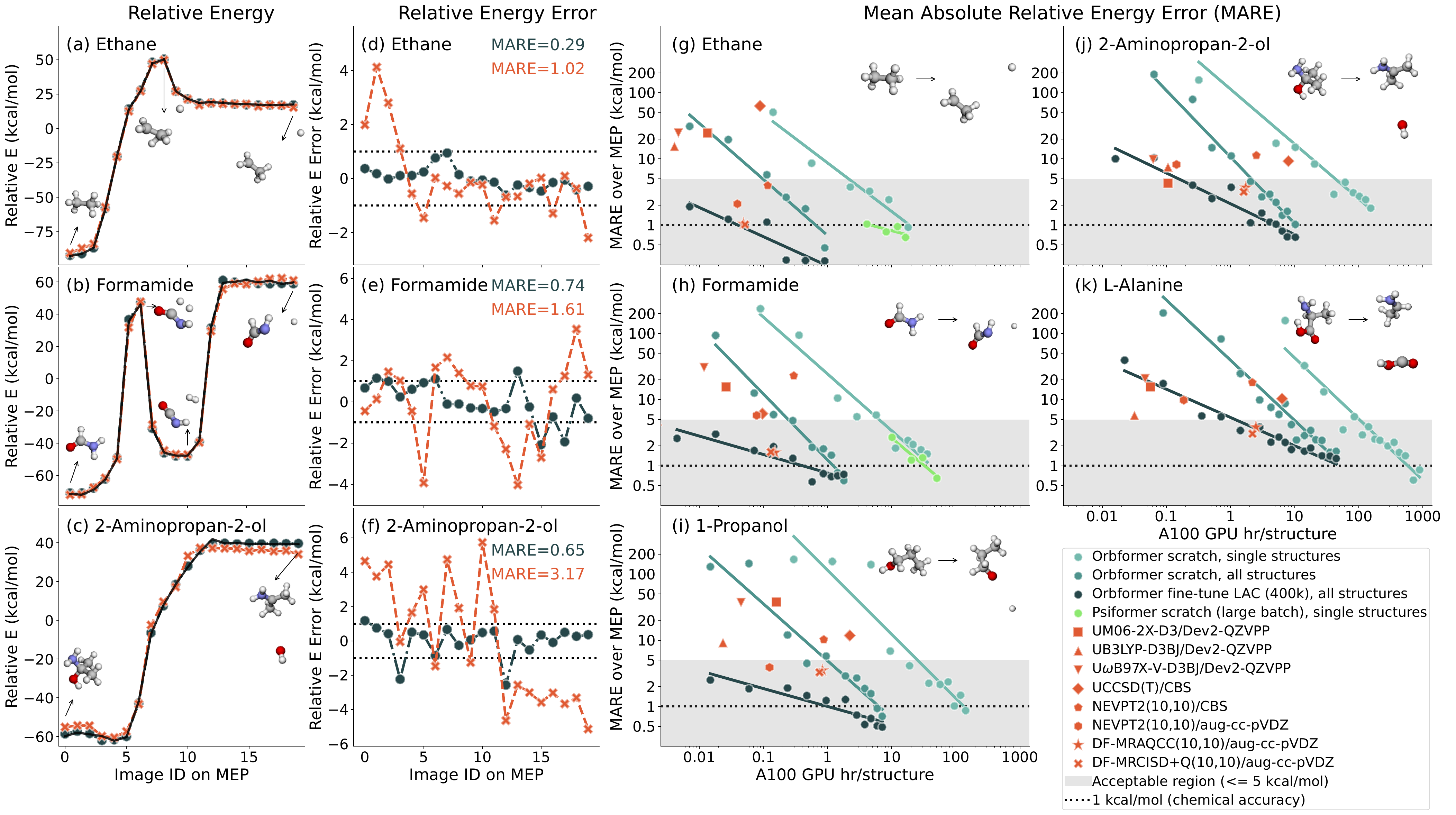}
    \caption{\textbf{Accuracy on minimum energy paths (MEPs) of five molecular dissociations.} The colour and marker for each method are the same across all panels (a)--(k) and are defined in the shared key. Left column: Reaction profile plotted as mean-centered relative energy for (a)~Ethane (b)~Formamide  and (c)~2-Aminopropan-2-ol. The solid black line shows the reference relative energies; the other two curves show the longest fine-tuned Orbformer prediction (Orbformer fine-tune LAC (400k)) and the DF-MRCISD+Q(10,10)/aug-cc-pVDZ result.
    Middle column: Relative energy error of each structure in the corresponding MEPs for (d)~Ethane (e)~Formamide  and (f)~2-Aminopropan-2-ol. 
    Panels (a)--(f) together show how the mean absolute error in relative energy is calculated across the reaction curve.
    Right column:  
    Mean absolute error in relative energy for a wide range of computational methods for (g)~Ethane (h)~Formamide (i)~1-Propanol (j)~2-Aminopropan-2-ol and (k)~L-Alanine. 
    The circles in blue/green colours with best fit lines represent results from different deep QMC variants. Traditional electronic structure theories are shown in orange with different marker symbols indicating the precise method used. The dashed line indicates chemical accuracy of 1 kcal/mol. The shaded area represents an acceptable accuracy region of 5 kcal/mol.  
    For a method that does not run on a GPU, we convert the computational cost using 1 GPU hour = 68.6 CPU core hours, computed from the power ratings of an A100 GPU and an AMD EPYC 7763 CPU (cf.~Supplementary Note~\ref{sec:app:bbmep_dataset}).
    Where a method uses multiple CPU cores or multiple GPU devices, this is accounted for in the cost.
    }
    \label{fig:bbmep_result}
\end{figure*}

\subsection{Cost--accuracy trade-offs of high accuracy multireference calculations on bond breaking}
\label{sec:bbmep}

For a more detailed examination of the cost--accuracy trade-off of Orbformer, we investigate relative energy errors at 20 points along a bond dissociation pathway in five different molecules with up to 48 electrons.
Fig.~\ref{fig:bbmep_result} (left column) illustrates some of the chosen dissociations, which are intended to give good coverage of the strongly correlated intermediate region.
For such detailed coverage of the potential energy surface, experimental references are not available.
Reference energies for this experiment were therefore obtained from independent, single-point deep QMC calculations run at very high cost to convergence, using a deliberately different and more expensive setup than the Orbformer calculations under test.
We validated this protocol against an established single-reference benchmark by comparing the difference in energy between the two dissociation endpoints (the subset of structures where single-reference methods are reliable) to literature (RO)CBS-QB3 \cite{BSE49} dissociation energies.
We found that the dissociation energies from our reference QMC protocol matched (RO)CBS-QB3 with a tolerance of 1 kcal/mol for all five dissociation curves (cf.~Supplementary Note~\ref{sec:app:bbmep_dataset}, Supplementary Tables~\ref{tab:bbmep_ref} and~\ref{tab:bbmep_dissociation} for further details).
Throughout this section, accuracy is judged with respect to these reference energies.

For each candidate method, we compute the relative energy error to the reference by removing any fixed offset between the energy profiles, and calculating the mean absolute relative error (MARE, defined in Supplementary Note~\ref{sec:metrics}) across all geometries (Fig.~\ref{fig:bbmep_result} middle column).
Comparison is made to the MRCISD, MRAQCC, NEVPT2, and CCSD(T) wavefunction methods, and several DFT functionals as these all have reasonable software support to treat systems with up to 48 electrons.
For the computational cost, we measure the machine hours consumed, and use the maximum power ratings of the respective devices as our conversion factor between CPU and GPU hours.

It is apparent  in each of Fig.~\ref{fig:bbmep_result}~(g)--(k) that Orbformer fine-tuned from LAC pretraining is either on, or significantly ahead of, the Pareto front formed by the set of orange points, i.e.~from cheaper density functional methods and increasingly costly complete active space-based wavefunction approaches. 
Even more striking is the fact that Orbformer exhibits a strict monotonic convergence with expended computational effort in all systems presented, while achieving systematic convergence with traditional methods is often difficult or impossible, see e.g.~Sec.~\ref{sec:diels-alder}.
We therefore believe that Orbformer represents the best available method for the modelling of strongly correlated molecular systems in dissociations of this nature.

The three Orbformer variants shown in Fig.~\ref{fig:bbmep_result}~(g)--(k) isolate the improvements brought about from joint finetuning and pretraining.
Optimizing the ansatz jointly across all geometries of a given minimum energy pathway (MEP) yields an approximately 20$\times$ efficiency gain over independent, single-point optimization,
amounting to almost perfect amortization of cost over the 20 point dissociation curve.
On top of this, starting the joint fine-tuning from the Orbformer model pretrained on the LAC dataset, one obtains additional gains in efficiency.
Here, we see that proximity to the pretraining distribution is important, with about a $16\times$ decrease in cost to reach chemical accuracy on ethane (in LAC dataset), a $6\times$ decrease on 1-propanol (somewhat larger than LAC). For L-Alanine (much larger than LAC), the pretrained model reaches a 5 kcal/mol error $8\times$ faster than its non-pretrained counterpart, but a 1 kcal/mol error only about $10\%$ faster.
Thus, pretraining is always beneficial, but decreases in importance with longer fine-tuning, as the network shifts to be more specialized to the fine-tuning task.
Nevertheless, the benefits from pretraining on out-of-distribution molecules persist down to chemical accuracy.

Not shown here is training without our engineering improvements, which would be up to $2.6\times$ slower across all Orbformer variants (see Supplementary Note~\ref{sec:app:timing}).
In totality, our approach recovers the results of independent, single-point deep QMC reference calculations at a fraction of the cost and so brings deep QMC to the cost--accuracy Pareto frontier formed by the range of classical methods that we have studied here.

\subsection{Comparison with existing deep QMC and the impact of pretraining}
\label{sec:existing-qmc}
\begin{figure*}[t]
    \centering
    \includegraphics[width=\textwidth]{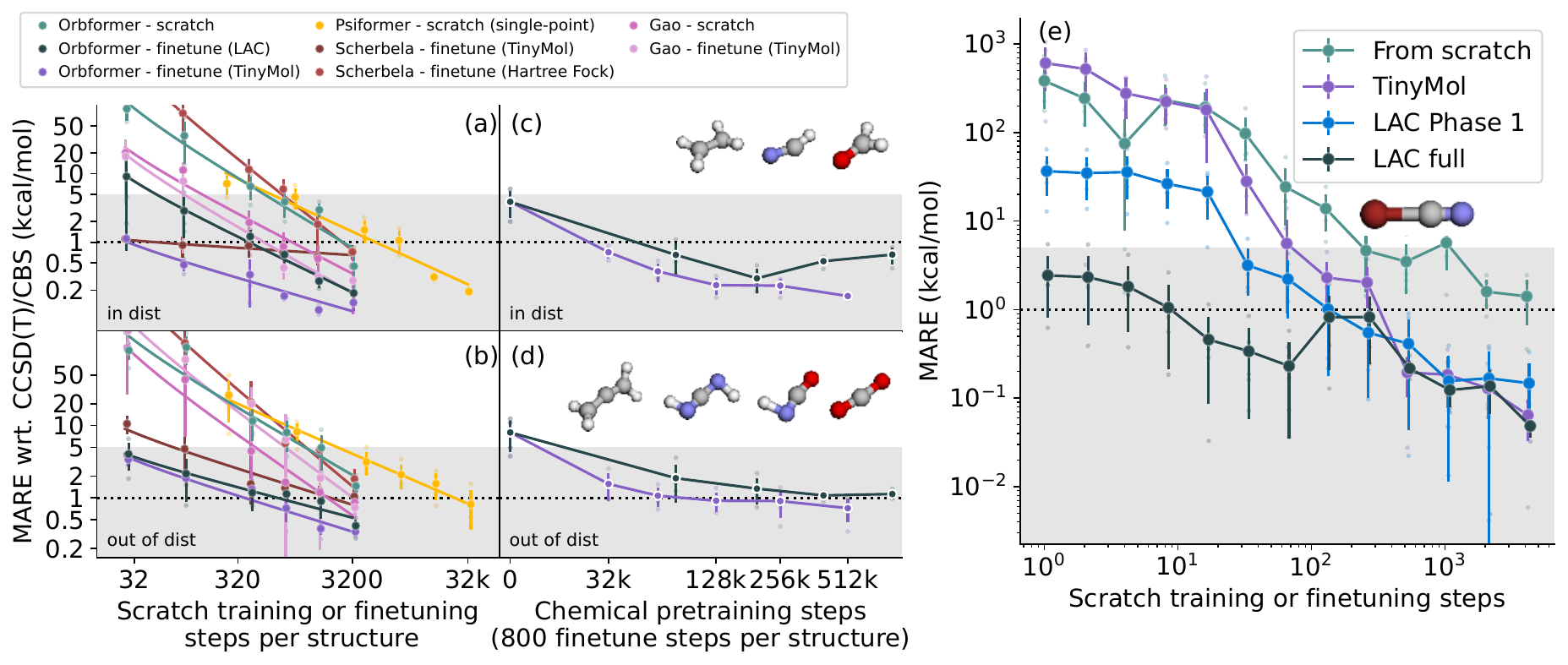}
    \caption{\textbf{Convergence with respect to pretraining and fine-tuning.} Mean absolute error in relative energy on the TinyMol in-distribution (a) and out-of-distribution (b) test sets. The error is computed with respect to reference energies obtained with CCSD(T) extrapolated to the complete basis set limit \cite{scherbela2024towards}.
    Each data point in panels (a)--(d) is averaged over ten geometries of three (in-distribution) or four (out-of-distribution) molecules; the error bars show the standard deviation across those molecules, and the individual per-molecule values are overlaid as faint points.
    The dotted horizontal line indicates chemical accuracy. 
    Panels (c) and (d) compare the two pretraining datasets used for Orbformer and the impact of the length of the pretraining, at 800 fine-tune steps per structure, on the in-distribution and out-of-distribution test sets, respectively.
    The effect of the pretraining dataset composition on the speed of fine-tuning on a new \ch{LiCN} test set is shown on panel (e), where each point is averaged over five \ch{LiCN} geometries, the error bars show the standard deviation across those geometries, and the individual per-geometry values are overlaid as faint points. Note that \ch{Li} is not present in the TinyMol pretraining dataset but is present in the LAC.
    }
    \label{fig:tinymol_relative_energies}
\end{figure*}

We now compare Orbformer to
previous attempts at chemical transferability within the deep QMC framework \cite{scherbela2024towards, scherbela2023variational, gao2023generalizing}.
For this, we employ the TinyMol dataset \citep{scherbela2024towards}, which has been used as a benchmark for medium scale chemical transferability \citep{gao2023generalizing}. 
This is an easier dataset without strong correlation and features CCSD(T)/CBS reference energies.
The TinyMol benchmark comes with its own small pretraining dataset, 
while the test set is divided into in-distribution (different geometries of pretraining molecules) and out-of-distribution (molecules not present during pretraining).
In addition to our main model pretrained from the LAC, we pretrained a separate Orbformer using the TinyMol pretraining set.

The obtained MAREs 
are compared to the results of Scherbela et al.~\cite{scherbela2023variational}, Gao et al.~\citep{gao2023generalizing} and a single-point Psiformer~\citep{glehn2022self} baseline, on the left panel of Fig.~\ref{fig:tinymol_relative_energies}.
We see clearly that Orbformer, pretrained on either the LAC or the TinyMol pretraining set, is the best performing model for both in- and out-of-distribution test sets.
On both test sets, the pretrained Orbformer reaches chemical accuracy at least an order of magnitude faster compared to training from scratch. %
The single-point Psiformer is accurate but 
more costly
because it does not exploit the joint fine-tuning methodology.
Compared to the baseline of \citet{scherbela2024towards} we see improvement, both in terms of accuracy at each fine-tuning iteration, as well as time to convergence to chemical accuracy.
Compared to \citet{gao2023generalizing}, we see that for out-of-distribution test structures, their model is actually harmed by more pretraining, whereas Orbformer still benefits from it.

On the center panel of Fig.~\ref{fig:tinymol_relative_energies}, we see in addition that longer fine-tuning of Orbformer is beneficial, although the rate of improvement slows beyond about 100k steps.
On the TinyMol test sets, pretraining using the TinyMol pretraining dataset is clearly better than pretraining using the much larger LAC dataset. 
We believe that this is because the TinyMol pretraining distribution is much more aligned to the TinyMol test distributions. %
To confirm this hypothesis, we created an additional small test set of five geometries of \ch{LiCN} undergoing bond stretching. 
This molecule is in the pretraining distribution of the LAC, is missing from Phase 1 LAC, while the TinyMol pretraining dataset doesn't even contain the \ch{Li} species.
As seen on the right panel of Fig.~\ref{fig:tinymol_relative_energies}, LAC pretraining is now the most effective, followed by LAC Phase 1.
This indicates, as expected, that molecules that are closer to the pretraining distribution can be fine-tuned faster, and is an encouraging sign that scaling up pretraining datasets will result in improved coverage of chemical space.

\subsection{Inner workings of the Orbformer model}
\label{sec:inner-workings}
\begin{figure*}[ht]
    \centering
    \includegraphics[width=\textwidth]{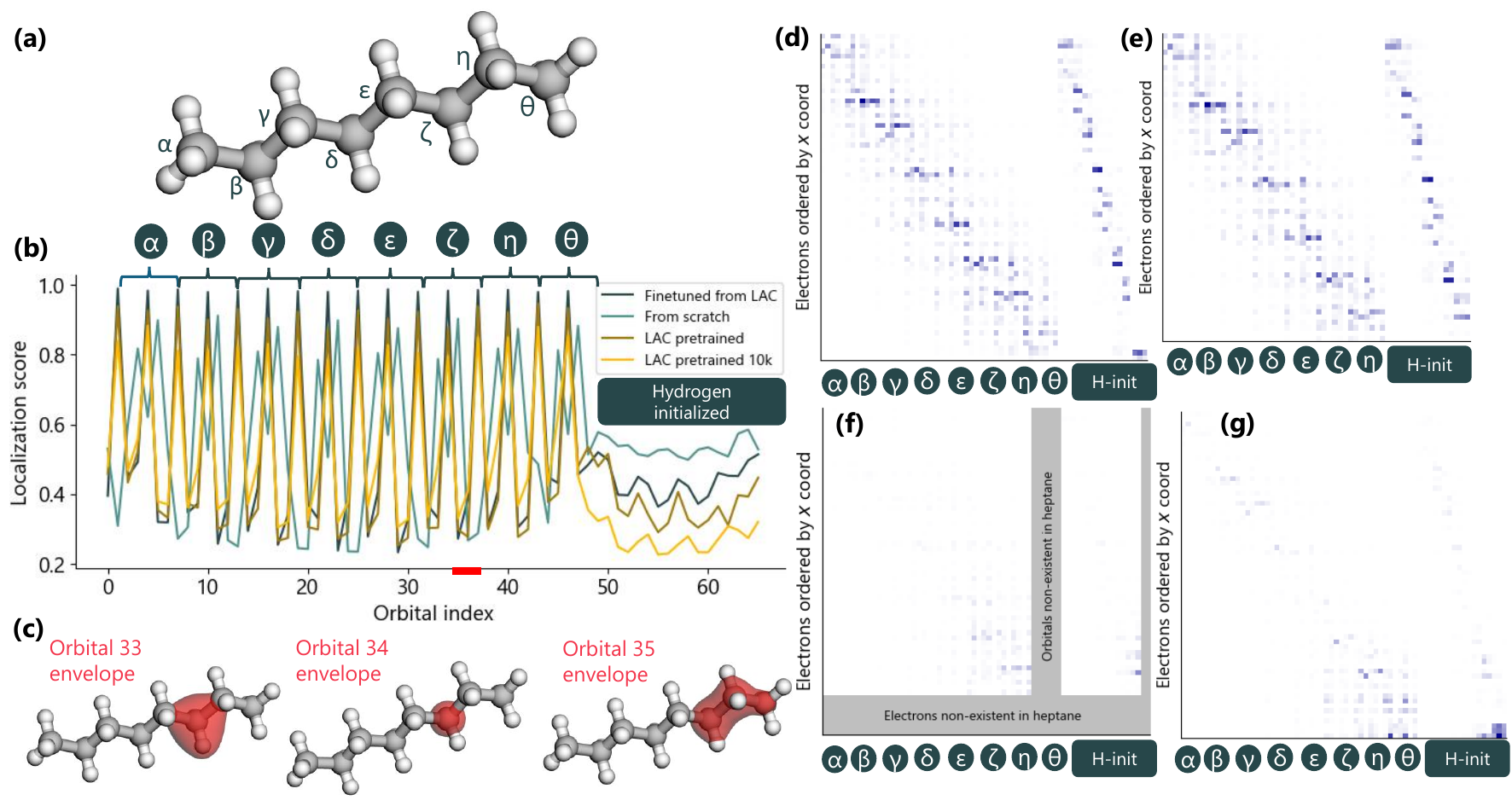}
    \caption{\textbf{Learning electronic structure patterns from data.} Orbformer was fine-tuned on a dataset of alkanes with 6 to 13 carbons inclusive. (a) Octane molecule with carbons labelled. (b) Orbformer orbital envelopes for octane ordered by the nucleus on which they are initialized: first all carbons, then all hydrogens. The localization score indicates whether a given orbital envelope places all weight on a single nucleus. The LAC pretrained and LAC pretrained 10k checkpoints were used as is on this molecule with no fine-tuning. (c) Isosurfaces of some of the envelopes shown above, here Orbital 34 has a localization score close to $1.0$. (d) One of the Orbformer Slater matrices for the octane molecule (using the LAC fine-tuned model), where we plot the pointwise absolute value, $|A^d|$, and white denotes $0$. The electrons are ordered by their $x$ co-ordinate which approximately corresponds to the order of the carbons from $\alpha$ to $\theta$. Orbitals are ordered exactly as in (b). (e) Using the same Orbformer parameters, we now compute the corresponding Slater matrix for the heptane molecule. Electron positions and spins are the same as for octane but with 8 electrons removed based on $x$ co-ordinate. (f) The pointwise absolute difference between the Slater matrices for heptane and octane, $|A^d(\cdot \mid \text{oct}) - A^d(\cdot \mid \text{hept})|$. The orbitals are arranged so that analogous orbitals that are initialized on the same nucleus in both molecules appear in the same place. Grey shading indicates electrons and orbitals which no longer exist in heptane. White colour denotes a difference of $0$ and the scaling is the same as in (d). (g) The absolute value of the gradient of the Slater matrix presented in (d) with respect to the $x$-position of the rightmost electron.}
    \label{fig:alkane_orbitals}
\end{figure*}

To analyze the properties of electronic structure learned by Orbformer, we fine-tune a single Orbformer model on a dataset of systematically generated alkane chains with 6 to 13 carbon atoms (transferable fine-tuning).
Several orbital envelopes---simple single-particle functions that control the spatial extent of Orbformer's generalized orbitals---are illustrated in Fig.~\ref{fig:alkane_orbitals}~(c).
The first intriguing property we found was that, without direct supervision, Orbformer learns to assign exactly two completely localized orbitals on every carbon atom, as seen in Fig.~\ref{fig:alkane_orbitals}~(b). It therefore appears that something akin to core electron orbitals are learned by the model for the carbon species, directly from transferable application of the variational principle.
We found that this orbital localization pattern actually emerges after as few as 10k Orbformer pretraining steps. 
When training from scratch, some localization does emerge, but not to the same degree as the model fine-tuned from LAC pretraining.
This is not surprising, considering that for a single molecule, properties of the determinant mean that there is no particular need for orbital localization, yet when attempting to generalize across similar molecules, the property appears highly beneficial. One may wonder whether orbital localization has the potential to harm performance on conjugated systems. In Supplementary Note~\ref{sec:app:additional-results:azulene} we fine-tuned Orbformer on azulene--naphthalene isomerization and showed that our reaction energy agrees very well with CCSD(T) at the complete basis set limit, indicating that the degree of localization can be adapted to the electronic structure required and does not practically limit the network's expressivity.

Returning to our systematic investigation of alkanes, we further investigate the extent to which the orbitals are transferable across molecules with the same local structure.
We extract the Slater matrices created by Orbformer for octane in Fig.~\ref{fig:alkane_orbitals}~(d), but also apply the same network to the smaller heptane molecule, using the octane electron positions after removing enough electrons to maintain electric neutrality, shown in (e). 
In Fig.~\ref{fig:alkane_orbitals}~(f) we now compute the pointwise difference between the octane Slater matrix in (d) and the heptane one in (e). %
What we see is a striking confirmation of the composability property of Orbformer. Namely, for electrons close to the $\alpha$-labelled carbon, which have essentially the same local environment in both octane and heptane, we see that the corresponding rows of the Slater matrix are identical for octane and heptane.
Electrons at the other end of the molecule see much larger variation in corresponding orbital values, because their local environments are not the same.
This provides firm evidence that orbitals in Orbformer are re-used between different molecules.
For an additional confirmation, panel (g) shows the gradient of the Slater matrix in (d) with respect to the position of the rightmost electron (nearer the $\theta$-labelled carbon).
It is clear that the Orbformer wavefunction is primarily local, with the strongest gradient occurring for the orbitals and electrons nearest to the perturbation. However, it is evident that the model does selectively retain a few longer-range interactions.
Overall, the requirement of chemical transferability forces the model to develop a theory of electronic structure in a way that is different from models that are optimized anew for every structure.

\section{Discussion}
\label{sec:discussion}
Orbformer demonstrates the possibility of large-scale amortization of computational cost in high-accuracy quantum chemistry through pretraining and transferable fine-tuning of a neural network wavefunction model.
We saw that Orbformer is competitive with traditional wavefunction methods, not only because it is eventually more accurate, but because it can achieve a favourable cost--accuracy trade-off.
This is only possible because of a drastic reduction in the per-structure computational cost brought about by learning to exploit recurring patterns in the electronic wavefunctions for different molecules and geometries.

For large-scale usage, we envisage Orbformer as a data generation method for a much cheaper model, such as a reactive force-field applicable to multireference systems.
With deep QMC approaching the cost of traditional methods, it can become a powerful and unique addition to the world of electronic structure.

\section{Methods}
\label{sec:method}
\subsection{Orbformer wavefunction architecture}
\label{sec:methods:architecture}

The Orbformer model introduced in Sec.~\ref{sec:orbformer} is a chemically transferable ansatz, meaning it accepts the molecular configuration $\molconf$ as an explicit input, rather than as a parameter which defines a particular optimization problem.
A trained Orbformer therefore represents the wavefunction for various molecules using a single set of neural network parameters.
The architectural changes required to implement a chemically transferable ansatz of this form are substantial when compared to single-molecule methods for molecular ground states with neural network ans\"{a}tze \citep{hermann2020deep,pfau2020ab}. 
In particular, a subnetwork that treats the molecular configuration input, which in Orbformer is the Orbital Generator, is now required.
The basic form of the Orbformer ansatz combines a scalar Jastrow factor, $J_\thetav$, with a sum over generalized Slater determinants,
\begin{equation}
\label{eq:methods-orbformer-dets}
    \Psi_\thetav(\xv_{1:\nelec} \mid \molconf) = e^{J_\thetav(\xv_{1:\nelec})}\sum_d \det[A^d_\thetav(\xv_{1:\nelec} \mid \molconf)],
\end{equation}
where each $A^d_\thetav$ is an all-electron $\nelec \times \nelec$ matrix. 
In contrast to traditional Slater determinants built from single-particle orbitals, these generalized Slater matrices contain permutation-equivariant many-body functions: each row depends not only on the position and spin of a single electron but on those of all electrons. The permutation equivariance then guarantees that $\Psi_\thetav$ is antisymmetric under electron exchange.

The Jastrow factor $J_\thetav$ enforces the electron--electron cusp conditions \citep{kato1957eigenfunctions} and is given by
\begin{equation}
\label{eq:methods-jastrow}
    J_\thetav(\xv_{1:\nelec}) = -\frac{1}{4}\sum_{e < e',\sigma_e = \sigma_{e'}}^\nelec \alpha_\text{par} e^{-\|\rv_e - \rv_{e'}\| / \alpha_\text{par}} -\frac{1}{2} \sum_{e<e',\sigma_e\ne \sigma_{e'}}^\nelec \alpha_\text{anti} e^{-\|\rv_e - \rv_{e'}\| / \alpha_\text{anti}},
\end{equation}
where $\alpha_\text{par},\alpha_\text{anti}$ are learnable parameters, electrons are indexed by $e=1,\dots,E$ and $\sigma_e$ denotes the spin of electron $e$. This exponentially decaying form allows the function $J_\thetav$ to become additive over well-separated subsystems.

Each generalized Slater matrix is formed as a pointwise product $A^d_\thetav = \Omega^d_\thetav \odot \Phi^d_\thetav$ of envelopes, $\Omega$, and projected electron representations, $\Phi$.
The envelopes dictate the spatial extent of each orbital. For each nuclear charge we associate a set of learnable exponents $\zeta^{Z}_{f}$ that define radially symmetric primitive envelopes $P_{enf} = e^{-\zeta^{Z_n}_f\|\rv_e - \Rv_n\|}$ on the electron--nuclei distances; these are combined using coefficient matrices $W^{\text{env},\sigma}_{donf}$ that are an output of the Orbital Generator,
\begin{equation}
\label{eq:methods-envelope}
    (\Omega^d_\thetav)_{eo} = \sum_{n=1}^N\sum_{f=1}^{F_\text{env}} P_{enf}W^{\text{env},\sigma_e}_{donf}.
\end{equation}
Here $d$, $o$, $n$ and $f$ index determinants, orbitals, nuclei and feature dimensions, respectively.
The projected electron representations are obtained from the permutation-equivariant electron representations $\elecrep_{ef}$ generated by the Electron Transformer together with coefficient matrices $W^{\text{repr},\sigma}_{dof}$ from the Orbital Generator,
\begin{equation}
\label{eq:methods-projection}
    (\Phi^d_\thetav)_{eo} = \sum_{f=1}^{F_\text{repr}} \elecrep_{ef}W^{\text{repr},\sigma_e}_{dof}.
\end{equation}

The electron representations are generated from an Electron Transformer. 
Our Electron Transformer network first pools incoming signals from each nucleus to each electron, to create a starting electron representation. These representations are then passed through a number of self-attention layers with an attention bias term that is computed from the electron--electron distances such that the attention logits decay to zero between very distant electrons.
Through the attention mechanism, every electron representation depends on the positions and spins of every other electron, and the set of electron representations is equivariant to a permutation of the electrons. The former property means that Orbformer ‘orbitals’ are a strict generalization of orbitals in classical quantum chemistry (which only depend on a single electron). The latter property ensures that the ansatz overall is antisymmetric under the exchange of two electrons, as required.
Orbformer is, to our knowledge, the first model that successfully combines a transformer \citep{vaswani2017attention} architecture with generalization across molecules.

The Orbital Generator subnetwork allows Orbformer to model the wavefunction for multiple molecules simultaneously. 
It produces the coefficient matrices $W^{\text{env},\sigma}_{donf}$ and $W^{\text{repr},\sigma}_{dof}$ that are used to form $\Omega$ and $\Phi$. The Orbital Generator depends only on the input molecular configuration, $\molconf$, and not on the electron positions and spins, $\xv$.
The Orbital Generator has two main stages: (i) initialization, and (ii) message passing.
In the initialization stage, orbital representations are instantiated that are completely localized to a single nucleus and depend only on the atom types that are present in the molecule. For example, in the \ch{LiH} molecule, the network will initialize three distinct orbitals on the \ch{Li} atom and one orbital on the \ch{H} atom.
Although crude, this scheme ensures that charges are locally balanced at initialization.
In the message passing stage, a sequence of message passing layers update the orbital representations to respond to the overall geometry of the molecule. 
This allows the network to see the local structure of the molecule and to respond accordingly, and it allows orbitals to delocalize away from their nucleus of initialization.

The property of composability is a constraint that we apply to the network architecture to encourage the network to learn generalizable features that can be re-used between different molecules that share similar sub-structure (e.g.~a functional group).
Composability guarantees size consistency for single-determinant Orbformer (but not when multiple determinants are used).
Consider electrically neutral molecules $\molconf_1,\molconf_2$ containing $E_1,E_2$ electrons respectively.
In the limit of infinite distance between $\molconf_1,\molconf_2$, a composable chemically transferable ansatz will have exactly $E_1$ orbitals localized to $\molconf_1$ and $E_2$ orbitals localized to $\molconf_2$; the value of the orbitals localized to $\molconf_1$ will be independent of the nucleus and electron positions at $\molconf_2$ and the same as would be obtained when running the ansatz on $\molconf_1$ alone, and vice versa; and the Jastrow factor for the combined system will be the sum of the Jastrow factors for the two separate systems.
In Supplementary Note~\ref{sec:app:arch:basic-principles}, we show that composability means that the Slater matrices $A^d$ for the combined system of $\molconf_1,\molconf_2$ can be brought to a block diagonal form by a sequence of column and row swaps.

Full details of the model architecture are given in Supplementary Note~\ref{sec:app:arch:elec-transformer}, and a detailed discussion of how Orbformer relates to existing transferable QMC methods can be found in Supplementary Note~\ref{sec:app:arch:related-work}.

\subsection{Accelerating and stabilizing the model}

Large-scale pretraining is only possible if the model is fast enough and stable enough to accommodate pretraining over a large, diverse dataset.
We found that many of the ansatz changes introduced to satisfy the composability constraint actually led to significant improvements in model stability as well.
In particular, ensuring that all particle--particle interactions decay to zero in the large distance limit, and initializing the orbitals to be partially localized appears to have been very beneficial for training stability.

As a consequence of this, we found that we could train Orbformer at TensorFloat-32 \citep{tf32} precision, a lower (hence faster) precision than the full Float-32 precision used for all previous deep QMC optimizations.
We also found that we could train with substantially smaller electron batch sizes than some previous work, for example we were able to train Orbformer with batch sizes that were at least $4\times$ smaller than electron batch sizes used for Psiformer \cite{glehn2022self}.
We also found that Orbformer behaved stably with very large distance separations (e.g.~testing size consistency), unlike some other existing models.

To further accelerate the model, we adopted the Forward Laplacian framework that was introduced by \citet{li2024computational} using the implementation in the \texttt{folx} package \citep{gao2023folx}.
Another avenue of acceleration was to switch to using FlashAttention \citep{dao2022flashattention} for the electron--electron attention layers.
FlashAttention both accelerates the attention layers and reduces their GPU memory consumption.
Unfortunately, there is no existing FlashAttention algorithm for the Laplacian of the attention layers. 

Rather than fall back on the default Laplacian, which is slower and more memory intensive, we introduce a FlashAttention-inspired algorithm to compute the Forward Laplacian updates for the electron--electron attention layers.
We implemented this using \texttt{pallas} \citep{jax2023pallas}, a kernel language for JAX. This kernel has now been made available in the \texttt{folx} package.
The precise details of our Laplacian implementation are given in Supplementary Note~\ref{sec:app:laplacian:flash-laplacian}.

\subsection{Generating electron configurations for chemically transferable training}
In single-molecule deep QMC training, the training data is generated from running Markov Chain Monte Carlo (MCMC) on the model itself and used to converge the ansatz towards the theoretical ground state wave function.
For Orbformer pretraining, we sample molecular configurations from our LAC dataset and apply data augmentation. 
This presents a problem, however, as we must be able to sample electron configurations from the distribution $|\Psi(\xv \mid \molconf)|^2$ for a molecule $\molconf$ that has not been previously encountered during pretraining.
This requires us to be able to burn in MCMC samplers for new molecules much faster than in previous works.

One natural MCMC algorithm for generating electron configuration data is Metropolis-Adjusted Langevin Algorithm (MALA) \citep{grenander1994representations}.
This combines a Langevin proposal of step-size $\tau$
\begin{equation}
    \xv_{1:\nelec}^{(\text{prop}, t+1)} =  \xv_{1:\nelec}^{(t)} + \tau^{(t)} \nabla_{\xv_{1:\nelec}} \log |\Psi_\thetav| + \sqrt{\tau^{(t)}} \xi^{(t)}
\end{equation}
with a Metropolis--Hastings accept/reject step.
However, we often observed that the step size $\tau$ must become very small in order to maintain a reasonable acceptance rate, leading to unacceptably slow burn-in.

Taking our inspiration from the use of Langevin samplers in the generative modelling community, we propose an initial sampling phase using Unadjusted Langevin Algorithm (ULA) that simply omits the accept/reject step. We combine ULA with step size annealing, and find that this algorithm very rapidly converges electron samples close to the desired distribution. 
We then switch to MALA towards the end of equilibration to converge to exactly the right distribution.
Supplementary Note~\ref{sec:app:sampling} contains an ablation study showing the much faster burn-in from using this ULA/MALA strategy.

\subsection{Multiple determinants}
Unlike much prior work in deep QMC, we do not use Hartree--Fock as a starting point for our ansatz.
We found that Hartree--Fock is unreliable for the multireference structures that we are interested in.
However, we noticed that training an ansatz from random initialization led to worse performance, even in the limit of long training.
We were able to trace this problem to `determinant collapse'---during early phases of training from random initialization, we saw that one determinant rapidly came to dominate the model, leading to an effective single-determinant model that persisted to the end of training.
Indeed, we verified that the worse energy that was obtained without initializing to Hartree--Fock was the same energy that was obtained from a single-determinant version of the model (whether or not Hartree--Fock initialization was used).

To rectify this problem without the use of a Hartree--Fock baseline, we introduce a penalty term that is minimized when all determinants contribute equally to the wavefunction value
\begin{equation}
    \mathcal{L}_\text{det} = -\lambda_\text{det}\mathbb{E}\left[ \sum_d \log \left( \frac{\det[A^d]}{\sum_{d'} \det[A^{d'}]}\right) \right].
\end{equation}
This penalty term can be directly backpropagated to compute the penalized training gradient.

Training with this penalty switched on, we discovered that starting from a randomly initialized network, we could recover the performance that was obtained when starting from a network initialized to the Hartree--Fock solution.
Full details of this ablation study are presented in Supplementary Note~\ref{sec:app:training:multi-det}.

\subsection{Light Atom Curriculum}
The Light Atom Curriculum (LAC) consists of molecules of up to 24 electrons containing the species \ch{H}, \ch{Li}, \ch{B}, \ch{C}, \ch{N}, \ch{O} and \ch{F}. We chose \ch{H}, \ch{C}, \ch{N}, \ch{O} and \ch{F} as a basic set of atoms that define a rich space of organic chemistry. We added \ch{Li} and \ch{B} to cover molecules beyond organic chemistry whilst keeping to light atoms.
This choice of atoms allows us to convincingly demonstrate that Orbformer can generalize across a large chemical space, whilst keeping the number of electrons low enough to avoid spiralling computational cost in the pretraining.
We note that our pretrained Orbformer checkpoint is only suitable for molecules containing these atomic species.
On the other hand, it is designed to generalize to systems much larger than 24 electrons, as shown in our experiments.

With the basic set of molecules defined, we sought to generate a more diverse set of geometries than previously considered---moving away from equilibrium geometries to more distorted structures where electronic structure is more likely to exhibit multireference character.
We created five distortion operations that could be randomly sampled without human intervention: bond bending, bond breaking, bond stretching, bond bending \& stretching, and bond bending \& breaking. 
Additionally, we apply on-the-fly data augmentation to all our structures that consists of (i) random rotation, (ii) adding small Gaussian noise to all atom co-ordinates.
We conducted analysis on the LAC to establish whether multireference electronic structure is indeed likely to occur.
On a small subset of our dataset, we computed the commonly used T1, D1 and $|\%\text{TAE}|$ diagnostics\citep{multiref_charac} using CCSD(T)/cc-pVDZ. 
We found that between 20\% and 45\% of structures were considered multireference by these diagnostics, with structures that feature bond breaking the most highly represented in the multireference subset across all diagnostics.

At pretraining time, we took a curriculum learning approach \citep{bengio2009curriculum} that operated in three distinct phases, rather than pretraining on the entire LAC from the start. This approach enables us to pretrain efficiently by first learning features on very small molecules, where gradient steps are cheapest, before moving on to somewhat larger molecules. Thanks to our model architecture, features that we learn on the smallest molecules generalize to much larger ones. Indeed, whilst the largest pretraining molecules have 24 electrons, we fine-tune the resulting model on much larger molecules (up to 106 electrons in this paper).
We also found that the curriculum learning approach improved pretraining stability.
Full details of the curriculum learning are given in Supplementary Note~\ref{sec:app:pretraining:curriculum} and details on the dataset composition and multireference diagnostics are presented in Supplementary Note~\ref{sec:app:datasets:lac-composition}.

\subsection*{Code availability}
The code that supports this project and the model checkpoint that serves as a starting point for users' calculations are available at \url{https://github.com/microsoft/oneqmc} \citep{oneqmc_repo}.

\begingroup
\setlength\bibitemsep{0pt}
\printbibliography
\endgroup

\subsection*{Acknowledgements}
\begingroup
{\footnotesize
We would like to thank C.\ Bishop for his overall leadership of the AI for Science organization, P.\ Gori-Giorgi for numerous helpful discussions about quantum chemistry, as well as M.\ Riechert, T.\ Vogels, and H.\ Schulz for invaluable support on GPU cluster management.}
\endgroup

\subsection*{Funding statement}
{\footnotesize
We acknowledge funding from Microsoft Research.
}

\subsection*{Author contributions statement}
{\footnotesize
J.H.~and F.N.~conceived the project. J.H., F.N.~and A.F.~managed the research. A.F., Z.S., P.B.S., L.C., J.K., G.C., N.G.~and J.L.~developed the JAX codebase, model architecture and training. A.F., Z.S., P.B.S. and L.C.~created the training data and ran the evaluations presented in the paper. A.F., Z.S., P.B.S., L.C.~and J.H.~wrote the paper.}

\appendix
\clearpage
\section*{Supplementary Information}

\section{Glossary of mathematical symbols}
\label{sec:app:glossary}
\begin{table}[h]
    \centering
    \begin{subtable}[t]{0.46\textwidth}
        \centering
        \resizebox{\textwidth}{!}{
        \begin{tabular}{ll}
        Symbol & Meaning \\
        \hline
        $\nelec$ & Number of electrons  \\
        $\nnuc$ & Number of nuclei \\
        $\rv$ & Electron spatial coordinates \\
        $\sigma$ & Electron $z$-axis spin \\
        $\xv$ & Electron spin-spatial coordinates, $(\rv, \sigma)$ \\
        $\Rv$ & Nucleus spatial coordinates \\
        $Z$ & Nucleus charge \\
        $\molconf$ & Molecular configuration, $(\Rv_{1:\nnuc},Z_{1:\nnuc})$ \\
        $\hat{H}$ & Hamiltonian \\
        $\energy$ & Energy \\
        $\Psi$ & Wave function \\
        $\thetav$ & Ansatz parameters \\
        $p(\cdot)$ & Probability density \\
        $A$ & Slater matrix \\
        $S_z$ & Total $z$-axis spin of electrons in a molecule \\
        $\norb$ & Number of orbitals \\
        $F$ & Number of features \\
        $H$ & Number of heads \\
        $J$ & Jastrow factor \\
        $\alpha$ & Jastrow parameters \\
        $\Omega$ & Envelope matrix \\
        $\Phi$ & Projected electron representation matrix \\
        $\zeta$ & Envelope exponent \\
        $P$ & Primitive envelopes, shape $[E, N, F_\text{env}]$ \\
        \end{tabular}
        }
    \end{subtable} 
    \hfill
    \begin{subtable}[t]{0.46\textwidth}
        \centering
        \begin{tabular}{ll}        
        Symbol & Meaning \\
        \hline
        $W$ & Linear weight matrix\\
        $\elecrep$ & Electron representations, shape $[E, F_\text{repr}]$ \\
        $\varphi$ & Featurization filter function \\
        $\beta, \gamma$ & Featurization parameters \\
        $\kappa$ & Attention distance bias weight \\
        $\orbrep$ & Orbital representations, shape $[O, N, F_\text{orb}]$ \\
        $\bm{\eta}$ & Initial orbital representations \\
        $\tau$ & MCMC step size \\
        $\xi$ & Standard Gaussian random variable~$\sim N(0,1)$ \\
        $B$ & Batch size \\
        $V$ & Potential \\
        $\mathcal{L}$ & Loss \\
        $\pi_\text{det}$ & Determinant relative probabilities \\
        $\lambda$ & Regularization weight \\
        $s$ & Standard deviation \\
        $\ell_\delta$ & Huber loss function, by default $\delta=1\,\text{Ha}$ \\
        $\bm{\mathcal{Q}}$ & Attention queries \\
        $\bm{\mathcal{K}}$ & Attention keys \\
        $\bm{\mathcal{V}}$ & Attention values \\
        $\bm{\mathcal{B}}$ & Attention edge biases \\
        $\bm{\mathcal{O}}$ & Attention outputs \\
        $\bm{\mathcal{S}}$ & Attention intermediate inner products
         \\
         $\bm{\mathcal{P}}$ & Attention weights
         \\
        \end{tabular}
    \end{subtable}
    \label{tab:glossary}
\end{table}

\section{Background}
\label{sec:app:background}

\subsection{Variational Monte Carlo (VMC)}
\label{sec:app:background:vmc}
We consider a molecular system comprising $\nelec$ electrons at positions $\rv_1,\dots,\rv_\nelec$ with spins $\sigma_1,\dots,\sigma_\nelec$, denoting their spatial-spin coordinates $(\rv_e, \sigma_e) = \xv_e$, and $\nnuc$ nuclei at positions $\Rv_1,\dots,\Rv_\nnuc$ with charges $Z_1,\dots,Z_\nnuc$, which we refer to as the molecular configuration $(\Rv_{1:\nnuc},Z_{1:\nnuc})=\molconf$.
If we make the Born--Oppenheimer approximation that the nuclei are fixed point charges, the stationary quantum states for the electrons satisfy the time-independent Schr\"{o}dinger equation $\hat{H}|\Psi \rangle = \energy |\Psi \rangle$ for the molecular Hamiltonian
\begin{equation}\label{eq:hamiltonian}
     \hat{H} := -\frac{1}{2} \nabla_{\rv_{1:\nelec}}^2 + \sum_{n=1}^{\nnuc}\sum_{n' < n} \frac{Z_n Z_{n'}}{\|\Rv_n - \Rv_{n'}\|} - \sum_{n=1}^{\nnuc}\sum_{e=1}^\nelec \frac{Z_n}{\|\Rv_n - \rv_e\|} + \sum_{e=1}^{\nelec}\sum_{e'<e} \frac{1}{\|\rv_e - \rv_{e'}\|}
\end{equation}
in atomic units.
Additionally, as electrons are fermions, electronic wave functions must be antisymmetric under the exchange of any two electrons.%

The ground state wave function is the eigenstate with the lowest eigenvalue.
In molecular variational Monte Carlo in first quantization, we propose an anti-symmetric, parametric, real-valued \citep{foulkes2001quantum} wave function ansatz $\Psi_\thetav(\xv_1, \xv_2, \dotsc, \xv_\nelec) \in \mathbb{R}$.  The ground state can then be approximated by minimizing the Rayleigh quotient with respect to the parameter $\thetav$
\begin{equation}
\label{eq:qmc}
	\thetav^\star = \min_\thetav \frac{\langle \Psi_{\thetav} | \hat{H} | \Psi_{\thetav} \rangle}{\langle \Psi_{\thetav} | \Psi_{\thetav} \rangle} = \min_\thetav \int \frac{|\Psi_\thetav(\xv_{1:\nelec})|^2}{\langle \Psi_{\thetav} | \Psi_{\thetav} \rangle}\frac{\hat{H}\Psi_\thetav(\xv_{1:\nelec})}{\Psi_\thetav(\xv_{1:\nelec})} d\xv_{1:\nelec}.
\end{equation}
The right-hand side of equation \eqref{eq:qmc} shows how this quotient can be approximated by Monte Carlo integration using samples from the distribution $p(\xv_{1:\nelec}) \propto|\Psi_\thetav(\xv_{1:\nelec})|^2$.
This stochastic approximation lifts the requirements of analytical integrability, allowing VMC ans\"{a}tze to explicitly model electronic correlation.

Traditional VMC uses ans\"{a}tze built from Slater determinants of single-particle orbitals. 
Explicit electron correlation is typically introduced by the inclusion of Jastrow factors, totally symmetric functions under the exchange of electrons, and back-flow transformations, totally symmetric displacements of the electron positions based on the remaining electrons \cite{needs2009continuum}.

More recently, neural network ans\"{a}tze were proposed \citep{carleotroyer2017}, including PauliNet \citep{hermann2020deep} and FermiNet \citep{pfau2020ab} as the first high-accuracy solutions for molecular ground states via the ab-initio neural network approach.
Various new network architectures and algorithmic improvements have since been proposed to further improve accuracy and efficiency \citep{spencer_better_2020,SchatzleJCP23,gao2024neural,li2024computational,gerard_2022,lin_explicitly_2023,kim2024neural, li_fermionic_2022,glehn2022self}.

From the perspective of deep learning, three salient points of difference between training (fermionic) neural network wave function ans\"{a}tze and training standard neural networks are
\begin{enumerate}
    \item the network architecture must enforce the anti-symmetry constraint,
    \item the training data is \emph{self-generated} \citep{hermann2023ab} using Markov Chain Monte Carlo (MCMC) to compute the loss \eqref{eq:qmc}, so no external source of data beyond the equations of physics is required, and
    \item the loss function involves the Laplacian of the wave function with respect to the electron co-ordinates.
\end{enumerate}

\subsection{Orbital localization and size consistency}
\label{sec:app:background:size-consistency}

In classical quantum chemistry, the term \emph{orbital} generally refers to a single-particle function $\psi(\xv_e)$.
In the context of molecular orbital theory \citep{pople1970molecular}, a localized orbital \citep{kleier1974localized} is a single-particle orbital that is significantly different from zero only when $\rv_e$ lies within a small region of space, e.g.~close to a specific nucleus or between a pair of nuclei.
By default, most classical quantum chemistry methods operate with non-localized orbitals that have support over the entire molecule \cite{helgaker2013molecular}.
However, exploiting the ``nearsightedness of electronic matter'', localized orbitals have been employed to scale classical quantum chemistry methods to very large systems \citep{pulay1983localizability,nakata2020large,szabo2023linear}.

The term \emph{size consistency} generally refers to two non-interacting subsystems \cite{hanrath2009concepts}, such as two molecules $\molconf_1$ and $\molconf_2$ separated by a large distance.
A method is said to be size consistent if the energy is additive for these non-interacting systems: $\energy(\molconf_1) + \energy(\molconf_2) = \energy(\molconf_1+ \molconf_2)$.
Since the interaction terms in the molecular Hamiltonian \eqref{eq:hamiltonian} tend to zero in the large distance limit, size consistency will hold if the wave function can be expressed as a tensor product of non-interacting wave functions of the subsystems $\ket{\Psi_{\molconf_1+\molconf_2}} = \ket{\Psi_{\molconf_1}}\otimes\ket{\Psi_{\molconf_2}}$, which can be suitably antisymmetrized. %
Perhaps the most straightforward approach to constructing a size consistent ansatz is to employ a single Slater-determinant built from orbitals \textit{localized} on either one of the two subsystems, with the correct number of orbitals assigned to each.
Then, if the subsystems are sufficiently distant, the determinant will factorize into the above-mentioned tensor product form.
We pick up on this thread in the next section when we discuss composability.

\section{Orbformer Architecture}
\label{sec:app:arch}
\subsection{Basic principles and desiderata for a transferable ansatz}
\label{sec:app:arch:basic-principles}

The original work on molecular ground states with neural network ans\"{a}tze \citep{hermann2020deep,pfau2020ab} trained a fresh network for each molecule and each molecular geometry, so $\molconf$ were treated as configuration parameters and not inputs to the network.
Given the remarkable capacity of neural networks to generalize, it makes intuitive sense to attempt the construction of a \emph{transferable ansatz}, $\Psi_\thetav(\xv_{1:\nelec} \mid \molconf)$, which takes the molecular configuration as an input.
However, the architectural changes required to implement a truly transferable ansatz are significant.

To discuss such issues in more detail, we introduce the basic form of the Orbformer wavefunction
\begin{equation}
\label{eq:orbformer-dets}
    \Psi_\thetav(\xv_{1:\nelec} \mid \molconf) =e^{J_\thetav(\xv_{1:\nelec})}\sum_d \det[A^d_\thetav(\xv_{1:\nelec} \mid \molconf)].
\end{equation}
In common with recent ans\"{a}tze \citep{hermann2020deep,pfau2020ab,glehn2022self,scherbela2024towards,gao2023generalizing}, Orbformer combines the sum of determinants of several all-electron $\nelec\times \nelec$ Slater matrices, $A_\thetav^d$, with a scalar Jastrow factor, $J_\thetav$.
In contrast to traditional Slater-determinants built from single particle orbitals, these generalized Slater matrices contain permutation equivariant many body functions, composed of orbital envelopes and a projected latent space encoding of electrons by deep neural networks. 

\paragraph{Differently sized inputs.}
For a chemically transferable ansatz, not only do the values of inputs $\Rv_{1:\nnuc}, Z_{1:\nnuc}$ and $\xv_{1:\nelec}$ change, but so do the number of those inputs $\nelec$ and $\nnuc$.
The network must therefore admit a variable number of inputs and construct Slater matrices with variable sizes.
To build such a model that can be JAX-compiled \citep{jax2018github}, we take an approach based on masking.
For a given compilation of the model, we first set maximum particle numbers $\nelec_\text{max}$ and $\nnuc_\text{max}$. Any non-existent particles are masked out, with the model adapting different operations accordingly.
For example, the $\nelec\times \nelec$ Slater matrices are promoted to $\nelec_\text{max} \times \nelec_\text{max}$ matrices, with non-existent electrons filled in with identity submatrices, thereby preserving the determinant value.

\paragraph{Parameter dimensions.}
We further require that no parameter size depends on the number of electrons or nuclei, nor on $\nelec_\text{max}$ or $\nnuc_\text{max}$. 
This is particularly beneficial when we train on a dataset of small molecules, setting $\nelec_\text{max}$ and $\nnuc_\text{max}$ small to reduce computation time, but then transfer those parameters to a larger molecule. 

\paragraph{Nuclei permutations.}\label{sec:particle-permutations}

Considering the molecular configuration as an explicit input to the neural network we design the architecture to respect the molecular symmetries, i.e.~permutations of the nuclei.
While not a strict necessity, the permutation invariance helps with generalization, removing arbitrary indexing of nuclei.

Orbformer is implemented in such a way that a permutation of the nuclei may, at worst, lead to an arbitrary permutation of the \emph{columns} of $A^d_\thetav$ (the same permutation across all electrons and all determinants), leading to a physically unimportant possible change of overall sign.

\paragraph{Composability.}
We use the term composability to refer to a property that is related to, yet distinct from, the classical concepts of orbital localization and size consistency (cf. Supplementary Note~\ref{sec:app:background:size-consistency}).

In Orbformer, single-particle orbitals are replaced with many-body orbitals, functions that depend on the positions of all electrons, constructed so that the rows of each Slater matrix are equivariant to electron permutations via the formula $\Av^{d, e}_\thetav(\xv_e, \{\xv_{\setminus e}\} \mid \molconf)$.
Extending the idea of orbital localization, a composable many-electron orbital with input $(\xv_e, \{\xv_{\setminus e}\})$ should be non-zero only when an electron $\rv_e$ is found in a small region of space, and additionally the dependence of its value on the positions of other electrons should decay as the distance between that electron, $\rv_e$, and other electrons increases.
Note that, by construction, electrons are indistinguishable and in practice there is no assignment of specific electrons to orbitals.

Without being fully prescriptive, we add a constraint to guarantee size consistency for a single-determinant Orbformer.
If we consider two electrically neutral molecules $\molconf_1$ and $\molconf_2$ that are separated by a large distance, then we require $\nelec_1$ orbitals to be localized to $\molconf_1$ and $\nelec_2$ orbitals to be localized to $\molconf_2$ in the combined system $\molconf_1+\molconf_2$.
We also require that the value of the orbitals localized to $\molconf_1$ is independent of the nucleus and electron positions at $\molconf_2$ and is the same as would be obtained when running the ansatz on $\molconf_1$ alone, and \textit{vice versa}.
To see explicitly that this gives size consistency, let us also assume that the first $\nelec_{1}$ electron samples are in the vicinity of molecule $\molconf_1$ whilst the latter $\nelec_{2}$ electrons are near to $\molconf_2$.
Then for a suitable ordering of the columns of $A^d_\thetav$ for the combined system, we will have
\begin{equation}\label{eq:composability}
\begin{split}
    A^d_\thetav(\xv_{1:E_{1}+E_{2}} \mid \molconf_1 + \molconf_2) = 
    \begin{pmatrix}
        A_1 & 0 \\
        0 & A_2
    \end{pmatrix}
    \end{split}
\end{equation}
where submatrices $A_1$ and $A_2$ should exactly match with what we would get if we evaluated Orbformer on either subsystem with the appropriate subset of electron inputs, i.e.~$A_1 = A^d_\thetav(\xv_{1:E_{1}} \mid \molconf_1)$ and $A_2 = A^d_\thetav(\xv_{E_1+1:E_{1}+E_2} \mid \molconf_2)$.
We also require that the Jastrow factor must be additive
\begin{equation}\label{eq:composable-jastrow}
    J_\thetav(\xv_{1:E_1+E_2}) = J_\thetav(\xv_{1:E+1})+J_\thetav(\xv_{E_1+1:E_1+E_2}).
\end{equation}
We stress that these result must hold for a single parameter $\thetav$ for the transferable ansatz evaluated on differently sized systems.
We also note that the block diagonal form \eqref{eq:composability} depends on the arbitrary ordering of the orbitals and on the ordering of the electrons, so column and row swaps may be required to arrive at the block diagonal form.

This composability requirement gives size consistency for a single determinant model, because the determinant of the block diagonal Slater, \eqref{eq:composability}, is a product of the determinants of the submatrices $A_1$ and $A_2$ and the exponential of the Jastrow factor, \eqref{eq:composable-jastrow}, is also a product, so that $\Psi_\thetav$ decomposes into the product of wavefunction values for subsystems $\molconf_1$ and $\molconf_2$.
Row and column swaps lead, at worst, to an overall change of sign.
Hence for fixed set of model parameters, single-determinant Orbformer \emph{is} size consistent.

Composability does not lead to size consistency for multi-determinant wavefunction models because the sum of products is not equal to the product of sums.
We were not able to find a simple way to obtain zero-shot size consistency whilst using a fixed number of determinants for differently sized systems.
To the best of our knowledge, traditional size consistent quantum chemistry methods either use a single determinant (e.g. Hartree--Fock), or use an exponential ansatz parametrization at the cost of losing direct access to the wavefunction.

Previous work on transferable deep QMC also discussed size consistency \citep{gao2023generalizing}.
We note that the conclusions of \citet{gao2023generalizing} are similar to ours in the sense that size consistency is only guaranteed for a single determinant model but they used a multi-determinant ansatz in practice.
Our approach to implementing localized orbitals differs from the approach of Globe---rather than anchoring each orbital to a single spatial location, from which localization can be defined, we allow our orbitals to spread across multiple nuclei; however, we limit the distance that messages can propagate to maintain localization.
We also note that \citet{gao2023generalizing} state all inter-particle interactions decay exponentially in their model; however, by the inclusion of the Jastrow factor of \citet{glehn2022self}, this is not true, since
\begin{equation}
    \exp\left(-c\,\frac{\alpha^2}{\alpha + r} \right) \sim 1 - c\,\frac{\alpha^2}{\alpha + r} \text{  as  } r \to \infty.
\end{equation}
An amended Jastrow similar to the one we use in Orbformer would be required to give an exponential decay rate.

\paragraph{Geometric invariances.}
Orbformer consumes only relative positions of particles, e.g.~$\rv_e - \Rv_n$, which grants us invariance to an overall translation of the system.
We consciously choose not to incorporate any special (equivariant) treatment of rotations into our ansatz.
While any observable of the molecule is indeed invariant under symmetry transformations of the nuclei accord to Curies principle, the wavefunction itself, not being an observable but a mathematical construct, \emph{cannot be treated} as invariant. 
The ground state wavefunction of the boron atom serves as an illustrative example. 
In this system, the equation $\hat{H}|\Psi\rangle = \energy|\Psi\rangle$ admits multiple, linearly independent solutions with the same ground state energy. Each solution corresponds to a particular physical orientation of the angular momentum vector of the electronic state. Thus, whilst the ensemble of all solutions is rotation invariant, training a network towards a specific solution requires us to be able to break rotation symmetry.
\paragraph{Spin assignment.}
Orbformer is a spin-assigned ansatz \citep{huang1998spin,szabo2024improved}, in which the $z$ component of the electron spins are fixed for a given system throughout training.
For transferability, we further simplify to the case in which the number of up- and down-spin electrons are equal ($E$ even) or there is one more up-spin electron ($E$ odd).
This latter simplification does not prevent us from expressing wavefunctions with arbitrary total spins, as wavefunctions with $S_z=0$ or $1$ exist for all possible total spin quantum numbers.

\paragraph{Limitation of our formulation.}
In our work, we treat the nuclear co-ordinates and charges as a complete description of the molecule. This means that we do not consider learning systems with an overall charge (ions), with a spin state different from $S_z = 0$ or $S_z = \tfrac{1}{2}$ (not a restriction on $S^2$), and excited states. None of these are fundamental limitations of the ideas we present, but we leave them as topics for future work.

\subsection{Architecture overview}
\begin{figure*}
    \centering
    \includegraphics[width=0.75\linewidth]{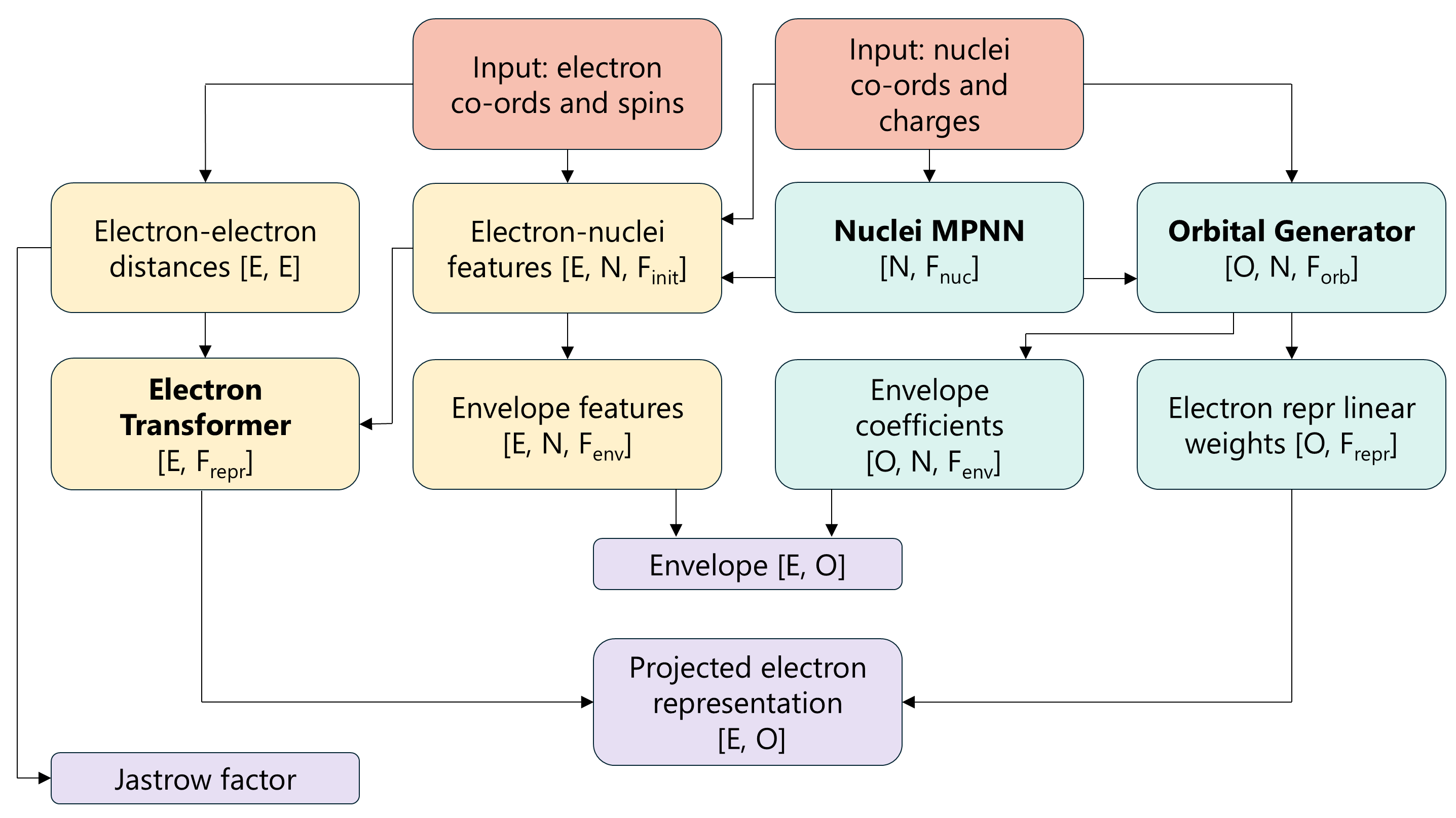}
    \caption{\raggedright \textbf{Orbformer overall architecture.} Shape hints are written in terms of the maximum number of Nuclei, Electrons, Orbitals and various Feature dimensions. Determinant dimensions are not included on this diagram. See Supplementary Figs. \ref{fig:electron-transformer} and~\ref{fig:orbital-generator} for detailed descriptions of the Electron Transformer, Orbital Generator and Nuclei MPNN subnetworks. Cyan denotes components of the network that do not depend on the electron inputs, $\xv_{1:E}$.
    }
    \label{fig:overall-architecture}
\end{figure*}

As discussed in Supplementary Note~\ref{sec:app:arch:basic-principles}, the Slater matrices $A^d_\thetav$ have \emph{rows} $\Av_\thetav^{d,e}$ that permute with the ordering of the input electrons $\xv_1,\dots,\xv_\nelec$, whilst the columns do not.
As a notational aid, we henceforth denote the shape of the Slater matrices as $\nelec \times \norb$, where $\norb$ refers to `orbitals' in a generalized sense. Whilst it must be the case that $\norb = \nelec$, this choice of notation helps to separate electrons and orbitals logically. 
The overall structure of Orbformer, using this notation and for a single determinant, is shown in Fig~\ref{fig:overall-architecture}.

\paragraph{Jastrow factor.}

The Jastrow factor is given by
\begin{equation}
    J_\thetav(\xv_{1:\nelec}) = -\frac{1}{4}\sum_{e < e',\sigma_e = \sigma_{e'}}^\nelec \alpha_\text{par} e^{-\|\rv_e - \rv_{e'}\| / \alpha_\text{par}} -\frac{1}{2} \sum_{e<e',\sigma_e\ne \sigma_{e'}}^\nelec \alpha_\text{anti} e^{-\|\rv_e - \rv_{e'}\| / \alpha_\text{anti}}
\end{equation}
where $\alpha_\text{par},\alpha_\text{anti}$ are learnable parameters. 
Like earlier works \citep{hermann2020deep,glehn2022self}, this Jastrow factor ensures the correct electron--electron cusps for the wavefunction \citep{kato1957eigenfunctions}.
Unlike earlier works, we use an exponentially decaying function as opposed to a harmonically decaying one.
Whilst both forms of the Jastrow factor do satisfy the composability requirement \eqref{eq:composable-jastrow}, we found that the exponential form was necessary for additivity of the Jastrow to also hold at finite distances where we would not expect an appreciable interaction to occur (e.g.~$12\text{\AA}$).

\paragraph{Slater matrices.}

The matrices $A^d_\thetav$ are constructed as a pointwise product $A^d_\thetav = \Omega^d_\thetav \odot \Phi^d_\thetav$ of envelopes, $\Omega$, and projected electron representations, $\Phi$. 

\paragraph{Envelopes.}
Our orbitals are a development of `simplified envelopes' \citep{gao2024neural}.
For each nuclear charge $1,\dots,Z_\text{max}$ we associate a set of learnable exponents $\zeta^Z_{1:F_\text{env}}$. The primitive envelopes are then formed as $P_{enf} = e^{-\zeta^{Z_n}_f\|\rv_e - \Rv_n\|}$.
As an output of the Orbital Generator network, we obtain coefficient matrices $W^{\text{env},\uparrow}_{donf},W^{\text{env},\downarrow}_{donf}$ where $d, o, n, f$ index the Slater matrices, orbitals, nuclei and primitive envelope features respectively. We then form the envelopes as a contraction
\begin{equation}
    (\Omega^d_\thetav)_{eo} = \sum_{n=1}^N\sum_{f=1}^{F_\text{env}} P_{enf}W^{\text{env},\sigma_e}_{donf}.
\end{equation}
We explored making the exponents $\zeta$ the output of either the Orbital Generator or the Nuclei MPNN, but found that this was prone to training instabilities. We also apply a transform that ensures $\zeta > 0.1$ throughout training.
Finally, we apply an initialization scheme for $\zeta$ inspired by single electron wavefunctions for atomic systems
\begin{equation}
    \zeta^Z_{f,\text{init}} = \max(1, Z / f) \text{ for } f = 1, \dots, F_\text{env}.
\end{equation}

\paragraph{Projected electron representations.}
As an output of the Electron Transformer we have representations $\elecrep_{ef}(\xv_{1:E} \mid \Rv_{1:N},Z_{1:N})$ where $\elecrep$ is constructed using a Transformer architecture \citep{vaswani2017attention} to be equivariant to permutations of the electrons, and $f=1,\dots,F_\text{repr}$.
From the Orbital Generator we obtain coefficient matrices $W^{\text{repr},\uparrow}_{dof},W^{\text{repr},\downarrow}_{dof}$ and then form
\begin{equation}
    (\Phi^d_\thetav)_{eo} = \sum_{f=1}^{F_\text{repr}} \elecrep_{ef}W^{\text{repr},\sigma_e}_{dof}.
\end{equation}

\subsection{Electron Transformer}
\label{sec:app:arch:elec-transformer}
\begin{figure*}
    \centering
    \includegraphics[width=0.94\linewidth]{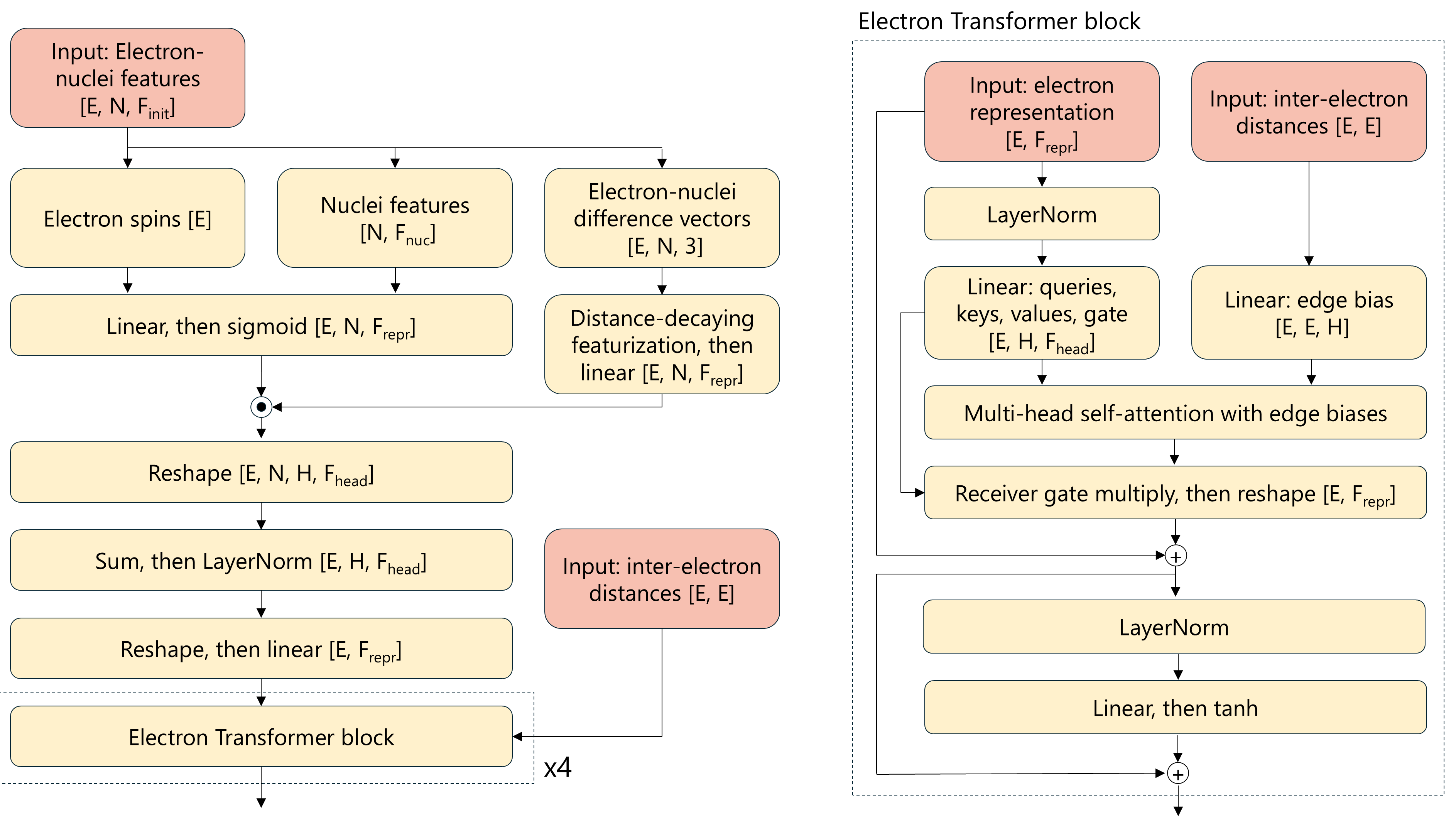}
    \caption{\textbf{Electron Transformer architecture.} Shape hints are given as in Supplementary Fig.~\ref{fig:overall-architecture}, and H denotes the number of heads. }
    \label{fig:electron-transformer}
\end{figure*}
The structure of this subnetwork is shown in Supplementary Fig.~\ref{fig:electron-transformer}.
Our use of a self-attention mechanism for electron representations is inspired by Psiformer \citep{glehn2022self}, but with several innovations necessary to satisfy the requirements laid out in Supplementary Note~\ref{sec:app:arch:basic-principles}.

\paragraph{Electron--nucleus feature extraction.}
We use distance-dependent filter functions $\varphi(\|\rv_e - \Rv_n\|)$ to weight electron--nuclei interactions.
In contrast to Psiformer, which uses a logarithmic rescaling, we focus on filters such that $\varphi(r) \to 0$ as $r \to \infty$. We found this made composability much easier to implement, because when summing signals from all nuclei we are guaranteed that distant nuclei contribute very little.

We use the following two families of filters
\begin{align}\label{eq:filter1}
    \varphi_\text{exp}(r\mid \beta) &= \beta e^{-\beta r}, \\
    \varphi_\text{sigmoid}(r\mid \gamma) &= \frac{1}{\gamma(1 + e^{r - \gamma})},\label{eq:filter2}
\end{align}
and we selected non-learnable values $\beta=1, 3; \gamma=2, 4$ Bohr. We found that it was important to incorporate filters that extend to a reasonable distance before decaying, which was the motivation for our sigmoid filters.
We experimented with learning the $\beta$ and $\gamma$ values, but found it was more unstable.
Importantly, our filters are non-zero at the origin with non-zero gradient, and thus enable the network to correctly model the nuclear cusp in the same way as the Psiformer features.

For each pair $(e, n)$, we then take the filter itself, along with the rescaled difference vector as features. %
To facilitate possible rotations based on the molecular geometry, we create matrices $W^\text{rot}_n$ as an output of the Nuclei MPNN of shape $6 \times 3$ and incorporate the 6-dimensional `rotated' features $W^\text{rot}_n(\rv_e - \Rv_n)$, suitably multiplied by our filter functions. There is no constraint for $W^\text{rot}_n$ to have orthonormal rows.
Features from different filters are stacked and then passed through a linear layer.

Separately, we create a spin--nucleus feature using the electron spin and nuclei features from the Nuclei MPNN for every $(e, n)$ pair, which are then passed through a single layer network.
This spin--nucleus feature is then multiplied by the filter-based featurization of the vector $\rv_e - \Rv_n$ to give our overall electron--nucleus feature.

\paragraph{Feature aggregation.}
The electron--nuclei features are summed over the nuclei axis to reduce them to electron feature vectors.
The sum operation combined with distance-decaying featurization ensures composability.
The exact summation procedure using multiple heads and LayerNorm \citep{lei2016layer} is shown in Supplementary Fig.~\ref{fig:electron-transformer}.

\paragraph{Electron transformer block.}

Here we only discuss the main differences between a standard Pre-LN transformer block \citep{wang2019learning} and our block.
First, we incorporate an edge bias based on the distance \citep{ying2021transformers} between electrons. Denoting the distance between two electrons as $r_{ee'} := \|\rv_e - \rv_{e'}\|$ and for attention head $h$, we include the bias
\begin{equation}
    \text{bias}_{ehe'} = -\log\left(1 + e^{r_{ee'} - 8}\right) - |\kappa_h r_{ee'}|.
\end{equation}
The first term ensures an asymptotic decay rate $\sim e^{-r_{ee'}
}$, guaranteeing that no information flows between distant electrons, which is essential for composability. The second uses a learnable coefficient $\kappa_h$ to control the distance bias of a given head in the non-asymptotic region. Thus $\kappa_h$ can be seen as controlling how `myopic' a given attention head is.

Second, unlike in standard transformers, the output of the self-attention computation is multiplied by a receiver gate. Given electron representations $\elecrep^{(\ell)}_{ef}$ at layer $\ell$, we apply self-attention over the electron axis, giving $\elecrep^{(\ell,\text{attn})}_{ef}$. We also apply a linear layer to $\elecrep^{(\ell)}_{ef}$ to give $\elecrep^{(\ell,\text{gate})}_{ef}$. The self-attention output is then reweighted by componentwise multiplication to give
\begin{equation}
    \elecrep^{(\ell,\text{gated attn})}_{ef} =  \elecrep^{(\ell,\text{attn})}_{ef} \odot \operatorname{sigmoid}\left(\elecrep^{(\ell,\text{gate})}_{ef}\right).
\end{equation}
We found that this gating leads to improvements in transferability performance.
Our hypothesis is that the receiver gate allows electrons that are very close to a nucleus to ignore incoming updates from other electrons.
Thirdly, we use a simplified tanh-activated MLP as in Psiformer,  but we combine it with Pre-LN.
See the right-hand side of Supplementary Fig.~\ref{fig:electron-transformer}  for a complete description.

\paragraph{FlashAttention.}
We identified the attention layers as the most computationally demanding component of the entire Orbformer network.
To accelerate the attention computations, we implement a variant of the FlashAttention algorithm \cite{dao2022flashattention} that is suitable for our architecture.
We used \texttt{pallas} \citep{jax2023pallas} for the implementation.
Significantly, as well as acceleration of the forward and backward passes, we also develop a FlashAttention-inspired implementation for the Laplacian, see Supplementary Note~\ref{sec:app:laplacian:flash-laplacian} for full details.

\subsection{Orbital Generator}
\label{sec:app:arch:orb}
\begin{figure}
    \centering
    \includegraphics[width=0.97\linewidth]{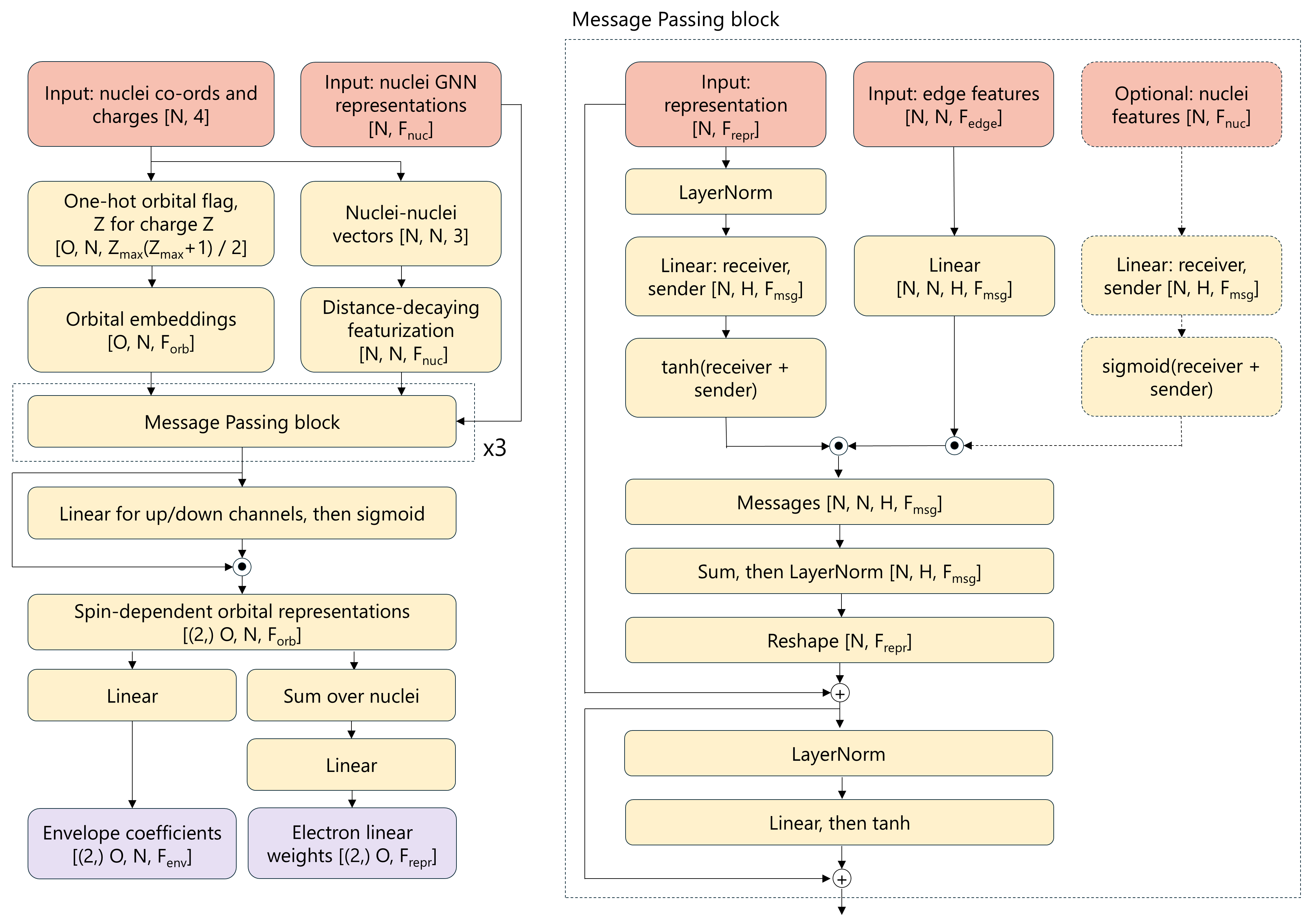}
    \caption{\raggedright \textbf{Orbital Generator architecture.} Shape hints are given as in Supplementary Figs.~\ref{fig:overall-architecture} and ~\ref{fig:electron-transformer}. The Message Passing layers are completely parallel over the Orbital dimension.
    The Nuclei MPNN is a simplified version of this subnetwork without the orbital dimension. In the Nuclei MPNN, initial features are simple embeddings of the nuclear charge, the third input (nuclei features) in the Message Passing blocks is not present, and there are no separate spin-dependent representations formed at the end.}
    \label{fig:orbital-generator}
\end{figure}
The structure of the Orbital Generator is presented in Supplementary Fig.~\ref{fig:orbital-generator}.
At a high level, the tensors that propagate through the Orbital Generator network have shape $[O, N, F_\text{orb}]$. 
This means that each feature vector corresponds to an orbital--nucleus pair. 
The advantage of this approach is that it allows us to control the spatial extent of different orbitals and enforce orbital localization. 
This comes at the price of a relatively sparse orbital representation, particularly for larger molecules.

\paragraph{Initial orbital embeddings.}
Under the assumption of a molecule with no overall charge, the number of electrons, and hence the number of orbitals, must equal the sum of nuclear charges $\sum_{n=1}^N Z_n$. 
To featurize these orbitals, we use an algorithm inspired by the linear combination of atomic orbitals (LCAO) approach.
Initially, the features of each orbital, which are of shape $[N, F_{\text{orb}}]$, are zero for all but one nucleus. We can therefore speak of an orbital being initialized `on' a certain nucleus. 
Using the assumption of electric neutrality, we initialize exactly $Z_n$ orbitals on nucleus $n$. 
Furthermore, we use a set of initial orbital embeddings that are determined solely by the atom type.
This is accomplished through a set of learnable initial feature vectors $\bm{\eta}^Z_i$ with $Z= 1, ... , Z_\text{max}$ defining the atom type, and $i=1,...,Z$ indexing that atom type's initial orbital embeddings.

Given an arbitrary ordering for the nuclei, we assign the first $Z_1$ orbitals to the first atom, the following $Z_2$ orbitals to the second, etc.
Supposing the $o$th orbital is initially assigned as the $i$th orbital of the $n$th atom, it's initial embedding of shape $[N, F_\text{orb}]$ will be
\begin{equation}
    \orbrep^{(0)}_{on'f} = \left(\eta_i^{Z_n}\right)_f \delta_{nn'} \text{ where } \begin{cases} n=1,\dots, \nnuc, \\ 
    i=1,\dots,Z_n, \\
    o = i + \sum_{m=1}^{n-1} Z_m, \\
    n' = 1, \dots, \nnuc, \\ 
    f = 1,\dots, F_\text{orb}
    \end{cases}
\end{equation}
where $\delta$ is the Kronecker delta, encoding the atom to which the given orbital is assigned and setting coefficients to zero on all other nuclei. %

We note that the ordering of the orbitals depends on the ordering of the nuclei.
It is this indeterminate ordering that gives rise to the possible permutation of the columns of the Slater matrices $A_\thetav^d$ that was discussed in Supplementary Note~\ref{sec:app:arch:basic-principles} `Nuclei permutations'.

Orbitals do not remain localized to a specific nucleus after initialization, instead we allow interaction through a sequence of message passing layers.
These layers allow the fixed initial embeddings to slowly adapt to the overall molecular geometry.
The message passing layers operate in parallel over the orbital dimension with information exchanged between different nuclei.
We tried some limited experiments incorporating orbital--orbital interaction, but we did not find that it improved performance.
Message passing steps on the nuclei axis utilize nucleus--nucleus edge features, which we now describe.

\paragraph{Nucleus--nucleus feature extraction.}
We re-use the filter-based featurization approach that was used for the electron--nuclei features in Supplementary Note~\ref{sec:app:arch:elec-transformer}.
Specifically, we compute filters $\varphi(\|\Rv_N - \Rv_{N'}\|)$ using the exponential and sigmoid functions in \eqref{eq:filter1} and \eqref{eq:filter2}. For the nuclei, we use non-learnable filter hyperparameters $\beta=1; \gamma=2, 4, 6$ Bohr. We set self-edges to 0.
As with the electrons, we concatenate the filter itself with the rescaled difference vectors .%
We then concatenate over the different filter functions. We do not use the rotated features that were used in the Electron Transformer.

\paragraph{Orbital LayerNorm}
Since orbital representations are presumed to be sparse, applying standard LayerNorm in parallel over the orbitals and nuclei axes is not appropriate.
Instead, we apply mean-centering as usual, and then rescale features using the maximum variance across nuclei within a given orbital
\begin{equation}
    Q^{(\ell,\text{LN})}_{onf} = Q^{(\ell)}_{onf}\left(\max_{n'} \left[\frac{1}{F_\text{orb}} \sum_{f=1}^{F_\text{orb}} \left(Q^{(\ell)}_{on'f}\right)^2\right] + \epsilon\right)^{-1/2}.
\end{equation}
All references to LayerNorm in Supplementary Fig.~\ref{fig:orbital-generator} refer to this variant.
Similarly, all linear layers have no bias unless otherwise specified, to retain sparsity.

\paragraph{Message passing block.}
We use a Message Passing Neural Network (MPNN)\citep{gilmer2017neural,gilmer2020message} to exchange information using the set of nuclei $\molconf$, in parallel over different orbitals.

Our message passing layer is designed so that the message between nuclei $n$ and  $n'$ on orbital $o$ tends to zero if either
\begin{itemize}
    \item the distance $\|\Rv_n- \Rv_{n'}\|$ is large, or
    \item the input features for both $n$ and $n'$ are zero on that orbital.
\end{itemize}
These restrictions mean that orbital information `propagates outward' from the initial single nucleus centre of a certain orbital.
To implement this concretely, we define the message $\mathbf{M}_{o,n\to n'}$ as the pointwise product of three terms
\begin{enumerate}
    \item a linear function of the nuclei--nuclei edge features which decay to zero in the large-distance limit,
    \item a linear function of the current representations of the receiver and sender with a tanh activation, \\$\tanh\left(W^{\text{sender}}\mathbf{Q}^{(\ell)}_{on} + W^{\text{receiver}}\mathbf{Q}^{(\ell)}_{on'}\right)$, which will be zero if both receiver and sender are zero, and
    \item a linear function of nuclei embeddings for the receiver and sender nuclei (derived from the Nuclei MPNN), with a sigmoid activation.
\end{enumerate}
Messages are split into multiple heads and summed across the sender dimension.
Additionally, we use the same tanh-activated MLP layer with Pre-LN as we use in the Electron Transformer. Note that this gives a zero update when the input is zero, as required.

\paragraph{Orbital-dependent coefficients}
As a final step of the Orbital Generator, we apply two different gating layers to produce separate representations for up- and down-spin electrons, denoted $Q^\sigma_{onf}$ where $\sigma=\uparrow,\downarrow$.
The orbital representations are fed back into the main ansatz by extracting envelope coefficients $W^{\text{env},\sigma}_{donf}$ and electron representation coefficients $W^{\text{repr},\sigma}_{dof}$.
We form $W^{\text{env},\sigma}_{donf}$ by applying linear layers to $Q^\sigma_{onf}$; we use different layers for the two spins, and for each determinant.  
For $W^{\text{repr},\sigma}_{dof}$, we first sum $Q^\sigma_{onf}$ over the nuclei axis, and then apply a linear layer, with multiple layers for spins and determinants as before.

We found that with standard training from a random initialization, a single determinant rapidly dominates the others, leading to an effective single-determinant wavefunction.
To partially mitigate this, we design an initialization scheme so that $W^{\text{env},\sigma}_{donf}$ and $W^{\text{repr},\sigma}_{dof}$ are (almost) equal for different determinants $d$ at initialization.
We defer a full discussion of this issue to Supplementary Note~\ref{sec:app:training:multi-det}.

\subsection{Nuclei MPNN}
The Nuclei MPNN uses the same architecture as the Orbital Generator, but without the presence of the orbital axis. 
The initial representations are simple embeddings of atom types.
We use standard LayerNorm because sparsity is not a requirement for this network. 
The Message Passing blocks use only two inputs: current nuclei representations and nucleus--nucleus edge features.
The final nuclei representations are returned after the last Message Passing block.
With these simplifications, we see that our Nuclei MPNN is similar to a standard MPNN for molecular graphs.

The matrices $W^\text{rot}_n$ that are fed from the Nuclei MPNN into the Electron Transformer are simple linear projections of the final nuclei representations.
In the Orbital Transformer, the nuclei representations are fed directly into each Message Passing layer as seen in Supplementary Fig.~\ref{fig:orbital-generator}.

\subsection{Single molecule mode}
\label{sec:app:arch:single_mol_mode}
The three points of exchange between the nuclei-only parts of the network and the electron-dependent part of the network are: the nuclei representations, the envelope coefficients $W^{\text{env},\sigma}_{donf}$ and the electron representation linear weights $W^{\text{repr},\sigma}_{dof}$.
When training or fine-tuning Orbformer on a single molecule, it is possible to make these tensors simple, trainable parameters.
Thus, the Orbital Generator and Nuclei MPNN are simply used to generate \emph{initializations} for these parameters, and these subnetworks can then be discarded.
We found that this results in more memory efficient single-point fine-tuning, but that the number of gradient steps required to converge could be larger than when fine-tuning the entire network on a single molecule. 
This mode was only used to generate single-point reference deep QMC energies for the ethane, formamide, 1-propanol, 2-aminopropan-2-ol, and L-alanine dissociation curves presented in the main text.
Single molecule mode is not suitable for training on  multiple molecules simultaneously.

\subsection{Table of architecture hyperparameters}
Architecture settings are shown in Supplementary Table~\ref{tab:app:architecture}.
\begin{table}[ht]
    \caption{Dimensions and parameter values used for the Orbformer architecture.}
    \label{tab:app:architecture}
    \centering
    \begin{tabular}{llr}
    Component & Parameter & Value \\
    \hline
    Electron Transformer & Attention  heads     &  8 \\
    & Featurization heads & 16 \\
    & Feature dimension (concat over heads), $F_\text{repr}$ & 256 \\
    & Attention layers & 4 \\
    & Featurization filters & 4 \\
    & Initial electron--nuclei feature dimension, $F_\text{init}$  & 40 \\
    Orbital Generator & Message passing heads & 4 \\
    & Feature dimension (concat over heads), $F_\text{orb}$ & 128 \\
    & Message passing layers & 3 \\
    Nuclei MPNN & Message passing heads & 2 \\
    & Feature dimension (concat over heads), $F_\text{nuc}$ & 64 \\
    & Message passing layers & 3 \\
    Envelopes & Exponents per nucleus, $F_\text{env}$ & 8 \\
    Determinants & Number of determinants & 16 \\
    \end{tabular}
\end{table}

\subsection{Relation to existing transferable QMC methods}
\label{sec:app:arch:related-work}

Orbformer builds on recent deep QMC work that has explored the possibility of transferability across molecules by taking the molecular configuration as input \citep{gao2021abinitio,scherbela2023variational}, most notably \citet{gao2023generalizing} and \citet{scherbela2024towards}.
Here we summarize how the design of Orbformer relates to, and differs from, these existing efforts.

\paragraph{Orbital construction.}
Comparing to the Globe model \citep{gao2023generalizing}, our orbital construction does not hardcode core, valence or bonding orbitals and avoids any discontinuities or jumps in the orbital representations.
Thus, when we discuss `core' orbitals in Sec.~\ref{sec:inner-workings} of the main text, these are an emergent phenomenon rather than an architectural design choice.
We achieve this by allowing a single atom type to define multiple initial orbitals, and relying on the message passing layers to identify `bonding' orbitals if and when they exist.
Our approach avoids assumptions that are associated with closed shell molecules, enabling us to better deal with bond dissociations.
Comparing to \citet{scherbela2024towards}, we do not use Hartree--Fock when setting orbital initializations.
This allows our network greater flexibility and avoids entangling our orbital generation with any problems of Hartree--Fock in multireference settings.
Our message passing algorithm also differs from all previous work in its specifics, such as the normalization layers and the formation of messages as a product of three terms (cf. Supplementary Note~\ref{sec:app:arch:orb}).

\paragraph{Composability and size consistency.}
Our approach to constructing localized orbitals is similar to \citet{gao2023generalizing}, who showed that the block diagonal form of a Slater matrix, \eqref{eq:composability}, can be achieved using localization constraints.
We prefer the term `composability' to emphasize the fact that these constraints do not lead to size consistency in the multi-determinant case.
We note some slight differences between the two formulations in Supplementary Note~\ref{sec:app:arch:basic-principles}.

\section{Sampling}
\label{sec:app:sampling}
\subsection{Self-generating electron configuration data for transferable training}
One of the most promising aspects of single-molecule VMC training is that training data is generated from the model itself and used to converge the ansatz towards the theoretical ground state wave function, as discussed in Supplementary Note~\ref{sec:app:background:vmc}.
This property of VMC marks a major departure from statistical learning paradigms that currently predominate in machine learning for natural science.

Here, we consider extending the self-generative training procedure of VMC to train our transferable Orbformer ansatz across a region of chemical space.
Specifically, we consider transferable training in which molecules $\molconf$ are sampled from a training distribution, $p_\text{train}$.
The capability to train against an arbitrary training distribution allows us to contemplate using very large datasets of molecular structures with data augmentation.
This is a significant change from previous work where Scherbela \textit{et al.} \citep{scherbela2023variational} using a dataset of 700 molecular structures was the largest molecular dataset yet considered in the field, to the best of our knowledge.

In order to train on a dataset of unlimited size, we must find a way to self-generate training data for new, unseen molecules.
That is, we must be able to sample $\xv_{1:\nelec} \sim |\Psi_\thetav(\xv_{1:\nelec} \mid \molconf)|^2$ for a molecule $\molconf$ that has not been previously encountered during training.
Previous work \citep{scherbela2024towards} side-stepped this problem by maintaining separate MCMC samplers for each training molecule in their training dataset, but this solution does not scale.

\paragraph{Multiple optimizer steps.}

The naive solution to this problem is to burn in new MCMC walkers from scratch at each training step. 
However, with the standard MCMC sampling approaches, the number of steps required to reach equilibrium is large and becomes even larger as molecule size increases.
As a first step to reduce the cost of re-equilibration, we fix the training molecules $\molconf$ for multiple optimization steps, allowing us to re-use MCMC walkers that have already burned in by running only a few additional MCMC steps between optimizer steps on the same molecules. 

\paragraph{Faster burn-in with Unadjusted Langevin Algorithm.}
Previous work  has used either Metroplis--Hastings \citep{metropolis1953equation} with Gaussian proposals or Metroplis--Adjusted Langevin Algorithm (MALA) \citep{grenander1994representations} to sample electron configurations.
MALA combines a Langevin proposal with step-size $\tau$
\begin{equation}\label{eq:langevin}
    \xv_{1:\nelec}^{(\text{prop}, t+1)} =  \xv_{1:\nelec}^{(t)} + \tau^{(t)} \nabla_{\xv_{1:\nelec}} \log |\Psi_\thetav| + \sqrt{\tau^{(t)}} \xi^{(t)}
\end{equation}
with a Metropolis--Hastings accept/reject step.

When using MALA, we observed that the step size $\tau$ must become very small in order to maintain a reasonable acceptance rate.
Sample rejection also tends to slow down mixing.
A large number of steps are therefore required to reach the equilibrium distribution.

In other fields of machine learning, Langevin sampling has seen remarkable success, for example, in the field of diffusion models \citep{song2019generative}. 
These applications use Unadjusted Langevin Algorithm (ULA) \citep{roberts1996exponential,roberts1998optimal}, i.e.~they do not apply an accept/reject step.
In the transferable VMC context, we therefore propose an initial sampling phase using ULA with step size annealing \citep{neal2001annealed,welling2011bayesian} to get close to the correct distribution.
We switch to MALA towards the end of sampling to reach perfect equilibrium, and we use MALA exclusively when updating samplers between multiple optimization steps on the same molecules.
The benefits of this strategy are illustrated in Supplementary Note~\ref{sec:app:ula-ablation}.

\paragraph{Improved Langevin sampler.}
We made several other modifications to the standard algorithm.
First, we apply norm clipping to the gradient term in \eqref{eq:langevin}.
Second, we modify the gradients when electron samples are very near to nuclei to prevent electron samples passing `through' a nucleus.
Third, we observed that it is possible for some MCMC chains to get stuck in low probability regions. This is common in partially dissociated molecules, where the probability of having charges imbalanced between the dissociated fragments becomes low as the wave function is trained.
Our solution to this problem is to simply remove walkers with very low $\log|\Psi_\thetav|$ values and replace them with a uniform sample from the surviving walkers.
Given $\log|\Psi_\thetav|$ values $L^{(b)}$ for a batch $b=1,\dots,B$, we let $L^{(\text{med})}$ denote the median value over the batch and $L^{(\text{spread)}} = \frac{1}{B} \sum_{b=1}^B |L^{(\text{med})} - L^{(b)}|$.
We prune the walkers where $L^{(b)} < L^{(\text{med})} - 4L^{(\text{spread)}}$.

\paragraph{Electron configuration sampling for fine-tuning}
When fine-tuning on single molecules or reaction curves, we use Metropolis--Hastings and we maintain separate MCMC chains for each molecule.

\subsection{Details of ablation study comparing ULA and MALA sampling}\label{sec:app:ula-ablation}
\begin{figure}
    \centering
   \includegraphics[width=0.5\textwidth]{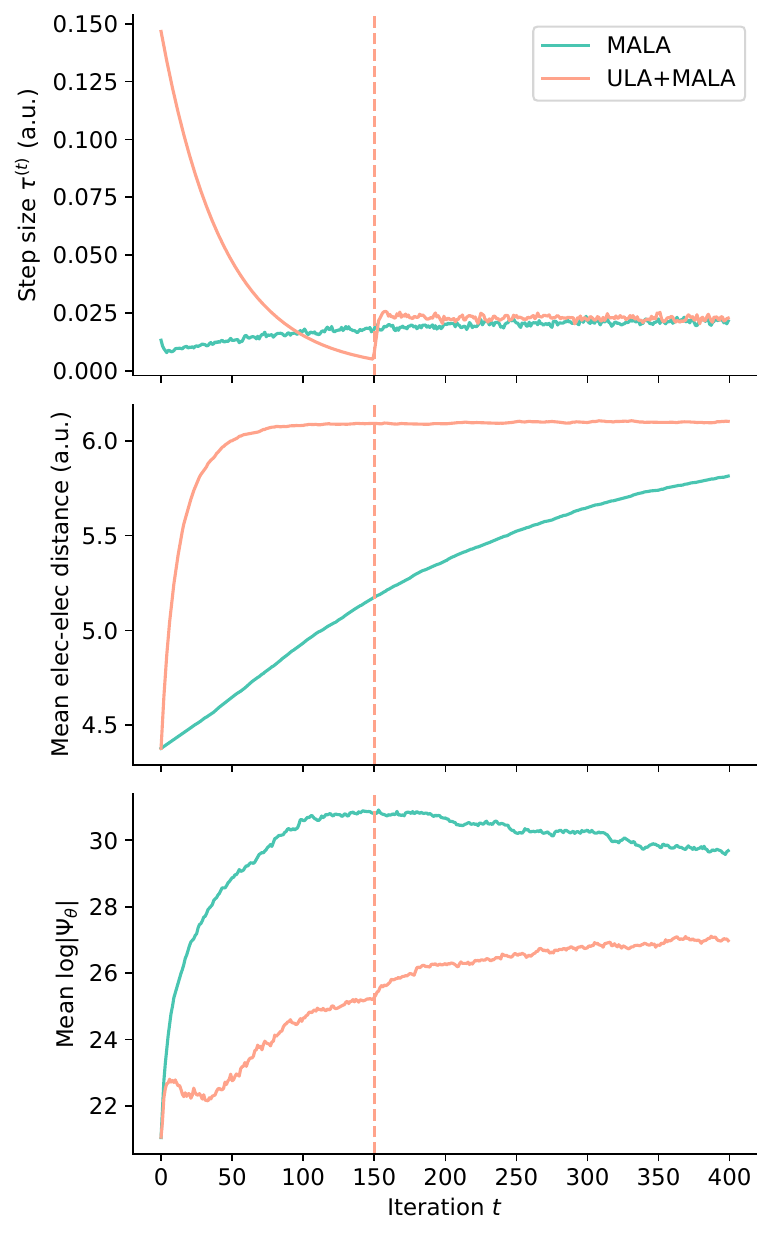}
    \caption{\raggedright \textbf{Illustrating the benefits of initial ULA sampling.} We ran Langevin sampling with 512 parallel chains on a randomly initialized Orbformer wave function for the benzene molecule at equilibrium geometry. We ran standard MALA for 400 steps with a target acceptance of 57\%, and we separately ran 150 steps of ULA followed by 250 steps of MALA. Distance units are Bohr (a.u.). 
    }
    \label{fig:double-langevin-study}
\end{figure}

For this study, we used a randomly initialized Orbformer network in single-molecule mode on the benzene molecule in its equilibrium geometry.
When running ULA, we annealed the step size from 0.15 to 0.005 using a geometric scheduler. We clipped the gradient norm to a maximum of 5.
As can be seen on Supplementary Fig.~\ref{fig:double-langevin-study}, a pure MALA sampler begins with a small step size and makes slow progress by first moving samples to a high probability region and then diffusing outwards. 
Initial ULA sampling very rapidly moves samples near to equilibrium, with MALA completing the equilibration.

\section{Variational Training}
\label{sec:app:training}
\subsection{The local energy}
\label{sec:app:training:local_energy}
Where single-molecule VMC training approaches a ground state wave function by minimizing the energy, we train our transferable ansatz to jointly represent the ground state wave function for an entire class of molecules by minimizing the expected value of the energy over the training distribution of molecules.
Following Supplementary Note~\ref{sec:app:background:vmc}, this expected energy can be written as a nested expectation
\begin{equation}
\label{eq:loss}
    \mathcal{L}_\text{energy} = \mathbb{E}\left[ \mathcal{E}_\text{loc}\left( \xv_{1:\nelec}, \molconf \right) \right]
\end{equation}
where the expectation involves sampling molecules $\molconf \sim p_\text{train}$ and then sampling electron configurations conditional on those molecules $\xv_{1:\nelec} \sim |\Psi_\thetav(\xv_{1:\nelec} \mid \molconf)|^2$ using the technique outlined in Supplementary Note~\ref{sec:app:sampling}.
The local energy is 
\begin{equation}
    \mathcal{E}_\text{loc}\left( \xv_{1:\nelec}, \molconf \right) = \frac{\hat{H}\Psi_\thetav(\xv_{1:\nelec} \mid \molconf)}{\Psi_\thetav (\xv_{1:\nelec} \mid \molconf)},
\end{equation}
which can be rewritten more explicitly using \eqref{eq:hamiltonian} as 
\begin{equation}
    \mathcal{E}_\text{loc}\left( \xv_{1:\nelec}, \molconf \right) = -\frac{1}{2}\frac{\nabla_{\rv_{1:\nelec}}^2 \Psi_\thetav(\xv_{1:\nelec} \mid \molconf) }{\Psi_\thetav(\xv_{1:\nelec} \mid \molconf)} + V(\xv_{1:\nelec},\molconf)
\end{equation}
where $\nabla^2$ is the Laplacian operator and $V$ is the Coulomb potential.
Whilst evaluating the potential given samples is fast, the computation of the Laplacian of the network $\Psi_\thetav$ with respect to the electron coordinate inputs is a significant computational bottleneck.
Our efforts to improve the efficiency of the Laplacian computation are discussed in Supplementary Note~\ref{sec:app:laplacian} in detail.

\subsection{Energy gradient computation}
\label{sec:app:local-energy-gradient}
To compute the gradient of the energy expectation value it is common to employ the ``gradient trick''. 
Utilizing the hermiticity of the Hamiltonian, the naive gradient of the energy is replaced with an unbiased estimator that involves at most second derivatives (instead of third derivatives arising due to taking the derivative of the kinetic energy operator with respect to the ansatz parameters).
For completeness we provide the derivation of the energy gradient 
\newcommand{\grad}{\nabla_{\thetav}}
\begin{align}
    \grad \braket{\hat{H}} &= \grad \frac{\int\Psi_\thetav \hat{H}\Psi_\thetav}{\int |\Psi_\thetav|^2}\\
    &= \frac{\int\grad\Psi_\thetav \hat{H}\Psi_\thetav}{\int |\Psi_\thetav|^2} - \frac{\int\Psi_\thetav \hat{H}\Psi_\thetav}{\int |\Psi_\thetav|^2}\frac{\int\grad\Psi_\thetav^2}{\int |\Psi_\thetav|^2} \\
    &= \frac{\int\Psi_\thetav \hat{H}(\grad\Psi_\thetav)+(\grad\Psi_\thetav)\hat{H}\Psi_\thetav}{\int |\Psi_\thetav|^2}- \braket{\hat{H}}\frac{\int2\Psi_\thetav(\grad\Psi_\thetav)}{\int |\Psi_\thetav|^2}\\
    &= \frac{2\int\Psi_\thetav \frac{\hat{H}\Psi_\thetav}{\Psi_\thetav}(\grad\Psi_\thetav)-2\energy\ \Psi(\grad\Psi_\thetav)}{\int |\Psi_\thetav|^2}\\
    &= \frac{2\int \Psi_\thetav^2(\frac{\hat{H}\Psi_\thetav}{\Psi_\thetav}-\energy)\grad\log\Psi_\thetav}{\int |\Psi_\thetav|^2} \\
    &= 2\mathbb{E}_{\Psi_\thetav^2}\Big[\big(\energy_\text{loc}-\energy\big)\grad\log\Psi_\thetav\Big]\,,
\end{align}
where 3~$\rightarrow$~4 utilizes the hermiticity of the Hamiltonian $\hat{H}$, 4~$\rightarrow$~5 adopts the equality $\grad\log\Psi=\tfrac{\grad \Psi}{\Psi}$ and 5~$\rightarrow$~6 introduces the local energy $\energy_\text{loc}=\tfrac{H\Psi_\thetav}{\Psi_\thetav}$. For better readability we use $\Psi_\thetav$ as shorthand for $\Psi_\thetav(\xv_{1:\nelec}\mid\molconf )$ and integrals are taken over the full domain of the wave function.

The generalization of the energy gradient to the transferable setting is straightforward. 
With the loss function being an expectation over contributions of molecular configurations $\molconf$, the gradient becomes
\begin{align}\grad\mathbb{E}_\molconf\braket{\hat{H}^\molconf}=2\mathbb{E}_{\molconf}\mathbb{E}_{\mathbf{r}\sim\Psi_\thetav^2}\Big[\Big(\energy_\text{loc}^\molconf-\energy^\molconf\Big)\grad\log\Psi_\thetav(\xv_{1:\nelec}\mid\molconf)\Big]\,,
\end{align}
where $\hat{H}^\molconf$ refers to the Hamiltonian implementing the nuclear potential of the molecule $\molconf$, $\energy^\molconf$ is the corresponding energy expectation value, $\energy_\text{loc}^\molconf$ is the local energy for that molecular configuration and the wave function is a function of (a varying number of) electrons with a parametric dependence on the molecular configuration.

\subsection{Multiple determinants}
\label{sec:app:training:multi-det}
Departing from almost all existing work in the area \citep{gao2023generalizing,gao2024neural,hermann2020deep,glehn2022self,scherbela2024towards}, we do not pretrain Orbformer towards a Hartree--Fock wave function.
We found that the Hartree--Fock solution can be very far from the correct wave function for molecules that are not near to their equilibrium geometry (e.g.~bond breaking).
Creating a Hartree--Fock pretraining mechanisms for a transferable ansatz is also significantly more burdensome than using it for single molecule training.

However, when training from a random initialization, we encountered the problem that a single determinant rapidly comes to dominate the wave function, leading to an effective single determinant model.
To overcome this, we first initialize all determinants to be equal as outlined in Supplementary Note~\ref{sec:app:arch:orb}, but we found that this alone was not sufficient to prevent determinant collapse.
The relative weighting of determinants can be described by the probability distribution
\begin{equation}
    \pi_{\text{det},d} = \frac{\det[A_\thetav^d]}{\sum_{d'}\det[A_\thetav^{d'}]}.
\end{equation}
To encourage all determinants to play a role in the wave function, we introduced the following regularization term to the training objective
\begin{equation}
    \mathcal{L}_\text{det} = -\lambda_\text{det}\mathbb{E}\left[ \sum_d \log \pi_{\text{det},d}(\xv_{1:\nelec} \mid \molconf) \right]
\end{equation}
where the expectation is with respect to the training data, as in \eqref{eq:loss}.
This regularization term is minimized when all determinants contribute equally to the wave function. 
Its gradient can be computed directly by automatic differentiation.

Surprisingly, we discovered that \emph{without Hartree--Fock pretraining} but using this multi-determinant penalty and an equal-determinant initialization we could obtain the same energy as was obtained when using Psiformer with Hartree--Fock pretraining.

\subsection{Ablation study on determinant collapse and Hartree--Fock pretraining}
\label{sec:app:det-collapse}
For this study, we trained from scratch on the \ch{CH4} molecule. We used an electron batch size of 4096 split across 4 GPUs. 
When Hartree--Fock pretraining was used, we ran 20k steps using the same settings as outlined in the original work introducing the Psiformer architecture \cite{glehn2022self}.
When the multi-determinant penalty was applied, we used the regularization weight $\lambda_\text{det} = 0.001$.
In Supplementary Fig.~\ref{fig:multi-det-ch4}, we see that training Psiformer without Hartree--Fock pretraining is equivalent to training a single determinant model (with or without pretraining).
It appears, therefore, that the benefit of Hartree--Fock pretraining has little to do with the quality of Hartree--Fock itself, but arises simply from preventing determinant collapse.
\begin{figure}
    \centering
    \includegraphics[width=0.55\linewidth]{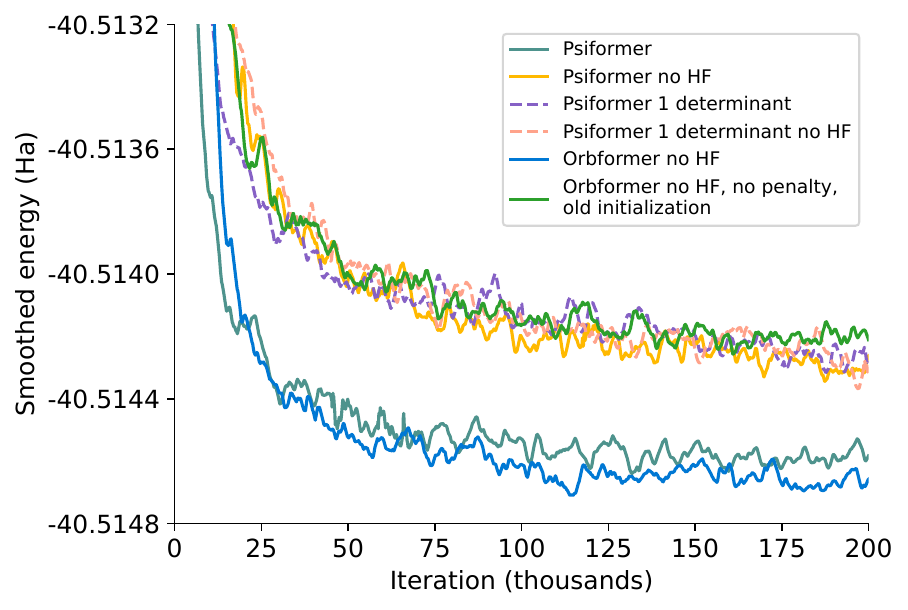}
    \caption{\raggedright \textbf{Determinant collapse.} We study the connection between Hartree--Fock pretraining and determinant collapse on the \ch{CH4} molecule at equilibrium geometry. We compare standard Psiformer training to variants with a single determinant or without Hartree--Fock (HF) pretraining. Orbformer is trained from scratch without Hartree--Fock but uses our multi-determinant penalty and initialization. 
    }
    \label{fig:multi-det-ch4}
\end{figure}

\subsection{Optimization with K-FAC}
The ansatz parameters $\thetav$ are updated to minimize the combined loss $\mathcal{L}_\text{energy} + \mathcal{L}_\text{det}$ using the K-FAC optimizer \citep{martens2015optimizing} implemented in the \texttt{kfac-jax} package \citep{kfac-jax2022github}.
Computing the gradient of $\mathcal{L}_\text{energy}$ with respect to $\thetav$ is analogous to the single molecule case as shown in Supplementary Note~\ref{sec:app:local-energy-gradient}.
We explored several alternatives to K-FAC but were unable to find one that matched its stability and speed of training.

We also apply clipping to the local energies, as this has shown itself to be essential in earlier VMC work with neural networks \citep{hermann2020deep,pfau2020ab,glehn2022self}. We apply clipping separately to local energies from different molecules.
Furthermore, for multi-molecule training, we observe that local energies may be in very different scales for different molecules.
During chemical pretraining, we therefore reweight the local energies in a similar manner to Globe \citep{gao2023generalizing} by computing the molecule-conditional local energy standard deviation, $s(\molconf)$, and rescaling the local energies for that molecule by $\min(2, 1 / s(\molconf))$; additionally we scale up the gradient of $\mathcal{L}_\text{det}$ by a factor of $s(\molconf)$.

\subsection{Tables of settings}
Training settings are shown in Supplementary Table~\ref{tab:app:training}.
\begin{table}[ht]
    \caption{Hyperparameter values used for the Orbformer training and sampling.}
    \label{tab:app:training}
    \centering
    \begin{tabular}{llr}
    Component & Parameter & Value \\
    \hline
    Local energy & Clipping width     &  5.0 \\
    Multi-determinant loss & Regularization weight $\lambda_\text{det}$ & 0.001 \\
    Optimizer (K-FAC) & Damping & 0.001 \\
    & Learning rate & 0.05 \\
    & Norm constraint & 0.001 \\
    Sampling & ULA steps at re-equilibration & 150 \\
    & ULA initial step size & 0.15 \\
    & ULA final step size & 0.005 \\
    & MALA steps at re-equilibration & 150 \\
    & MALA target acceptance rate & 57\%  \\
    & Decorrelation steps & 60 \\
    & Langevin gradient norm clipping & 5.0 \\
    & Pruning width & 4.0 
    \end{tabular}
\end{table}

\section{Efficient wavefunction Laplacians}
\label{sec:app:laplacian}
The widespread success of foundation models is due in large part to the outstanding computational efficiency with which they can be trained on increasingly larger datasets.
For chemically transferable VMC to even approach the scaling behaviour of foundation models, reducing the compute time required to evaluate the Laplacian is essential.

\subsection{Computing the Laplacian in a single forward pass}
The Forward Laplacian framework is a new approach to computing the Laplacian of a neural network which can result in VMC training that is significantly faster than a naive implementation \cite{li2024computational}. The 
essential principle of the Forward Laplacian is encapsulated in a chain rule for the Laplacian
\begin{equation}
    \frac{\partial^2 v}{\partial x^2} = \frac{\partial v}{\partial u}\frac{\partial^2 u}{\partial x^2} + \frac{\partial u}{\partial x}\frac{\partial^2 v}{\partial u^2}\frac{\partial u}{\partial x}.
\end{equation}
where $v = v(u(x))$.
The Forward Laplacian framework propagates the value, gradient and Laplacian forwards through the computational graph of a neural network.
For further details, see 
\citet{li2024computational}.
We use the Forward Laplacian implementation in the \texttt{folx} package \cite{gao2023folx}.

\subsection{Forward Laplacian of edge-biased attention}
\label{sec:app:laplacian:flash-laplacian}
For most layers of Orbformer, we were able to rely on standard Forward Laplacian update rules implemented in \texttt{folx}.
However, we early on identified the electron--electron attention layers (cf.~Supplementary Note~\ref{sec:app:arch:elec-transformer}) as some of the most demanding computations in our model.
Our use of a variant of FlashAttention \cite{dao2022flashattention} accelerated the forward and backward passes through the attention layers.
However, 
no existing FlashAttention implementation exists for the Laplacian.

We therefore introduce a FlashAttention-inspired algorithm that computes the forward Laplacian update rule of the attention layers.
This was possible by making use of the \texttt{pallas} kernel language in JAX \citep{jax2023pallas}.
The resulting kernel evaluates the Laplacian update rule for multi-head attention significantly faster than a na\"{i}ve Forward Laplacian implementation applied to an un-optimized attention layer.
Our kernel is publicly accessible in the \texttt{folx} package, while full details of the update rule are given in the following.

We parallelize all computations over the batch and head dimensions of the attention computation, distributing these to separate streaming multiprocessors. 
Here, we therefore deal only with the attention algorithm for a single head and without any batch dimensions.

\subsubsection{Forward pass}
We denote the number of electrons as $E$ and the feature dimension as $F$. 
The forward pass of edge-biased attention takes as input: queries ${\bm{\mathcal{Q}}} \in \mathbb R^{E\times F}$, keys ${\bm{\mathcal{K}}} \in \mathbb R^{E\times F}$, values $\bm{\mathcal{V}} \in \mathbb R^{E\times F}$, and an edge bias $\bm{\mathcal{B}} \in \mathbb R^{E \times E}$ and computes the output $\bm{\mathcal{O}} \in \mathbb{R}^{E\times F}$ as follows
\begin{align}
    \mathcal{S}_{qk} &= \sum_f \mathcal{Q}_{qf}\mathcal{K}_{k f} +\mathcal{B}_{qk},\\
    \mathcal{P}_{qk} &= \frac{\exp(\mathcal{S}_{qk})}{\sum_{k'} \exp(\mathcal{S}_{qk'})},\\
    \mathcal{O}_{qf} &= \sum_k \mathcal{P}_{qk} \mathcal{V}_{kf}.
\end{align}

\subsubsection{Forward gradient}
We use the notation $\dot{x}$ to represent the gradient of any tensor with respect to a scalar input. 
Our forward gradient operates in parallel over the $3E$ coordinates of the electron inputs.
\begin{align}
    \dot{\mathcal{S}}_{qk} &= \sum_f \dot{\mathcal{Q}}_{qf}\mathcal{K}_{k f} +\sum_f \mathcal{Q}_{qf}\dot{\mathcal{K}}_{k f} + \dot{\mathcal{B}}_{qk},\\
    \dot{\mathcal{P}}_{qk} &= \sum_{k'} \frac{\partial \mathcal{P}_{qk}}{\partial \mathcal{S}_{qk'}} \dot{\mathcal{S}}_{qk'} = \sum_{k'} \mathcal{P}_{qk'}(\delta_{kk'} - \mathcal{P}_{qk}) \dot{\mathcal{S}}_{qk'} = \mathcal{P}_{qk}\left(\dot{\mathcal{S}}_{qk} - \sum_{k'} \mathcal{P}_{qk'}\dot{\mathcal{S}}_{qk'} \right), \\
    \dot{\mathcal{O}}_{qf} &= \sum_k \dot{\mathcal{P}}_{qk} \mathcal{V}_{kf} + \sum_k \mathcal{P}_{qk} \dot{\mathcal{V}}_{kf}.
\end{align}

\subsubsection{Forward Laplacian}
The second derivatives $\ddot{x}$ can be calculated similarly
\begin{align}
    \ddot{\mathcal{S}}_{qk} &= \sum_f \ddot{\mathcal{Q}}_{qf}\mathcal{K}_{k f} +\sum_f \mathcal{Q}_{qf}\ddot{\mathcal{K}}_{k f} + \ddot{\mathcal{B}}_{qk} &&+ 2 \sum_f \dot{\mathcal{Q}}_{qf}\dot{\mathcal{K}}_{k f}, \\
    \ddot{\mathcal{P}}_{qk} &= \mathcal{P}_{qk}\left(\ddot{\mathcal{S}}_{qk} - \sum_{k'} \mathcal{P}_{qk'}\ddot{\mathcal{S}}_{qk'} \right) &&+ \dot{\mathcal{P}}_{qk}\left(\dot{\mathcal{S}}_{qk} - \sum_{k'} \mathcal{P}_{qk'}\dot{\mathcal{S}}_{qk'} \right) - \mathcal{P}_{qk}\sum_{k'} \dot{\mathcal{P}}_{qk'}\dot{\mathcal{S}}_{qk'}, \\
    \ddot{\mathcal{O}}_{qf} &= \sum_k \ddot{\mathcal{P}}_{qk} \mathcal{V}_{kf} + \sum_k \mathcal{P}_{qk} \ddot{\mathcal{V}}_{kf} &&+ 2 \sum_k \dot{\mathcal{P}}_{qk} \dot{\mathcal{V}}_{kf}.
\end{align}
The Laplacian is then defined as the sum of second derivatives with respect to every electron coordinate
\begin{equation}
    \Delta x = \sum_{e=1}^E \sum_{i=1}^3 \frac{\partial^2 x}{\partial r_{ei}^2}.
\end{equation}
In our algorithm, we separate the parts of the Laplacian that can be computed directly from $\Delta  \mathcal{Q}, \Delta  \mathcal{K}, \Delta  \mathcal{V}, \Delta \mathcal{B}$ from the parts of the Laplacian involving some terms from the Jacobian.

To accelerate this algorithm using \texttt{pallas}, we note that every operation of the Forward Laplacian update is fully parallel over the query axis (indices labelled as $q$). This means that we can break inputs into blocks along the $q$-axis and distribute them to separate streaming multiprocessors.
Thus, intermediate tensors such as $\bm{\mathcal{S}}$, $\bm{\mathcal{P}}$ and their derivatives are not materialized explicitly, leading to significant performance improvements.

We note that we have not explored writing a \texttt{pallas} kernel for the cross-attention mechanism that was introduced by \citet{li2024computational}, this has the potential to be even faster than our current set-up, whilst deviating further from standard transformers.

\subsubsection{Backward gradient}
We also implemented a simplified version of the backward gradient that is found in FlashAttention.
The exact calculations, for a terminal scalar loss denoted $\mathcal{L}$, are 
\begin{align}
    \frac{\partial \mathcal{L}}{\partial \mathcal{V}_{ka}} &= \sum_q \mathcal{P}_{qk} \frac{\partial \mathcal{L}}{\partial \mathcal{O}_{qa}}, \\
    \frac{\partial \mathcal{L}}{\partial \mathcal{P}_{qk}} &= \sum_a \mathcal{V}_{ka} \frac{\partial \mathcal{L}}{\partial \mathcal{O}_{qa}}, \\
    \frac{\partial \mathcal{L}}{\partial \mathcal{S}_{qk}} &= \sum_{k'} \frac{\partial \mathcal{L}}{\partial \mathcal{P}_{qk'}}\frac{\partial \mathcal{P}_{qk'}}{\partial \mathcal{S}_{qk}} \\
    &= \sum_{a} \frac{\partial \mathcal{L}}{\partial \mathcal{O}_{qa}}\mathcal{V}_{ka} \mathcal{P}_{qk} - \sum_{a} \frac{\partial \mathcal{L}}{\partial \mathcal{O}_{qa}}\mathcal{O}_{qa} \mathcal{P}_{qk}, \\
    \frac{\partial \mathcal{L}}{\partial \mathcal{B}_{qk}} &= \frac{\partial \mathcal{L}}{\partial \mathcal{S}_{qk}}, \\
    \frac{\partial \mathcal{L}}{\partial \mathcal{Q}_{qa}} &= \sum_k \frac{\partial \mathcal{L}}{\partial \mathcal{S}_{qk}}\mathcal{K}_{ka}, \\
    \frac{\partial \mathcal{L}}{\partial \mathcal{K}_{ka}} &= \sum_q \frac{\partial \mathcal{L}}{\partial \mathcal{S}_{qk}}\mathcal{Q}_{qa}.
\end{align}
For implementation, the primary difference here is that we \emph{cannot} parallelize the backward pass over the query axis for every input, because we require a summation over this axes for $\partial \mathcal{L}/ \partial \mathcal{V}$ and $\partial \mathcal{L}/ \partial \mathcal{K}$.
For the backward gradient of ${\bm{\mathcal{Q}}}$, we parallelize over the query axis and run an inner summation over the key axis (label $k$).
For the backward gradient of $\bm{\mathcal{K}}, \bm{\mathcal{V}}$, we parallelize over the key axis and run an inner summation over the query axis.
The backward gradient of $\bm{\mathcal{B}}$ is parallel over both axes.

\section{Large scale chemical pretraining}
\label{sec:app:pretraining}
\subsection{Curriculum learning for chemical space}
\label{sec:app:pretraining:curriculum}
Our objective is to pretrain a transferable wave function ansatz across a region of chemical space defined by a distribution over molecular structures, $p_\text{train}$.

\paragraph{Size-based curriculum.}
It is well known that \emph{gradient steps are much cheaper for smaller molecules}.
The computation of the Slater determinants, for instance, scales as $O(\nelec^3)$ in the number of electrons.
The design of Orbformer outlined in Supplementary Note~\ref{sec:app:arch} placed significant emphasis on `composability'. 
We expect that enforcing composability in our model will allow us to learn generalizable features on small molecules that will transfer to larger ones. 
Given this, it seems sensible to take as many optimization steps on the smallest possible molecules, where they are cheapest, before training on larger molecules. This led us to use a curriculum learning \citep{bengio2009curriculum} approach where we train Orbformer on datasets of increasingly large molecules.
We created two nested curriculum levels with up to 10 and 24 electrons respectively.

\paragraph{Choice of atomic species and molecules.}
We train Orbformer on a dataset containing the atomic species \ch{H}, \ch{Li}, \ch{B}, \ch{C}, \ch{N}, \ch{O}, and \ch{F}.
We first selected \ch{H}, \ch{C}, \ch{N}, \ch{O}, and \ch{F} as a basic set of atoms that together define a rich chemical space including a significant fraction of all organic molecules \cite{clayden2012organic}.
We also elected to include \ch{Li} and \ch{B} since they are light atoms that allow us to cover varied and well-studied chemistry beyond standard organic chemistry \citep{wietelmann2018200,grams2024rise}.
This choice of species allows us to convincingly demonstrate wave function generalization across chemical space and allows us to explore a variety of interesting applications, whilst keeping the number of electrons (and hence training cost) to a minimum. 
Our pretrained Orbformer is limited to operate on this set of light atoms and is not expected to work on other atom types.

Given this set of atoms, we identified a rich set of molecules of up to 24 electrons that contain any of these species.
We sought to include all common organic groups \cite{clayden2012organic} as well as commonly studied \ch{Li}- and \ch{B}-containing molecules \cite{wietelmann2018200, grams2024rise}.
We  restrict ourselves to molecules with no overall charge and with a spin state that is $S_z=0$ or $S_z=\tfrac{1}{2}$, as discussed in Supplementary Note~\ref{sec:app:arch:basic-principles}.
Several molecules were set aside as an out-of-distribution test set.
Supplementary Table~\ref{tab:oc_composition} lists all molecules selected.

\paragraph{Geometry sampling.}
As discussed in Sec.~\ref{sec:introduction}, the most promising application area for a chemically transferable wave function model is to molecular structures with `multireference character'.
It is calculations on these structures where the absence of a \emph{de facto} standard method from classical quantum chemistry \cite{multiref_charac,T1} leaves a space for neural network VMC to fill.
We expect that geometries that are highly perturbed relative to the equilibrium geometry of a molecule, e.g.~by bond breaking, are more likely to exhibit multireference character (we shortly confirm this for our own geometries). 
Yet the majority of existing work using neural network ans\"{a}tze for molecules has focused on equilibrium or near-equilibrium geometries \citep{hermann2020deep,pfau2020ab,glehn2022self,scherbela2024towards}.
We also expect that generalization is unlikely to occur from training on a few geometries per molecule---a significantly larger number of structures than has previously been considered is likely needed before the benefits of generalization will manifest.

We therefore generate a large number of perturbed geometries programmatically. Our starting point is the Z-matrix \citep{parsons2005practical} of the equilibrium geometry or a known stereoisomer.
We apply probabilistic distortion to the angles and bond lengths of the Z-matrix to create new off-equilibrium geometries. When simulating bond-breaking, we allow a single distance entry of the Z-matrix to be distorted by a much larger degree than the others.
Finally, we apply rejection sampling \citep{gilks1992adaptive} to prevent collision between any pair of atoms.
For each molecule, we generate 100--300 geometries using the five sampling procedures outlined in Supplementary Note~\ref{sec:app:datasets:lac-composition}.
In addition, we generate a number of hypothetical reaction mechanisms for small molecules (see Supplementary Table~\ref{tab:oc_composition}).

The resulting Light Atom Curriculum (LAC) contains 22350 structures; its molecular composition along with the proportion of potentially multireference structures are detailed in Supplementary Note~\ref{sec:app:datasets:lac-composition}.

We divide LAC into two curriculum levels.
Structures with up to 10 electrons with bend, stretch, and break geometries are included in LAC Level 1.
The full dataset is referred to as Level 2 and contains all of Level 1.

Finally, we apply on-the-fly data augmentation to all of our structures by (i) applying a random rotation, (ii) adding independent Gaussian noise of standard deviation $0.1$ Bohr to every atom coordinate.

\subsection{Infrastructure and training phases}
We trained Orbformer using up to 16 NVIDIA A100 PCIe GPUs.
Thanks to improvements in model stability, we were able to use TensorFloat-32 \citep{tf32} precision for matrix multiplication during training.
We conducted three sequential phases of pretraining.
We found that further dividing LAC Level 1 training into two separate phases, the first using only bend and stretch geometries, led to more stable training.
During Level 1 training, we used 8 GPUs with one molecule per GPU; for Level 2 training we increased to 16 GPUs.
If we encountered a NaN value during training, we wound the model back to the last checkpoint, drew a new batch of training molecules, and re-equilibrated electron samples to continue training. We allowed up to 200 restarts per phase.

\section{Fine-tuning and Evaluation}
\label{sec:app:finetune}
\subsection{Transferable fine-tuning}
We found that it was efficient to fine-tune similar structures together using a single set of parameters and simultaneous training. This allows us to use the transferable nature of Orbformer twice---once to pretrain the model, and again to share compute when fine-tuning.
For example, we fine-tune different geometric structures of the same molecule together, and for the Diels--Alder reaction, we fine-tuned reactants, products, and transition states together.
This can often lead to an order of magnitude reduction in cost over and above the benefits of pretraining.

We recommend always fine-tuning Orbformer on target structures, as we found that at least a few steps were always necessary. 
Even with no fine-tuning, energy evaluation of the Orbformer wave function requires running MCMC and computing local energies on target structures. 
Thus, in practical terms, if the amount of fine-tuning is of the same order of magnitude as the number of evaluation steps, fine-tuning does not significantly increase the cost of using Orbformer.

\subsection{Energy evaluation}
\label{sec:app:eval}
To correctly assess the energy expectation value of the wave function, we re-equilibrate new MCMC chains after the fine-tuning and compute the local energy over many MCMC steps. 
The final energy estimate should be the average over the chains and time steps, however, we found in practice that a few extreme values could lead to significant deterioration in the estimates of the mean.
We therefore estimate a \emph{robust mean} of the evaluation data by minimizing the Huber loss \citep{huber1992robust} over the set of evaluated local energies
\begin{align}
\ell_\delta(x) &:=  \begin{cases}
    \frac{1}{2} x^2 & \text{if } |x|<\delta \\
    \delta (|x| - \frac{1}{2}\delta) &\text{ otherwise},
\end{cases}\\
    \energy_\text{robust} &= \min_x \sum_b \sum_t \ell_\delta(x - \energy_{bt})
\end{align}
where $b$ and $t$ respectively indicate the batch of MCMC samplers and the time step (with suitable decorrelation).
The variance between energy estimates from different chains gives an estimate of the statistical error in our energy estimate.

\subsection{Metrics}
\label{sec:metrics}
The primary metric we use to compare Orbformer energies to a reference is mean absolute relative error (MARE). 
Suppose we have a reaction curve with geometries indexed by $i=1,\dots,n$. Suppose that we have corresponding energy curves $(E_i)_{i=1}^n$ and $(E'_i)_{i=1}^n$. Let $\bar{E}=\frac{1}{n}\sum_{i=1}^n E_i$ and $\bar{E'} = \frac{1}{n}\sum_{i=1}^n E'_i$. Then the mean absolute relative error is defined by $\text{MARE}(E_1,\dots,E_n;E'_1,\dots,E'_n) = \frac{1}{n} \sum_{i=1}^n |(E_i - \bar{E}) - (E'_i - \bar{E'})|$.

\section{Structure generation and dataset composition}
\label{sec:app:datasets}
\subsection{Light Atom Curriculum dataset}\label{sec:app:datasets:lac-composition}
\begin{table}
    \centering
    \caption{\raggedright Composition of the Light Atom Curriculum (LAC) dataset of 22350 structures. Formulas marked with a star contribute two stereo-conformers. The excluded molecules are the test molecules from the TinyMol dataset \citep{scherbela2024towards}. }
    \resizebox{\textwidth}{!}{
    \begin{tabular}{clrc|clrc}
        Subset                    & Formula       & $N_\text{elec}$ & Geometries                                                                                               & Subset                    & Formula       & $N_\text{elec}$ & Geometries \\ \hline \hline
        \multirow{10}*{Dimers}    & \ch{H2}       & 2               & \multirow{10}*{\makecell{50 stretch \\ 50 break}}                                                         & \multirow{16}*{6 atoms} & \ch{LiBH4}    & 12              & \multirow{16}*{\makecell{50 bend \\ 50 stretch \\ 50 break \\ 50 stretch \& bend \\ 100 break and bend}} \\
                                  & \ch{LiH}       & 4              &                                                                                                          &                           & \ch{C2H4}    & 16              & \\
                                  & \ch{HF}       & 10              &                                                                                                          &                           & \ch{CH3OH}*   & 16              & \\
                                  & \ch{C2}       & 12              &                                                                                                          &                           & \ch{LiCHCH2}   & 18              & \\
                                  & \ch{LiF}       & 12              &                                                                                                          &                           & \ch{N2H4}*   & 18              & \\
                                  & \ch{CO}       & 14              &                                                                                                          &                           & \ch{H2BCCH}  & 20              & \\
                                  & \ch{N2}       & 14              &                                                                                                          &                           & \ch{HBCCH2}    & 20              & \\
                                  & \ch{NO}       & 15              &                                                                                                          &                           & \ch{LiOCH3}     & 20              & \\
                                  & \ch{O2}      & 16               &                                                                                                          &                           & \ch{CH2BOH}     & 22              & \\
                                  & \ch{F2}      & 18              &                                                                                                          &                           & \ch{CH2CHF}  & 22              & \\ \cline{1-4}
        \multirow{13}*{Trimers}   & \ch{CH2}      & 8              & \multirow{13}*{\makecell{50 bend \\ 50 stretch \\ 50 break \\ 50 stretch \& bend \\ 100 break \& bend}}  &                           & \ch{CH3BO}    & 22             & \\
                                  & \ch{H2O}      & 10              &                                                                                                          &                           & \ch{CH3CN} & 22              & \\
                                  & \ch{LiOH}       & 12              &                                                                                                          &                           & \ch{CH3NO}  & 24              & \\
                                  & \ch{CNH}      & 14              &                                                                                                          &                           & \ch{HBO2H2}*   & 24              & \\
                                  & \ch{HCN}      & 14              &                                                                                                          &                           & \ch{HCONH2}   & 24              & \\
                                \cline{5-8}  & \ch{Li2O}      & 14              &                                                                                                          &     
        \multirow{9}*{7 atoms}   & \ch{CH3BH2}      & 16              & \multirow{9}*{\makecell{50 bend \\ 50 stretch \\ 50 break \\ 50 stretch \& bend \\ 100 break \& bend}} \\
                                  & \ch{CHF}      & 16              &                                                                                                          &                           & \ch{CH3NH2}*     & 18              & \\
                                  & \ch{HNO}      & 16              &                                                                                                          &                           & \ch{LiCH2NH2}      & 20              & \\
                                  & \ch{LiCN}      & 16               &                                                                                                          &                           & \ch{CH2BNH2}    & 22              & \\
                                  & \ch{LiOF}      & 20              &                                                                                                          &                           & \ch{CH3BNH}    & 22              & \\
                                  & \ch{FCN}     & 22              &                                                                                                          &                           & \ch{CH3NBH}     & 22              & \\
                                  & \ch{N2O}     & 22              &                                                                                                          &                           & \ch{C2H4O}     & 24              & \\
                                  & \ch{O3}     & 24              &                                                                                                          &                           & \ch{CH2CHOH}*      & 24              & \\
                                  \cline{1-4}
        \multirow{15}*{4 atoms} & \ch{BH3}    & 8              &  \multirow{15}*{\makecell{50 bend \\ 50 stretch \\ 50 break \\ 50 stretch \& bend \\ 100 break \& bend}} &                           & \ch{CH3CHO}*     & 24              & \\
                                  \cline{5-8} & \ch{NH3}     & 10              &                                                                                                          &               \multirow{4}*{8 atoms}            & \ch{C2H6}*    & 18 &      \multirow{4}*{\makecell{50 bend, stretch, break \\ 50 stretch \& bend \\ 100 break \& bend}}         \\
                                  & \ch{LiNH2}    & 12              &                                                                                                          &                           & \ch{LiCH2CH3}    & 20              & \\
                                  & \ch{C2H2} & 14              &                                                                                                          &                           & \ch{BH2CHCH2}  & 22              & \\
                                  & \ch{BH2F} & 16              &                                                                                                          &                           & \ch{CH3BCH2}    & 22              & \\
                                  \cline{5-8} & \ch{CH2O}   & 16              &                                                                                                          &         \multirow{7}*{Reactions}            & \ch{H4 ->} 2\ch{H2}   & 4              &  \multirow{7}*{50 structures}  \\
                                  & \ch{H2N2}*   & 16              &                                                                                                          &                           & \ch{C2H4 -> C2H2 + H2}  & 16              &  \\
                                  & \ch{H2O2}*   & 18              &                                                                                                          &                           & \ch{H2O2 -> H2O + O.}   & 18              & \\
                                  & \ch{LiCCLi}   & 18              &                                                                                                          &                           & \ch{CH3OH -> CO} + 2\ch{H2} & 18              &  \\
                                  & \ch{LiCHO}  & 18              &                                                                                                          &                           & \ch{N2H4 -> H2N2 + H2}   & 18              & \\
                                  & \ch{NH2F}   & 18              &                                                                                                          &                           & \ch{C2H6 -> C2H4 + H2}  & 18              & \\
                                  & \ch{LiOOLi}    & 22              &                                                                                                          &                           & 2\ch{HF -> H2 + F2}     & 20              & \\
                                  \cline{5-8} & \ch{BHF2}   & 24             &                                                                                                          &      \multirow{3}*{H-bonding}                     & 2\ch{H2O}    & 20              & 50 structures \\
                                  & \ch{CHOF}    & 24              &                                                                                                          &                           & \ch{H2O + NH3}    & 20              & \multirow{2}*{100 structures} \\
                                  & \ch{HNO2}*      &24               &                                                                                                          &                           & \ch{H2O + HF}     & 20              & \\
                                  \cline{1-8}
        \multirow{12}*{5 atoms} & \ch{CH4}    & 10              &  \multirow{12}*{\makecell{50 bend \\ 50 stretch \\ 50 break \\ 50 stretch \& bend \\ 100 break \& bend}} &     \multirow{4}*{Excluded}                      & \ch{C3H4}     & 22              & \multirow{4}*{n/a} \\
                                & \ch{LiCH3}   & 12             &                                                                                                          &                           & \ch{CN2H2}    & 22              & \\
                                  & \ch{CH2BH}    & 14              &                                                                                                          &                           & \ch{CNOH}    & 22              & \\
                                  & \ch{CH2NH}      & 16               &                                                                                                          &                           & \ch{CO2}     & 22              & \\
                                  & \ch{CH3F}    & 18              &                                                                                                          &                           &     &               & \\
                                  & \ch{HBCBH}      & 18               &                                                                                                          &                           &      &               & \\
                                   & \ch{NH2OH}    & 18              &                                                                                                          &                           &     &               & \\
                                  & \ch{LiCH2F}      & 20               &                                                                                                          &                           &      &              & \\
                                  & \ch{LiNHOH}    & 20              &                                                                                                          &                           &    &               & \\
                                  & \ch{LiONH2}      & 20               &                                                                                                          &                           &      &               & \\
                                  & \ch{CH2BF}    & 22              &                                                                                                          &                           &     &               & \\
                                  & \ch{HCOOH}      & 24               &                                                                                                          &                           &      &               & \\
    \end{tabular}
    }
    \label{tab:oc_composition}
\end{table}

\begin{figure*}
    \centering
    \includegraphics[width=0.9\linewidth]{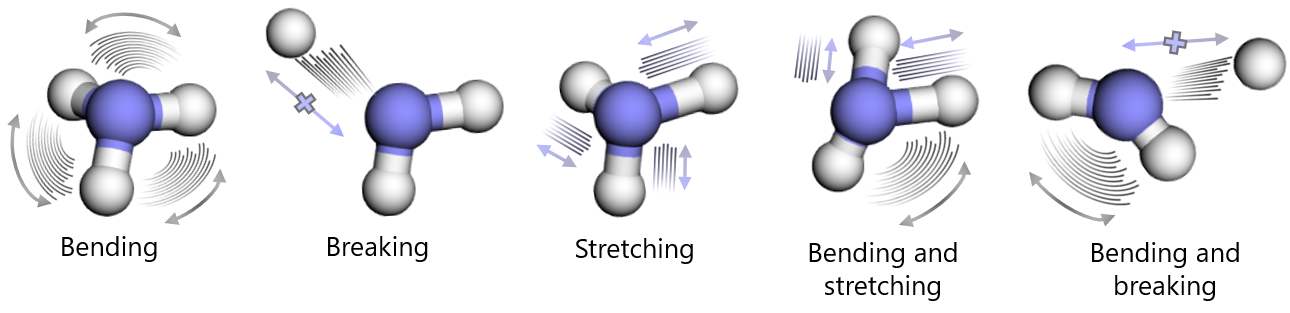}
    \caption{\raggedright \textbf{Geometry sampling methods used for programmatic generation of the LAC dataset.} We generated 50 structures using bending, breaking, stretching and bending-with-stretching for every molecule, and 100 structures using bending-with-breaking. Note that rotation is applied separately as a data augmentation.}
    \label{fig:lac-geometry-sampling}
\end{figure*}

Supplementary Table~\ref{tab:oc_composition} details the composition of the LAC dataset and Supplementary Fig.~\ref{fig:lac-geometry-sampling} illustrates the basic geometry distortion types used to generate a diverse set of configurations for the molecules of Supplementary Table~\ref{tab:oc_composition}.

\begin{figure}
    \centering
    \includegraphics[width=0.55\linewidth]
    {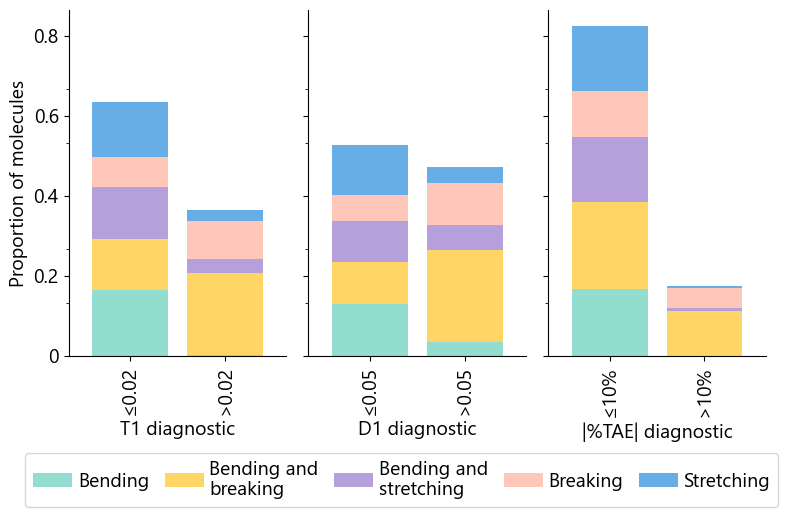}
    \caption{\raggedright \textbf{Multireference diagnostics for the LAC.} The T1 D1, and $|\%\text{TAE}|$ diagnostics computed using CCSD/cc-pVDZ wave functions for all the geometries of \ch{BH3}, \ch{LiCH3}, \ch{C2H4} and \ch{NH2F} that are present in the LAC dataset. These established diagnostics indicate that LAC dataset contains a significant fraction of structures with multireference character. 
    }
    \label{fig:diagnostic-bars}
\end{figure}

To assess the multireference character of our molecules, we compute T1, D1 and the percentage of the contribution of (T) to the total atomization energy ($|\%\text{TAE}|$) diagnostics \cite{multiref_charac} using CCSD(T)/cc-pVDZ calculations via \textsc{PySCF 2.7.0} \cite{sun2018pyscf} for structures of 5 selected molecules with different electron numbers in the LAC dataset (1500 structures in total). The $|\%\text{TAE}|$ diagnostic is computed using
\begin{equation}
\label{eq:tae}
    |\%\text{TAE}| = \frac{|\text{TAE}[\text{CCSD(T)}] - \text{TAE}[\text{CCSD}]|}{\text{TAE}[\text{CCSD(T)}]},
\end{equation}
where TAE[CCSD(T)] and TAE[CCSD] are the total atomization energies computed via CCSD(T)/cc-pVDZ and CCSD/cc-pVDZ, respectively.
We apply the criteria of T1 $\geq$ 0.02, D1 $\geq$ 0.05, and $|\%\text{TAE}|$ $\geq$ 10\% to indicate if a structure might exhibit multireference character \cite{multiref_charac}. %
Supplementary Fig.~\ref{fig:diagnostic-bars} displays the proportion of structures with multireference character for each geometry distortion. 
There are a significant number of structures showing multireference character, particularly with geometries distorted via bond breaking. 
We repeated this analysis for the much smaller TinyMol dataset, finding that very few, if any, structures exhibit multireference character in that dataset. 

\subsection{Bond breaking MEP}\label{sec:app:bbmep_dataset}
We take the BSE49 \citep{BSE49} dataset as our inspiration when constructing our own BBMEP dataset.
BSE49 is a diverse dataset that benchmarks the homolytic dissociation energies of various molecules (\ch{A-B}) by computing the closed-shell molecules and their corresponding radical fragments (\ch{A.} and \ch{B.}). However, it remains challenging to compute structures and energies along the entire dissociation curve due to the multireference character of those structures. In this study, we select five closed-shell molecules with different electron sizes from BSE49 \cite{BSE49} dataset (Supplementary Table~\ref{tab:bbmep_sys}) and construct their corresponding minimum energy paths (MEPs), i.e.,~transition paths between initial and final images, of the homolytic bond breaking reactions (\ch{A-B -> A. + B.}), which is referred to as the BBMEP dataset. The \ch{A-B} reactant structures (initial images) are directly extracted from BSE49 dataset and the product structures (final images) are constructed by combining the two radicals along its original bond direction at 6 \AA. The bonds that are broken are shown in the Figure~\ref{fig:bbmep_result} in the main text. The Nudged Elastic Band (NEB) \cite{NEB} method is a commonly used approach to find the MEPs for homolytic cleavage reactions, and we apply the climbing-image \cite{CI-NEB} AutoNEB \cite{AutoNEB} implementation in \textsc{ASE 3.22.1} \cite{ASE} with 20 total images relaxed by PBE \citep{PBE} / def2-TZVP \cite{def2-basis1} with D3BJ \cite{D3,D3BJ} corrections computed using \textsc{ORCA 5.0} \cite{ORCA5}. The initial and final images are provided to the AutoNEB and the initial guesses of the intermediate images are generated by image dependent pair potential (IDPP) \cite{IDPP} approach and then optimized with FIRE \cite{FIRE} optimizer and a maximum force of 0.05 (fmax=0.05) along the NEB path. The final collected MEP structrues are included in our GitHub repository.
Please note that the endpoint structures that we use in our 20-point dissociation pathways are \emph{not} the same as the BSE49 endpoint structures, but are produced from the AutoNEB procedure.

\begin{table}
    \centering
    \caption{Composition of the bond breaking minimum energy path (BBMEP) dataset. Each path contains 20 total images, including the initial and final images.}
    \label{tab:bbmep_sys}
    \begin{tabular}{cccc}
    \hline
    System & Reaction & Dissociated bond & $N_\text{elec}$ \\ \hline \hline
    Ethane & \ch{CH3CH3}\ch{->}\ch{CH3CH2. + H.} & \ch{C-H} & 18 \\
    Formamide &  \ch{HCONH2}\ch{->}\ch{HCONH. + H.} & \ch{N-H} & 24\\
    1-Propanol &  \ch{CH3CH2CH2OH}\ch{->}\ch{CH2CH2CH2O. + H.} & \ch{O-H} & 34\\
    2-Aminopropan-2-ol & \ch{(CH3)2NCH2OH}\ch{->}\ch{.CH2N(CH3)2 + .OH } & \ch{C-O} & 42 \\
    L-Alanine & \ch{HOOCCH(CH3)NH2}\ch{->}\ch{HOOC. + .CH(CH3)NH2} & \ch{C-C} & 48\\ \hline
    \end{tabular}
\end{table}

\textbf{Deep QMC reference energies.}
The deep QMC reference energies that we computed for our BBMEP dissociation curves are presented in Supplementary Table~\ref{tab:bbmep_ref}. The protocol used to compute these values is described in Supplementary Table~\ref{tab:experiment_hp}. Note that we use much larger batch sizes for the reference calculations, ran separate single-point calculations for every geometry. To reduce similarity between our own deep QMC calculations and the references, we used the Psiformer ansatz for Ethane and Formamide. However, we struggled to converge Psiformer for the larger molecules. We therefore used Orbformer in single-molecule mode for these reference calculations---this version of the ansatz has no Orbital Generator network making it closer in spirit to the Psiformer while being more stable.

We perform two validations on these reference energies. First, Supplementary Table~\ref{tab:experiment_hp} shows the statistical errors in the Monte Carlo evaluation of the energy that we performed for each of these calculations. All statistical errors are well below 1 mHa.
Secondly, we validate our \emph{protocol} by running it on the original BSE49 endpoints. 
We run our protocol in two different ways---first by running separate calculations on the two dissociated fragments, and second by running on a combined system of the two fragments separated by 1000 \AA. 
We can then compare these values to the literature dissociation energies. Since the overall dissociation energy depends only on the endpoints, it can be computed more reliably with single-reference methods, as was done in \citet{BSE49} 
In Supplementary Table~\ref{tab:bbmep_dissociation}, we see that the two deep QMC references are in excellent agreement with each other, this gives and also agree well with the literature values.
With these validations, we believe that our reference energies for our BBMEP structures, including the multireference structures, are the most accurate obtainable reference values.

\textbf{Energy conversion calculations.} A single NVIDIA A100 GPU is 300 W and a 64 core AMD EPYC 7763 CPU is 280W, therefore the GPU/CPU power conversion factor is $300/(280/64) = 68.571$.
\begin{table}
    \centering
    \caption{Reference deep QMC absolute energies of five MEP curves in Hartree. The stochastic errors are also included.}
    \label{tab:bbmep_ref}
    \begin{tabular}{cccccc}
    \hline
    Image ID & Ethane                 & Formamide              & 1-Propanol             & 2-Aminopropan-2-ol  & L-Alanine              \\ \hline
    0  & -79.82485(164) & -169.90316(12)  & -194.35847(2)  & -249.72372(12)  & -323.76290(22)  \\
    1  & -79.82256(55)  & -169.90378(1)   & -194.35793(4)  & -249.72086(9)   & -323.77085(26)  \\
    2  & -79.81550(57)  & -169.89904(10)  & -194.35672(7)  & -249.72220(10)  & -323.76719(29)  \\
    3  & -79.77022(40)  & -169.88887(9)   & -194.34606(10) & -249.72366(9)   & -323.76160(28)  \\
    4  & -79.70922(50)  & -169.86811(8)   & -194.31957(9)  & -249.72745(5)   & -323.76803(33)  \\
    5  & -79.65521(5)   & -169.73234(9)   & -194.23575(9)  & -249.72454(11)  & -323.77344(10)  \\
    6  & -79.63422(74)  & -169.71585(4)   & -194.20289(3)  & -249.69502(10)  & -323.75211(29)  \\
    7  & -79.60235(143) & -169.83813(6)   & -194.18426(4)  & -249.63980(8)   & -323.75864(3)   \\
    8  & -79.59743(121) & -169.86235(5)   & -194.17617(4)  & -249.61615(12)  & -323.76477(25)  \\
    9  & -79.63422(41)  & -169.86518(9)   & -194.17345(7)  & -249.59906(8)   & -323.76331(23)  \\
    10 & -79.64304(153) & -169.86527(6)   & -194.17316(5)  & -249.58461(7)   & -323.75992(11)  \\
    11 & -79.64785(146) & -169.85041(6)   & -194.17274(5)  & -249.57176(14)  & -323.76476(15)  \\
    12 & -79.64688(145) & -169.73853(4)   & -194.17312(5)  & -249.56164(13)  & -323.75962(11)  \\
    13 & -79.64784(161) & -169.69446(8)   & -194.17262(9)  & -249.56518(14)  & -323.76217(5)   \\
    14 & -79.64879(50)  & -169.69352(4)   & -194.17324(9)  & -249.56474(26)  & -323.66576(19)  \\
    15 & -79.64920(17)  & -169.69160(8)   & -194.17328(7)  & -249.56572(29)  & -323.60473(8)   \\
    16 & -79.64993(97)  & -169.69411(14)  & -194.17306(7)  & -249.56542(8)   & -323.52371(5)   \\
    17 & -79.65030(78)  & -169.69243(18)  & -194.17617(13) & -249.56586(20)  & -323.60698(9)   \\
    18 & -79.65005(126) & -169.69549(12)  & -194.17523(17) & -249.56580(13)  & -323.60705(35)  \\
    19 & -79.64967(84)  & -169.69411(6)   & -194.17625(3)  & -249.56581(14)  & -323.61342(9)  \\ \hline
    \end{tabular}
\end{table}

\begin{table}
    \centering
    \caption{Comparison of dissociation energies for reactant and product structures from literature for five BBMEP curves (kcal/mol) using literature high-quality electronic structure calculations and our deep QMC reference protocol for the BBMEP experiment with large batch sizes and long training times. Note that the underlying geometries are distinct from the geometries in our BBMEP experiment; here we use geometries from the BSE49 dataset.}
    \label{tab:bbmep_dissociation}
    \begin{tabular}{cccc}
    \hline
    System & Literature (RO)CBS-QB3 \cite{BSE49} & deep QMC reference & deep QMC reference 1000 \AA \\ \hline
    Ethane &  109.29 & 108.80 & 108.80\\ 
    Formamide & 124.50 & 124.50 & 124.47\\ 
    1-Propanol & 113.00 & 112.33 & 112.44 \\ 
    2-Aminopropan-2-ol & 98.43$^\dagger$ & 97.45 & 97.68 \\ 
    L-Alanine & 90.26 & 89.79 & 89.83 \\ \hline
    \multicolumn{4}{l}{\small $^\dagger$ This value is reported in the BSE49 dataset ``BES49\_Existing\_1327" but labelled as ``Dimethylaminomethanol"}
    \end{tabular}
\end{table}

\subsection{Diels-Alder reaction dataset}
We aim to characterize the two possible pathways along which the cycloaddition of ethene and trans-butadiene might proceed.
The concerted and step-wise mechanisms naturally share their reactants (ethene, butadiene) and product (cyclohexene).
The concerted pathway proceeds through a single transition state structure, while the step-wise mechanism involves two transitions states flanking an intermediate structure.
Of the two step-wise transition states, here we only consider the first one, as it has the higher energy of the two \cite{lischka2004}, and therefore presumably determines the reaction rate along the pathway.
In total, this amounts to six considered molecular structures (ethene, butadiene, cyclohexene, concerted transition state, first step-wise transition state, step-wise intermediate).
The geometries of these species are taken from the work of Lischka \cite{lischka2004}, where they are optimized at the MRAQCC(6,6) level of theory.
Zero-point vibrational and thermal corrections for all species are also taken from the same work, computed using the B3LYP density functional and the 6-31G** basis set.

\section{Hyperparameters of each experiment}
\label{sec:app:experiment_hyperparams}
The hyperparameters of each experiment presented in the main text are summarized in Supplementary Table~\ref{tab:experiment_hp}.
\begin{table}[t]
    \centering
    \caption{Training hyperparameters for all the experiments. \textsuperscript{\emph a}Orbformer trained in single molecule mode (see Supplementary Note~\ref{sec:app:arch:single_mol_mode}). \textsuperscript{\emph b}Trained across 32 GPUs using the ULA/MALA sampling scheme: the same molecule will be sampled on multiple GPUs but with an independent batch of electron samples.}
    \label{tab:experiment_hp}
    \resizebox{\textwidth}{!}{%
    \begin{tabular}{ccccccccc}
    \hline
     Experiment & Dataset & Ansatz & \makecell{Training\\mode} &  \makecell{Electron\\batch size} & \makecell{Molecule\\batch size} & \makecell{Hartree--Fock \\ train steps} & \makecell{Training\\steps} & \makecell{Evaluation \\steps per\\ structure} \\
     \hline
     \multirow{3}*{TinyMol} & TinyMol pretrain & Orbformer & chemical pretraining & 1024 & 4 & - & 512 k & - \\
     & TinyMol test sets & Orbformer & \makecell{geom-transferable \\ fine-tuning} & 512 & 4 & - & 32 k  & 8k \\
     & TinyMol test sets & Psiformer & \makecell{single-point \\ training} & 2048 & 1 & 20 k & 32 k  & 2k \\
     \hline
     \multirow{4}*{\makecell{BBMEP \\ reference }}& Ethane & Psiformer & \makecell{single-point \\ training} & 2048 & 1 & 20 k & 200 k & 10 k\\
     & Formamide & Psiformer & \makecell{single-point \\ training} & 4096 & 1 & 20 k & 250 k & 10 k\\
     & \makecell{1-Propanol \& \\ 2-Aminopropan-2-ol}& Orbformer\textsuperscript{\emph a} & \makecell{single-point \\ training} & 4096 & 1 & - & 250 k & 10 k\\
     & L-Alanine & Orbformer\textsuperscript{\emph a} & \makecell{single-point \\ training} & 4096 & 1 & - & 300 k & 10 k\\
    \hline
     \multirow{3}*{\makecell{BBMEP \\ Psiformer scratch \\ (large batch), \\ single structures }}    &  Ethane & Psiformer & \makecell{single-point \\ training} & 1024 & 1 & - & 40 k & 6 k\\
    &  Formamide & Psiformer & \makecell{single-point \\ training} & 2048 & 1 & - & 50 k & 6 k\\ \\
     \hline
    \multirow{4}*{\makecell{BBMEP \\ Orbformer \\ scratch, single \\ structures }} &  Ethane & Orbformer & \makecell{single-point \\ training} & 1024 & 1 & - & 32 k & 6 k\\
    & Formamide & Orbformer & \makecell{single-point \\ training} & 1024 & 1 & - & 100 k & 6 k\\
    & 1-Propanol & Orbformer & \makecell{single-point \\ training} & 1024 & 1 & - & 120 k & 6 k\\
    & \makecell{2-Aminopropan-2-ol} & Orbformer & \makecell{single-point \\ training} & 1024 & 1 & - & 160 k & 6 k\\
    & L-Alanine & Orbformer & \makecell{single-point \\ training} & 1024 & 1 & - & 500 k & 6 k\\
    \hline
     \multirow{4}*{\makecell{BBMEP \\ Orbformer \\ scratch or \\ LAC fine-tuning, \\ all structures }} &  Ethane & Orbformer & \makecell{geom-transferable \\training/fine-tuning} & 256 & 4 & - & 32 k & 6 k\\
    & Formamide & Orbformer & \makecell{geom-transferable \\ training/fine-tuning} & 256 & 4 & - & 100 k & 6 k\\
    & 1-Propanol & Orbformer & \makecell{geom-transferable \\ training/fine-tuning} & 256 & 4 & - & 120 k & 6 k\\
    & \makecell{2-Aminopropan-2-ol} & Orbformer & \makecell{geom-transferable \\training/fine-tuning} & 256 & 4 & - & 160 k & 6 k\\
    & L-Alanine & Orbformer & \makecell{geom-transferable \\training or fine-tuning} & 256 & 4 & - & 500 k & 6 k\\
     \hline
     \multirow{2}*{\makecell{Diels--Alder}} & all structures & Orbformer & \makecell{single-point \\ training} & 4096 & 1 & - & 200k & 8k \\
    & all structures & Orbformer & \makecell{geom-transferable \\ fine-tuning} & 512 & 8 & - & 200k & 8k \\
    \hline
    Alkanes & Alkanes C6--13 & Orbformer & \makecell{molecule-transferable \\ fine-tuning} & 32 & 32\textsuperscript{\emph b} & - & 180k & 20k  \\
    \hline\hline
    \end{tabular}
    }
\end{table}

\section{Additional results}
\label{sec:app:additional-results}
\subsection{Additional results on the Diels--Alder reaction}
\label{sec:app:additional-results:da-results}
\begin{figure}[t]
    \centering
    \includegraphics[width=0.60\linewidth]{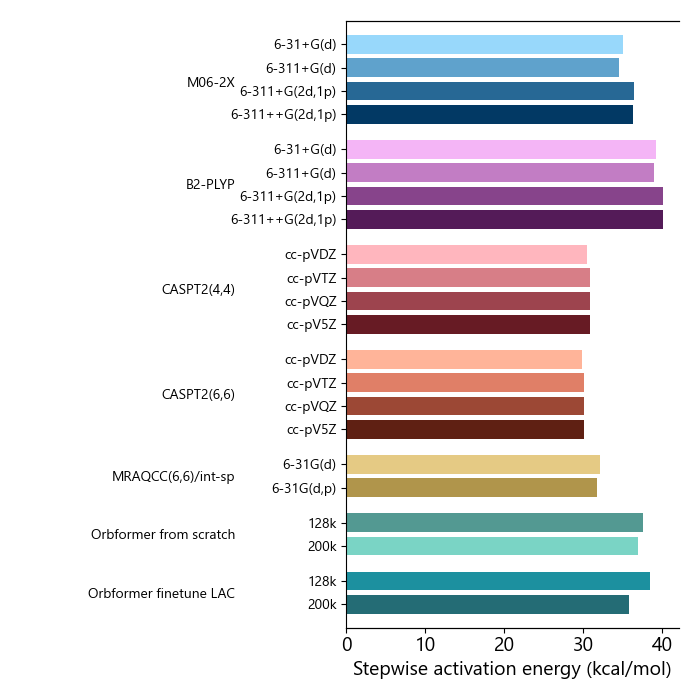}
    \caption{\raggedright \textbf{Stepwise pathway of the Diels--Alder reaction.} Activation energies of the Diels--Alder reaction studied in Sec.~\ref{sec:diels-alder} via the \emph{stepwise} pathway. All details are the same as in Fig.~\ref{fig:diels-alder-activation}. The experimental activation energy of 25.8 kcal/mol is below all results presented here, strongly indicating that the stepwise mechanism is not energetically favoured. 
    }
    \label{fig:diels-alder-stepwise-bar}
\end{figure}

After establishing the highly satisfactory match between Orbformer and experimental results on the iconic Diels--Alder reaction energy and concerted activation barrier, we turn our attention to the intricate step-wise mechanism, involving biradical transition states and intermediate structure.
Supplementary Fig.~\ref{fig:diels-alder-stepwise-bar} summarizes theoretical estimates for the first activation energy of this pathway.
As no experimental data is available for this quantity, theoretical results can only be compared among themselves.
The Orbformer result agrees best with the density functional estimates, predicting an activation energy somewhere between the M06-2X and B2-PLYP values.
Compared to these CASPT2 and MRAQCC predict a significantly lower energy barrier, just as for the concerted reaction pathway.
Overall, Orbformer's prediction that the concerted pathway is favoured to the step-wise one by about 10 kcal/mol is in good agreement with the currently available theoretical and experimental data \cite{cui2014thorough,uchiyama1964thermal}.

\subsection{Comparing to experimental reference for dinitrogen dissociation}
\label{sec:app:additional-results:n2}
\begin{figure}[t]
    \centering
    \includegraphics[width=0.55\textwidth]{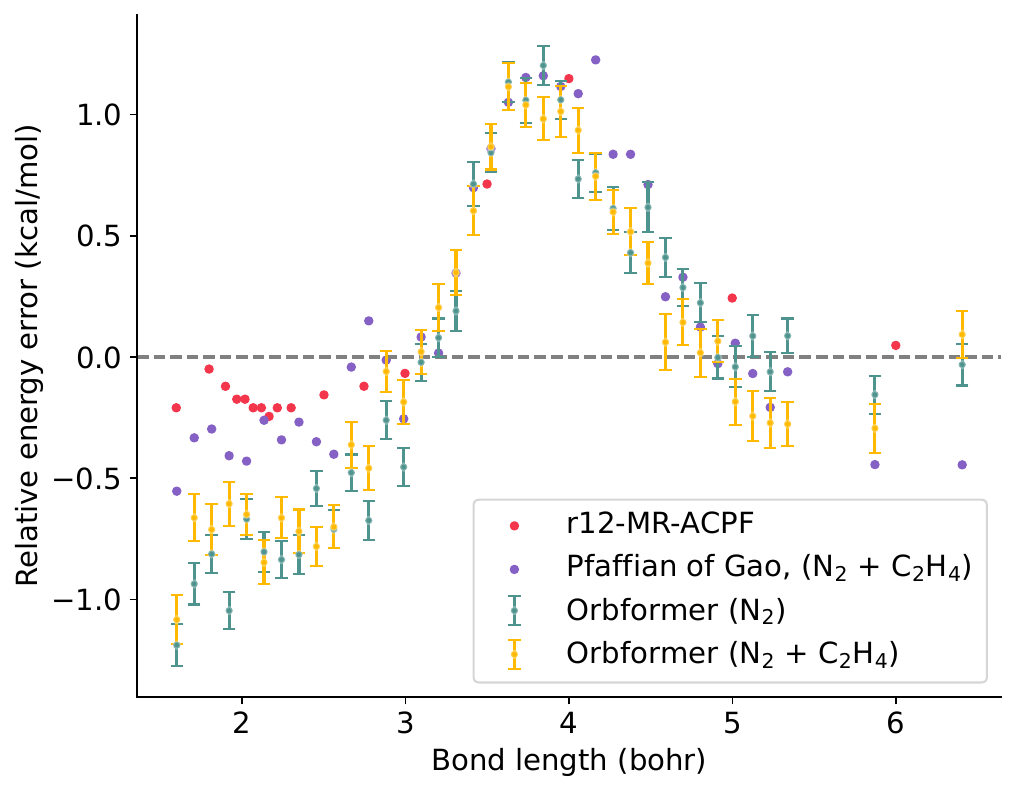}
    \caption{\textbf{Dinitrogen dissociation with and without distractors.} Relative energy of the \ch{N2} dissociation with and without ethene as a distractor molecule in the fine-tuning dataset.
    Additionally, results obtained with the r12-MR-ACPF reference theoretical method are also shown \cite{gdanitz1998accurately}.  The usual mean-shifting is applied to all energies. Error bars show the statistical error (one standard deviation of the estimates from independent MCMC chains, cf.~Suppl.~Note~\ref{sec:app:eval}). %
    }
    \label{fig:N2_distractor}
\end{figure}

Another valuable feature of Orbformer is the capacity to simultaneously model several molecules, with very different electronic structures, without loss in accuracy.
To demonstrate this property we consider the PES of the \ch{N2} dissociation and compare (geometrically) transferable training on a dataset of 38 \ch{N2} geometries with (chemically) transferable training on the same 38 \ch{N2} geometries and an additional ten geometries of ethene (\ch{C2H4}), replicating the setting of \citet{gao2023generalizing}.
This also gives us an opportunity to compare with one of the very few experimental references for the interior of a bond dissociation curve \cite{le2006accurate}.
Both trainings used a total of 200k iterations split evenly across the datasets of 38 and 48 molecules, respectively.
With \ch{N2} forming a triple bond and ethene containing double and single bonds only, these systems show very different characteristics albeit being of similar size.
Despite their dissimilarity, our results in Fig~\ref{fig:N2_distractor} show that training Orbformer with ethene as a distractor does not have a negative effect on relative energies and both runs are in perfect agreement with previous results \cite{gao2024neural}.
Moreover, these results confirm that Orbformer can match very well with an experimental reference, including in the strongly correlated regime. We note that the original experimental reference has uncertainty above 1 kcal/mol. The close agreement between deep-learning QMC and r12-MR-ACPF \cite{gdanitz1998accurately} (two completely unrelated theoretical methods) gives us further confidence in the adequacy of our theoretical description.
Comparing total energies of the geometrically transferable \ch{N2} optimization and the optimization on the distractor dataset, we obtain a small increase of about 0.2~mHa.
We attribute this to the reduced number of training iterations per geometry.
Orbformer's capability of representing distinct molecules is all the more remarkable in the context of its generalization properties in chemically transferable pretraining.

\subsection{Comparing iteration timing}
\label{sec:app:timing}
\begin{figure}[t]
    \centering
    \includegraphics[width=0.75\linewidth]{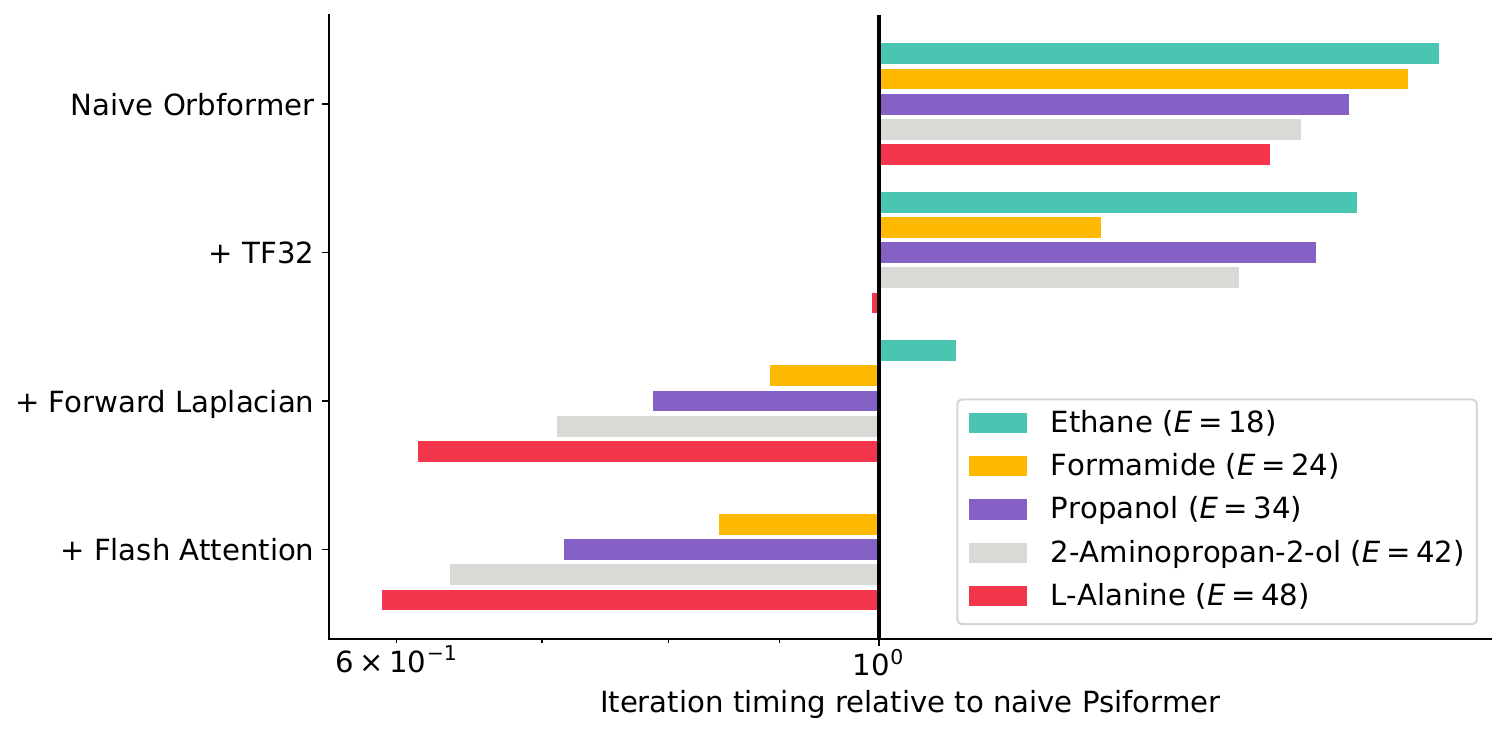}
    \caption{\textbf{Variational training iteration time reduction from engineering enhancements.} We show the relative iteration time of Orbformer compared to an unimproved Psiformer implementation, isolating the effects of TensorFloat-32 precision (TF32), Forward Laplacian and FlashAttention. 
    }
    \label{fig:timing}
\end{figure}

We compare iteration timing between different versions of Orbformer compared to a naive Psiformer implementation.
In this study, we used a batch size of 256 and ran on a single geometry of each of the molecules used in the bond-breaking experiments.
We clearly see that the larger molecules benefit more from the engineering improvements that we considered, which is consistent with \citet{li2024computational}.
Overall, these improvements give about a $1.7\times$ speed-up compared to a naive Psiformer and a $2.6\times$ speed-up compared to a naive Orbformer.

\subsection{Alkanes}
\label{sec:app:alkanes}

The largest molecule in this dataset contains 106 electrons, making it significantly larger than the LAC and larger than most structures previously tackled using deep QMC.
The alkanes dataset was constructed systematically with perfectly tetrahedral angles, with \ch{CC} bond length of $1.5210$\AA \ and \ch{CH} bond lengths of $1.0919$\AA.
This construction ensures that, when examining composability, the local environments are identical between different alkanes.

Beyond the model introspection presented in the main paper, we also sought to establish that our model had indeed converged on this dataset including some large molecules up to 106 electrons.
We fine-tuned the model across 32 GPUs with 32 electron samples per GPU. For this training, we used the ULA/MALA sampling scheme that we used for pretraining in order to get a better coverage of electron space.

Given our construction of the molecules, we would expect that the total energy scales almost exactly linearly in the number of carbon atoms. We verified that CCSD(T) indeed gives a linear energy scaling in energy for these molecules.
In Supplementary Fig.~\ref{fig:linear-scaling} (left) we fitted a linear model to the total energies of the eight alkanes and saw residuals of order no more than 1 kcal/mol, with an MSE of 0.92. This indicates that the model has converged well, and furthermore it gives an indication that, with joint fine-tuning, it is not the case that smaller molecules converge faster. However, it is also clear that linear scaling of the energy is not an automatically satisfied property for Orbformer. We assume this may be because 16 determinants are used for each molecule, which hampers exact properties like size consistency and size extensivity being present.
In Supplementary Fig.~\ref{fig:linear-scaling} (right), we looked at the difference between Orbformer total energies and CCSD(T)/CBS energies. Now it becomes apparent that, whilst both models achieve a linear scaling of energy with system size, the Orbformer total energies are significantly lower across the board, indicating that Orbformer does a much better job of capturing electron correlation.

\begin{figure}
    \centering
    \includegraphics[width=0.75\linewidth]{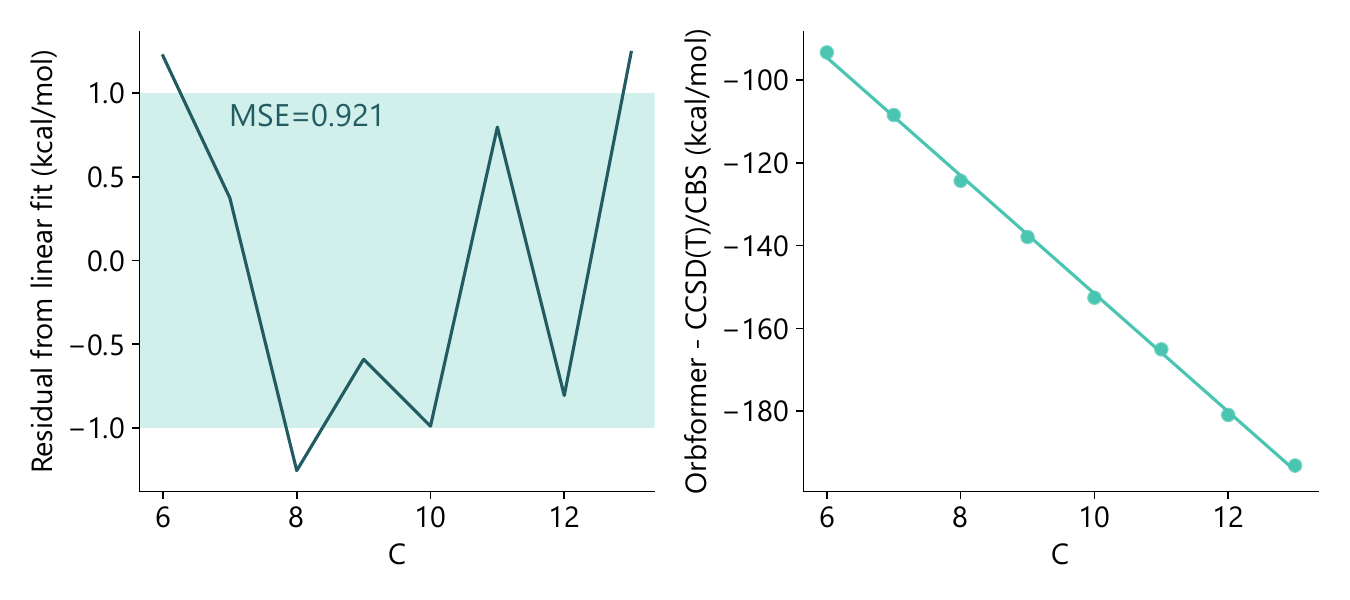}
    \caption{\raggedright \textbf{Evaluating model convergence on the alkanes dataset.} We use linear scaling of energy as a criterion for convergence. Left: residuals from a linear fit of a model fine-tuned from the LAC for 180k steps. Right: the difference between the aforementioned Orbformer model and CCSD(T)/CBS total energies, with a linear fit line. Here we see that Orbformer obtains much lower energies, whilst both observe a linear scaling in the molecule size. 
    }
    \label{fig:linear-scaling}
\end{figure}

\subsection{Azulene--naphthalene isomerization}
\label{sec:app:additional-results:azulene}
\begin{figure}
    \centering
    \begin{tabular}{c|c|c}
            \includegraphics[width=0.3\linewidth]{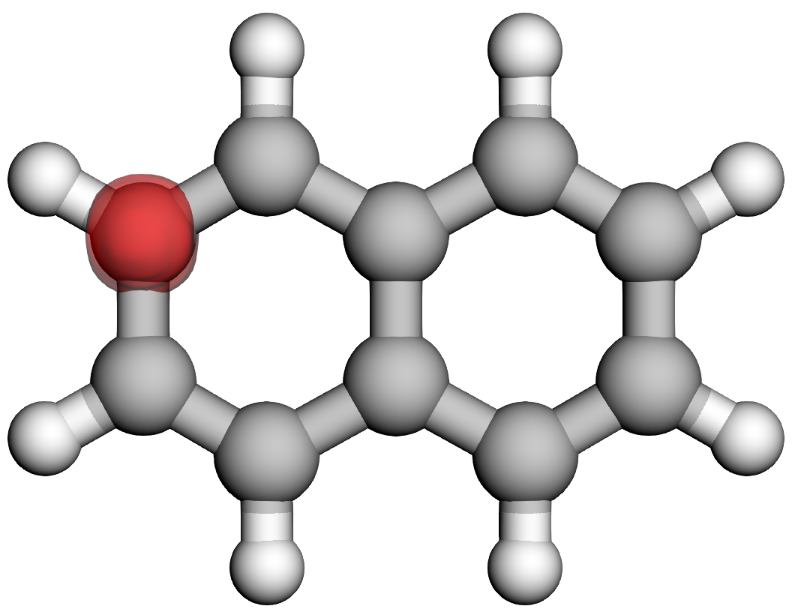} & \includegraphics[width=0.3\textwidth]{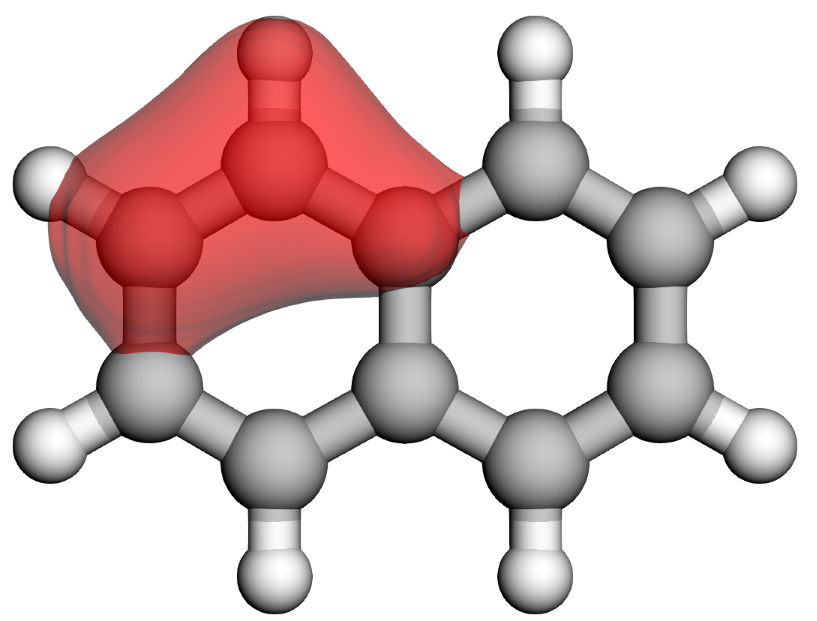} & \includegraphics[width=0.3\textwidth]{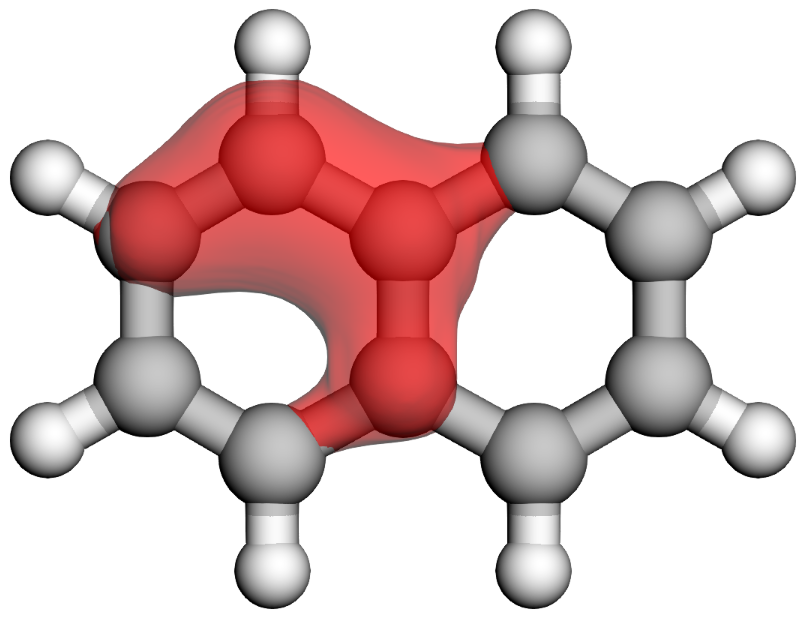}\\
    \end{tabular}
    \caption{Visualizations of three selected one-particle envelopes of naphthalene from the Orbformer model finetuned on azulene and naphthalene.}
    \label{fig:naphthalene-orbitals}
\end{figure}

To showcase the ability of the proposed approach to accurately describe aromaticity and large-scale conjugation, we evaluated the reaction energy of the azulene--naphthalene isomerization using an Orbformer model fine-tuned from the LAC checkpoint.
Equilibrium geometries for naphthalene and azulene were obtained from Mirzaei et al.~\cite{mirzai2022azulene}.
Fine-tuning was carried out jointly for the two geometries, for 80k steps.
The reference reaction energy value was obtained with the W1 composite approach \cite{martin1999towards} at the CCSD(T)/CBS level of theory.
Orbformer's predicted reaction energy converges convincingly within chemical accuracy of the reference 36.8 kcal/mol after 50k fine-tuning iterations.
Sampling errors were not obtained, due to the use of the robust mean in computing the energy expectation values.
Instead the four separate evaluations along the fine-tuning trajectory are intended to demonstrate the reliability of both sampling and training convergence.
The fact that Orbformer can appropriately describe the extended delocalized $\pi$-system of naphthalene indicates that our composable architecture does not hinder modelling the behaviour of electron delocalization at this scale.
In further support of this, Supplementary Fig. \ref{fig:naphthalene-orbitals} visualizes three selected one-particle envelopes of the fine-tuned Orbformer model for naphthalene.
The three orbitals, localized to one, three and four carbon atoms respectively, illustrate how Orbformer can operate on the different length-scales required for the description of the electronic structure of naphthalene.
Further, we recall that Orbformer orbitals are pointwise products between envelopes and transformer representations.
Transformer representations, unlike classical orbitals, are functions of all electron positions and spins rather than of one electron's position. 
To guarantee composability for large separations of subsystems, the distance-weighted attention mechanism employed in Orbformer introduces a finite perceptive field for the interactions.
However, the electron-electron correlation is not directly mediated by the orbital envelope.
This means that, even with a rather localised envelope, some long-range electron correlation can be captured by the transformer representations.

\begin{figure}
    \centering
    \includegraphics[width=0.5\linewidth]{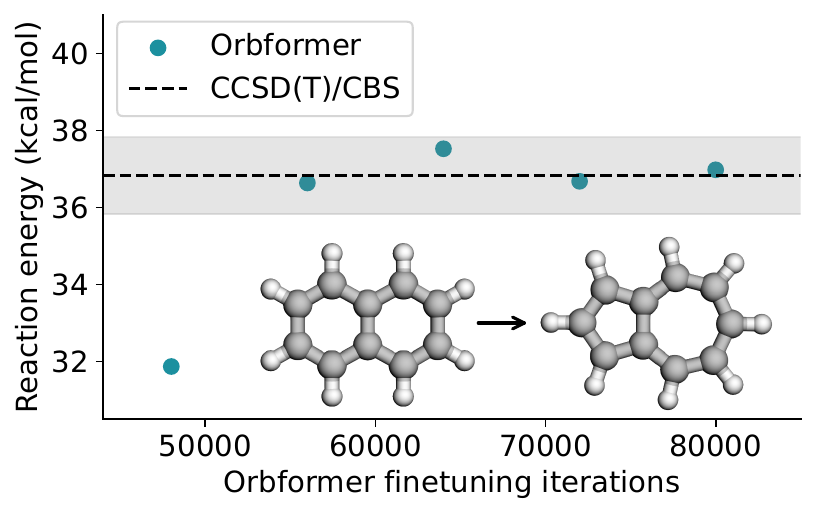}
    \caption{Comparison of the predicted reaction energy for the azulene-naphthalene isomerization between high-level electronic structure methods Orbformer and CCSD(T) at the complete basis set limit.
    The coupled cluster result was computed in-house, using the W1 thermochemical protocol \cite{martin1999towards}. The Orbformer checkpoint pretrained on LAC was finetuned on azulene and naphthalene jointly. Orbformer energies were obtained by computing the robust mean of energies obtained in independent evaluation runs. 
    }
    \label{fig:naphthalene-azulene-convergence}
\end{figure}

\subsection{Atomization energies of small molecules}
\label{sec:app:additional-results:atomization}
\begin{figure}
    \centering
    \includegraphics[width=0.5\linewidth]{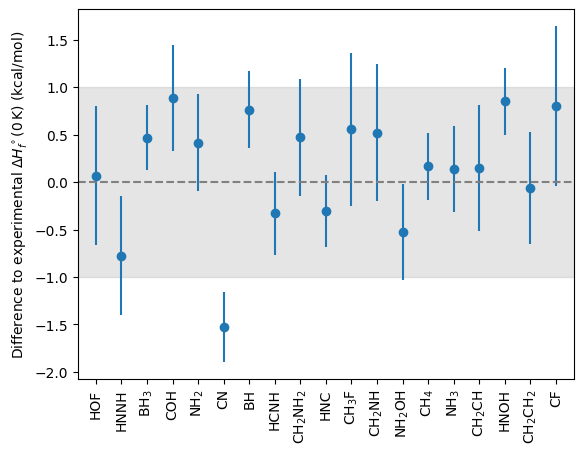}
    \caption{Deviations of the zero-point corrected Orbformer total atomization energies from experimental reference values \cite{ruscic2005active}. Error bars show the statistical error (one standard deviation of the estimates from independent MCMC chains, cf.~Suppl.~Note~\ref{sec:app:eval}). %
    }
    \label{fig:atct}
\end{figure}

While other experiments of this manuscript mainly compare the performance of Orbformer against other theoretical methods, here we provide a comparison of Orbformer's predicted atomization energies against experimental references.
Reference data was collected for 19 molecules from the active thermochemical tables (ATcT) \cite{ruscic2005active}, by subtracting the standard gas phase heat of formation of the molecule from those of the constituent atoms, at 0 K.
Geometries and corrections for zero-point vibrational effects were obtained from the theoretical study benchmarking the W4 composite method \cite{karton2006w4}.
Orbformer was finetuned from the LAC checkpoint independently for each considered molecule.
The deviations of the zero-point corrected Orbformer total atomization energies (TAE) from experimental references are displayed on Supplementary Fig. \ref{fig:atct}.
All but one of the Orbformer values fall within $\pm 1$ kcal/mol of the experimental value. %
The MAE of 0.51 kcal/mol across the 19 molecules engenders confidence in the quality of Orbformer results as compared to experiment. %

\section{Traditional electronic structure calculation details}
\label{sec:app:reference_methods}
\subsection{Energy calculations}
To compare the efficiency and accuracy of Orbformer, we perform traditional electronic structure calculations at various costs and accuracy using commonly used theories. Without human expert design and intervention, all the calculations are run under a practitioner perspective to have a more fair comparison. Without any further specification, all the SCF calculations, including KS-DFT \cite{ks-dft}, restricted open-shell determinant (ROHF) \cite{rohf} and active space SCF method (CASSCF) \cite{CASSCF}, are computed using an energy convergence threshold of $10^{-10}$ a.u. and gradient convergence threshold to $10^{-6}$ a.u.~with a maximum SCF cycle of 500. All the post-SCF calculations, including unrestricted coupled-cluster with single, double and a perturbative treatment of triple excitations (UCCSD(T)) \cite{rohf-ccsdt1}, the second-order $n$-electron valence state perturbation theory (NEVPT2) \cite{NEVPT2}, and density-fitted uncontracted multireference configuration interaction with single and double excitations and Davidson size-extensivity correction (DF-MRCISD+Q) \cite{mrci},
are computed using an energy convergence threshold of $10^{-10}$ a.u.~and a maximum iteration of 600.

In this study, all the complete basis set (CBS) extrapolations are performed with a two-point extrapolation strategy used in \textsc{ORCA 5.0} \cite{ORCA5} in equations~\ref{eq:scf_extrap} and \ref{eq:corr_extrap}. The final total energy is computed as $\energy^{\infty}_{\text{total}}=\energy^{\infty}_{\text{SCF}}+\energy^{\infty}_{\text{corr}}$
\begin{equation}
\label{eq:scf_extrap}
\energy^{X}_{\text{SCF}} = \energy^{\infty}_{\text{SCF}} + A\exp(-\alpha\sqrt{X}),
\end{equation}
where $X$ is the basis set cardinal number and in this study we use aug-cc-pVDZ ($X=2$) and aug-cc-pVTZ ($X=3$) basis sets \cite{aug-cc-basis}, $A$ is a parameter to be computed, $\alpha=4.3$ is a basis-set dependent constant \cite{CBS_alpha_beta}, and $\energy^{\infty}_{\text{SCF}}$ is the corresponding SCF energy at CBS limit. We compute the extrapolated correlation energy as
\begin{equation}
\label{eq:corr_extrap}
\energy^{\infty}_{\text{corr}}= \frac{X^\beta \energy^{X}_{\text{corr}}-Y^\beta \energy^{X}_{\text{corr}}}{X^\beta -Y^\beta},
\end{equation}
where $\beta=2.51$ is a basis-set dependent constant \cite{CBS_alpha_beta}, $X=2, Y=3$ are the basis set cardinal numbers for aug-cc-pVDZ and aug-cc-pVTZ, respectively, in this study, and $\energy^{\infty}_{\text{corr}}$ is the corresponding correlation energy at CBS limit.

The computational details for each theory at specific basis sets are as follows:

\begin{itemize}
    \item UB3LYP-D3(BJ)/Def2-QZVPP: The spin-unrestricted B3LYP\cite{B3LYP1,B3LYP2,B3LYP3,B3LYP4} plus D3 dispersion correction with Becke–Johnson damping \cite{D3,D3BJ} with the Def2-QZVPP basis set  \cite{def2-basis2} calculations are performed by \textsc{PySCF 2.7.0} \cite{sun2018pyscf} with \textsc{PySCF-dispersion 1.3.0} plug-in, which interferences \textsc{Simple-DFTD3 1.2.1} \cite{simpled3}. 
    \item UM06-2X-D3/Def2-QZVPP: The spin-unrestricted M06-2X \cite{M06-2X} plus D3 dispersion correction \cite{D3,M06-2X-D31} with the Def2-QZVPP basis set \cite{def2-basis2} calculations are performed by \textsc{PySCF 2.7.0} \cite{sun2018pyscf} with \textsc{PySCF-dispersion 1.3.0} plug-in, which interferences \textsc{Simple-DFTD3 1.2.1}\cite{simpled3}. Note that M06-X has already included mid-range dispersion correction in the parametrization, but literature study suggests a D3 correction could still benefit the calculation \cite{M06-2X-D32}.
    \item U$\omega$B97X-V-D3(BJ)/Def2-QZVPP: The spin-unrestricted $\omega$B97X-V \cite{wB97XV1,wB97XV2} plus D3 dispersion correction with Becke–Johnson damping \cite{D3,D3BJ} with the Def2-QZVPP basis set \cite{def2-basis2} calculations are performed by \textsc{PySCF 2.7.0} \cite{sun2018pyscf} with \textsc{PySCF-dispersion 1.3.0} plug-in, which interferences \textsc{Simple-DFTD3 1.2.1} \cite{simpled3}. 
    \item UCCSD(T)/CBS: The UCCSD(T) with aug-cc-pVDZ and aug-cc-pVTZ basis set calculations are performed using ROHF wavefunctions \cite{rohf-ccsdt1,rohf-ccsdt2}. The two-point CBS extrapolation then is performed by separately extrapolating the SCF and correlation contributions via equations~\ref{eq:scf_extrap} and \ref{eq:corr_extrap}, respectively.
    \item NEVPT2/CASSCF(10,10)/CBS: In this study, NEVPT2 calculations are computed \cite{NEVPT2} using the strongly contracted (SC) internal contraction scheme \cite{SC-NEVPT21,SC-NEVPT22} implemented in \textsc{PySCF 2.7.0} \cite{sun2018pyscf} with a CASSCF wavefunction. CASSCF is solved by the FCI solver \cite{FCI}  with 10 electron and 10 active orbitals in the active space and the initial guess of the CASSCF is using ROHF wavefuction. To automatically select an active space with a consistent active space size, we use the initial guess strategy introduced in PySCF, which is not the best but a practical approach to select active space. After ROHF calculations, we localize orbitals for occupied and virtual spaces separately via Pipek-Mezey (PM) localization \cite{PM}. The orbitals then are sorted by their MO coefficients and the top 10 MOs are picked as the active space.
    \item DF-MRCISD+Q/aug-cc-pVDZ: We note that due to inefficient memory for L-Alanine dissociation MEP calculations with aug-cc-pVTZ basis set, only calculations using aug-cc-pVDZ basis set were performed. All MRCISD+Q calculations are performed using frozen-core uncontracted MRCI solved by a Davidson solver implemented in \textsc{ORCA 5.0} \cite{ORCA5}. As is common practice in electronic structure calculations, we apply the density fitting (DF) approximation, also known as the resolution of identity (RI), with the automatically generated auxiliary basis set \cite{autojk} to speed up the calculations. CASSCF is solved by the default solver and active space selection strategy implemented in \textsc{ORCA 5.0} with 10 active electrons and 10 active orbitals in the active space and a DF-ROHF wavefuction as initial guess. 
    \item DF-MRAQCC/aug-cc-pVDZ: Similar to the DF-MRCISD+Q/aug-cc-pVDZ calculations, we perform frozen-core uncontracted MRAQCC calculations solved by DIIS solver via \textsc{ORCA 5.0} \cite{ORCA5} with CASSCF wavefunctions initialized by ROHF solutions. Active space selection is performed by the default strategy in \textsc{ORCA 5.0} an active space with 10 active electrons and 10 active orbitals.
\end{itemize}

\section{Extended discussion}
\label{sec:app:extended_discussion}
Looking to future developments that may follow on from Orbformer, we saw that pretraining reduces the amount of fine-tuning required to achieve chemical accuracy, often very substantially, yet in very few cases was it possible to evaluate energies directly from the pretrained model with no fine-tuning.
We believe that with a further improved and much larger model than Orbformer, it will be possible to converge to chemical accuracy across a dataset as large as the LAC without the need for fine-tuning.
Yet increasing model capacity without significantly slowing down training and remaining within memory constraints of current hardware presents a serious challenge.
One direction would be to do more processing that is independent of the electron positions, since any computations here bypass the Laplacian.
It would also be beneficial to improve the single-determinant version of the model to match the current 16-determinant Orbformer, since a single-determinant model is size consistent by construction, and this may lead to improvements in generalization.
Engineering development of the K-FAC \citep{martens2015optimizing} optimizer would also be necessary for further scaling of this technology.
Another axis of development would be to extend Orbformer to operate on charged systems and systems with a total $z$-spin that is different to 0 or $\tfrac{1}{2}$.
Here, the localization of Orbformer orbitals presents a conceptual challenge, since it is unclear where in space any orbitals should be added or removed when forming an ion. For certain molecules, say \ch{R-COO-}, we can form a good guess as to where the negative charge will be localized. However, for a more obscure ion such as any \ch{R-}, the new orbital may not be localized at all.
The case of spin is likely to be somewhat easier to handle, as the total number of orbitals does not change. We envisage that communicating the overall spin of the molecule to all the molecule-dependent parts of the network would allow it to correctly adapt to changes in the overall molecular spin.

\end{document}